\documentclass[twocolumn,twoside,english,aps, nofootinbib, prd, lengthcheck, superscriptaddress,tightenlines,showkeys,
showkeywords,amsmath,amssym]{revtex4-2}
\usepackage[switch,columnwise]{lineno}
\usepackage{lipsum} 
\usepackage{fancyhdr}
\fancypagestyle{firstpage}{%
  \lhead{\it{\footnotesize{Astrophysical Bulletin, 2023, Vol. 78, No. 2}}}
  \chead{\footnotesize{Preprint \today}}
  \rhead{\footnotesize{ISSN 1990-3413}}
}

\usepackage[T2A]{fontenc}
\usepackage[utf8x]{inputenc}
\usepackage{multirow}
\usepackage{xcolor}

\definecolor{graylink}{rgb}{0.0, 0.4, 0.0}
\definecolor{lightciteblue}{rgb}{0.0, 0.7, 0.85}
\usepackage[colorlinks=false,linktocpage=true,linkbordercolor=red,citebordercolor=lightciteblue,urlbordercolor=blue]{hyperref}

\usepackage{verbatim}
\usepackage{bm}
\usepackage{amsmath}
\usepackage{amssymb}
\usepackage{graphicx}
\usepackage{slashed}
\usepackage{wrapfig}
\usepackage{bbm}
\usepackage{dcolumn}% Align table columns on decimal point
\usepackage{bm}% bold math

\makeatletter
\def\figurename{Figure}
\def\tablename{Table}

\renewcommand \thefigure{\@arabic\c@figure}
\renewcommand \thetable{\@arabic\c@table}
\def\fnum@table{{\bfseries \tablename~\thetable}}%
\def\fnum@figure{{\bfseries \figurename~\thefigure}}

\def\bibsection{%
 \@ifx@empty\refname{%
  \par
 }{%
  \section*{\refname}%
 }%
}%

\def\p@subsection {}
\def\p@subsubsection {}

\makeatother

\usepackage[margin={5pt,5pt},font=small,labelsep=period]{caption}
\usepackage{aas_macros}
\setcitestyle{authoryear,round,aysep={,},semicolon}
\begin{document}


%\titlerunning{SPECKLE INTERFEROMETRY WITH  CMOS DETECTOR}
%\authorrunning{STRAKHOV et al.}
%\toctitle{Speckle Interferometry with  CMOS Detector}

%\tocauthor{I.~A.~Strakhov, B.~S.~Safonov, and D.~V.~Cheryasov}

\title{Speckle Interferometry with  CMOS Detector}

\author{\firstname{I.~A.}~\surname{Strakhov}}
 \email{strakhov.ia15@physics.msu.ru}
 \affiliation{Sternberg Astronomical Institute of
 Lomonosov Moscow State University,
Moscow, 119234 Russia}

\author{\firstname{B.~S.}~\surname{Safonov}}
 \affiliation{Sternberg Astronomical Institute of
 Lomonosov Moscow State University,
Moscow, 119234 Russia}

\author{\firstname{D.~V.}~\surname{Cheryasov}}
 \affiliation{Sternberg Astronomical Institute of
 Lomonosov Moscow State University,
Moscow, 119234 Russia}

\received{February 14, 2023} \revised{March 20, 2023} \accepted{March 23, 2023}

\begin{abstract}
In 2022 we carried out an upgrade of the speckle polarimeter
(SPP)---the facility instrument of the 2.5-m telescope of the
Caucasian Observatory of the SAI MSU.
During the overhaul, CMOS Hamamatsu ORCA-Quest qCMOS
C15550-20UP was installed as the main detector, some drawback of the previous version of
the instrument were eliminated. In this paper, we
present a description of the instrument, as well as study some
features of the CMOS detector and ways to take them into account
in speckle interferometric processing. Quantitative comparison of
CMOS and EMCCD in the context of speckle interferometry is
performed using numerical simulation of the detection
process. Speckle interferometric observations of 25 young variable
stars are given as an example of astronomical result. It was
found that BM\,And is a binary system with a separation of
273~mas. The variability of the system is dominated by the
brightness variations of the main component. A binary system was
also found in NSV\,16694 (TYC 120-876-1). The separation of this
system is 202~mas.

\end{abstract}

\maketitle
\thispagestyle{firstpage}
\section{INTRODUCTION}
Passive methods for achieving diffraction limited angular resolution on
ground-based telescopes, such as speckle interferometry, lucky imaging,
differential speckle polarimetry, operate with a large
number of short-exposure images of an object distorted by
atmospheric turbulence \citep{Labeyrie1970,Tokovinin1988}. The need
to ``freeze'' an image that changes with a characteristic time
scale of $\tau_0$, called the atmospheric coherence time,
determines the characteristic exposures that are applied:
$(3$--$5)\tau_0=10$--$30$~ms.

In order to apply passive high angular resolution techniques to the
observation of astronomical objects that typically have very low
fluxes, it is critical to use detectors that have high quantum
efficiency, high readout speed, and low readout noise. The last
two qualities seem to be mutually exclusive, since faster output
amplifier operation means higher readout noise, all other things
being equal \citep{Howell2000}.

Various technologies have been used to overcome this
contradiction, but a breakthrough was made with the invention in
the late 1990s-early 2000s of the Electron Multiplying
Charge--Coupled Device (EMCCD) \citep{Basden2004}. \mbox{EMCCDs}
differ from conventional CCD detectors by the presence of a special
 electron multiplication register between the horizontal register and the output amplifier. The electron multiplication register is a sequence of $N$ cells between
which an increased potential is created. There is a probability
\mbox{$p_{\mathrm{EM}}\approx10^{-3}$} that a photoelectron, passing
between two successive cells of the electron multiplication register, will knock out
one more electron. Passing through $N\approx600$ cells of the electron multiplication register, the photoelectron turns into
$G_\mathrm{EM}=(1+p_{\mathrm{EM}})^N$ electrons, on average, where
$G_\mathrm{EM}$ is the so-called electron multiplication factor, which for
commercially available EMCCDs can vary from 1 to 1000.

%fig1
\begin{figure}[b]
%\setcaptionmargin{5mm} \onelinecaptionsfalse \captionstyle{normal}
%\captionstyle{normal}
\includegraphics[width=0.935\linewidth] {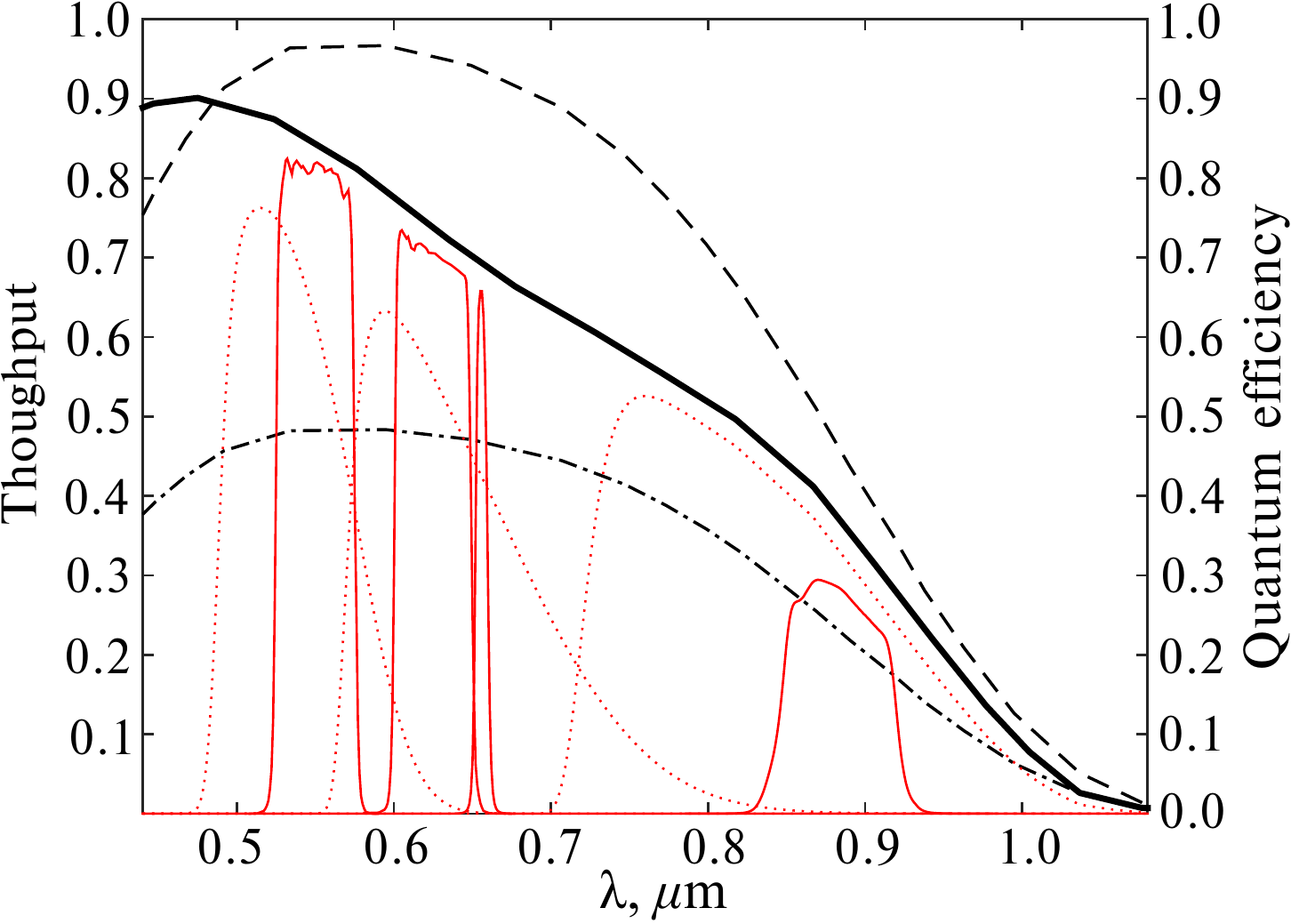} %{SpeckleCMOS_figs/bands/ORCA_bands.eps} }
\caption{Quantum efficiency of detectors: EMCCD vs. CMOS, and transmission of filters. The dashed line
is the quantum efficiency curve of
the EMCCD Andor iXon 897 detector. The dash-dotted line is the same, but
the quantum efficiencies are halved to account for amplification noise. The
thick solid line is the quantum efficiency curve of the Hamamatsu
ORCA-quest CMOS detector. The red lines are the transmission curves
of the filters used in the speckle polarimeter multiplied by the
Hamamatsu ORCA-quest quantum efficiency curve. The red dotted
lines are broadband filters, from left to right: $V$,
$R_\mathrm{c}$, $I_\mathrm{c}$. The red solid lines are midband
filters: 550, 625, 656.3 (H$_\alpha$), 880~nm.}
\label{fig:1}
%\vspace{-20pt}
\end{figure}

\renewcommand{\baselinestretch}{1.0}
\begin{table*}[]
%\setcaptionmargin{0mm}\onelinecaptionstrue \captionstyle{normal}
\caption{
Parameters of detectors for speckle interferometry. The first
column is EMCCD Andor iXon 897, a detector used in the speckle
polarimeter from October 2015 to July 2022. The second and third
columns are standard and ultra-quiet modes of the Hamamatsu
ORCA-quest CMOS detector installed in the speckle polarimeter,
starting from August 2022.
\label{tab:1} }
\centering
\vspace{5pt}
\renewcommand{\arraystretch}{1.6} % Default value: 1
\begin{tabular}{l|c|cc}
\hline
\multicolumn{1}{c|}{\multirow{2}{*}{Parameter}}& Andor   & \multicolumn{2}{c}{Hamamatsu ORCA-quest}  \\
\cline{3-4}
                                              & iXon 897& \multicolumn{1}{c|}{~\,\,\;standard\,\,\;~}  & \multicolumn{1}{c}{ultra-quiet}  \\
\hline
Technology         &   EMCCD        &   \multicolumn{2}{c}{qCMOS} \\
Dimensions, px     & $512\times512$ & \multicolumn{2}{c}{$4096\times2304$} \\
Pixel size, microns& 16             & \multicolumn{2}{c}{4.6}              \\
RMSD of readout noise, $e^{-}$&      48        &  \multicolumn{1}{c|}{0.43}           & 0.27   \\
Effective RMSD of readout noise, $e^{-}$ &      0.048    &  \multicolumn{1}{c|}{0.43}           & 0.27   \\
CIC, $e^{-}$/px &   0.045  &   \multicolumn{2}{c}{0}     \\
EM gain factor $G_\mathrm{EM}$ &   1--1000    &  \multicolumn{2}{c}{1}      \\
Conversion factor, $e^{-}$/ADU &   11.57     &  \multicolumn{2}{c}{0.107}   \\
Register size, ADU    &   16\,384        &  \multicolumn{2}{c}{65\,536} \\
Potential well depth, $e^{-}$&  180\,000        &  \multicolumn{2}{c}{7000} \\
Frame rate (full frame), Hz       &   35           &  \multicolumn{1}{c|}{120}  & 5      \\
Frame rate (region $512\times512$~px), Hz &   35       &  \multicolumn{1}{c|}{532}  & 22     \\
\hline
\end{tabular}
\renewcommand{\arraystretch}{1.0} % Default value: 1
\end{table*}
\renewcommand{\baselinestretch}{1.0}

Due to the fact that the signal is amplified before being
digitized, the effective readout noise is reduced by a factor of
$G_\mathrm{EM}$. In practice, the EMCCD readout noise is reduced
to negligible values: \mbox{$0.01$--$0.1e^{-}$}. Strong
suppression of readout noise makes it possible to use the output
amplifier at frequencies that are orders of magnitude higher than
the readout frequencies of classical CCDs: 10--30~MHz. The EMCCD
is usually based on a back-illuminated chip and has a quantum
efficiency of about 90\% over a very wide wavelength range (see
example in Fig.~\ref{fig:1}).

All of these properties have made EMCCD extremely popular for
implementing speckle interferometry and similar methods \citep{Law2006,Hormuth2008,Oscoz2008,Maksimov2009,Tokovinin2010,Scott2018}. We also
used the EMCCD Andor iXon 897 as the main detector of the speckle
polarimeter, the facility instrument of the 2.5-m telescope of the
Caucasian Observatory of the SAI MSU \citep{Safonov2017}, the detector parameters are given in Table~\ref{tab:1}.

EMCCDs have some known disadvantages. \linebreak \mbox{Firstly}, this is the
so-called amplification noise: the amplification process is
stochastic in nature, and the actual electron multiplication factor
for each particular photoelectron is a random variable. This is expressed
as a twofold increase in the photon noise variance \citep{Harpsoe2012}. For conditions where photon noise is the dominant
contributor to the total noise, this means that twice as many photons must be
accumulated to achieve the same $S/N$ ratio.

Secondly, when the charge is clocked to the  electron multiplication register, there
is a non-zero probability that a spurious
 electron will appear
(Clock-induced charge---CIC). This electron will be amplified and
it will no longer be possible to distinguish it from the
photoelectron knocked out by the source photon. For the iXon 897
detector used in the speckle polarimeter, the probability of
registering a  spurious electron is $0.045$/px.

Another significant disadvantage of EMCCDs is their low dynamic
range. For example, when working with \mbox{$G_\mathrm{EM}=500$},
the maximum number of photons that can be registered by a given
pixel before saturation is 350. The problem is aggravated by the
fact that the use of a detector in conditions where  at
least one pixel is saturated leads to accelerated degradation of
the electron multiplication register. Thus, the use of the standard CCD photometry
strategy when bright stars are saturated, but this does not spoil
the photometry of faint stars, turns out to be impossible for
EMCCD.

Note that although in speckle interferometry we usually deal with
one source per frame, the range of brightness in the speckle
pattern is quite large. The value of the EM gain has to be
chosen so that the brightest speckle is not saturated. If this
gain turns out to be low, then for faint speckles, which are the majority,
 the detector operates with readout noise which is much larger than subelectron.

As an alternative to CCD detectors, CMOS detectors are used, which
in recent years have been approaching, and in some cases even
surpassing, CCDs in terms of performance. Technologically,
CMOS, unlike CCD, reads each pixel individually. As a result, each
specific ADC (Analog-to-Digital Converter) can operate relatively slowly, and, accordingly, have
low readout noise.

Examples of the use of CMOS detectors in speckle interferometry
can be found in \cite{Genet2016} and \cite{Wasson2017}.
However, the detectors used by the authors have a read noise of
about $1{e}^{-}$ and are inferior to EMCCD.

%fig2
\begin{figure*}
 %%\setcaptionmargin{5mm} \onelinecaptionsfalse \captionstyle{normal}
\vspace{-0.1cm}
\includegraphics[width=0.85\linewidth, bb= 80 450 7900 5400,clip]{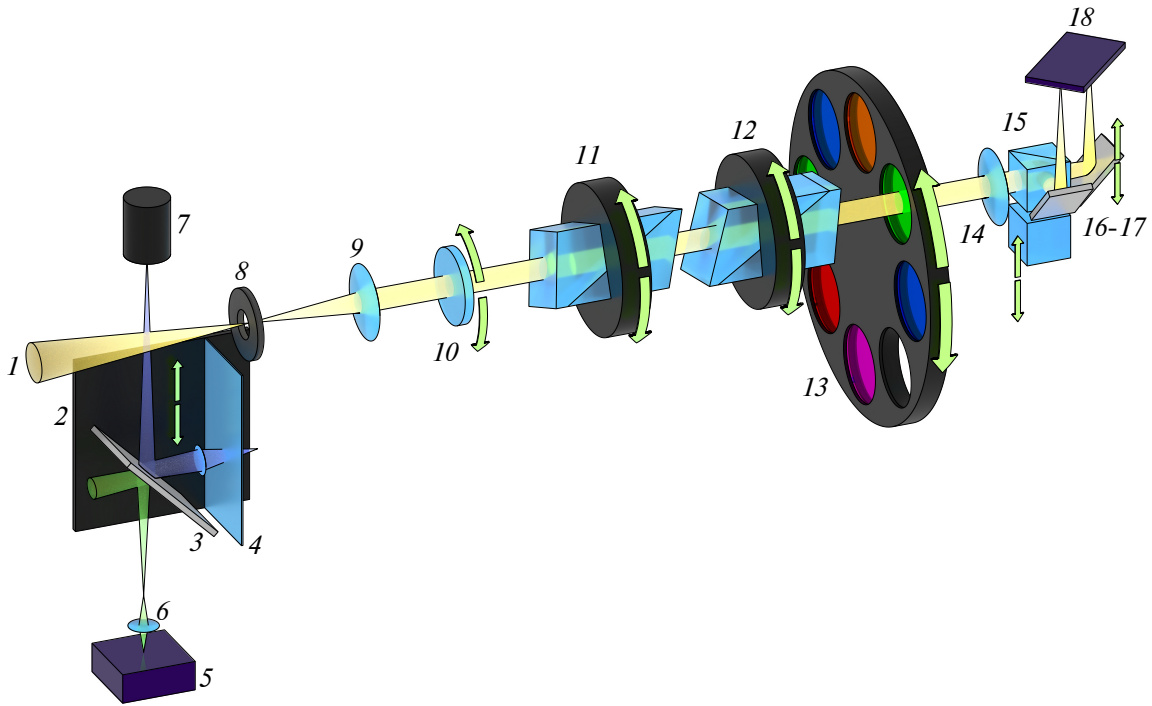}% {SpeckleCMOS_figs/scheme/principalSchemeBlender_compressed.pdf} }
\caption{Instrument layout. {\it 1}---Beam from a telescope.
The prefocal unit {\it2--8}:
the movable carriage {\it 2}, %carries
mirrors {\it 3}, one of the mirrors relays  light from the
telescope to the auxiliary camera ({\it5--6}), another mirror
relays light from the calibration source ({\it7}) to the
detector, {\it4}---the linear polarizer, {\it8}---the field
aperture; the collimated beam unit {\it9--14}: {\it9}---the
collimator, {\it10}---the rotating half-wave plate, {\it11,
12}---the prisms of the atmospheric dispersion compensator
(the wedge angles are increased for clarity), {\it13}---the
filter wheel, {\it14}---the lens; the camera unit
{\it15--18}: {\it15}---the beam splitter, {\it16, 17}---the
relay mirrors, {\it18}---the detector. The green arrows mark
the motorized degrees of freedom. }
%\vspace{-20pt}
\label{fig:2}
\end{figure*}

In Table~\ref{tab:1}, we list the parameters of the Hamamatsu ORCA-Quest
C15550-20UP CMOS detector (hereinafter simply Hamamatsu
ORCA-Quest). It can be seen that the key characteristics important
for speckle interferometry---read speed and readout noise---are
comparable to EMCCD. Sub-electronic readout noise is achieved
without amplification and therefore without the problems
associated with it: amplification noise and low dynamic range.

In Fig.~\ref{fig:1} we compare the quantum efficiency curves of Andor iXon
897 and Hamamatsu ORCA-Quest. It can be seen that the CMOS quantum
efficiency curve is slightly lower than that of CCD, but this
advantage disappears when the presence of amplification noise in the latter
is taken into account.

The Hamamatsu ORCA-Quest detector was chosen by us as the main
detector of the speckle polarimeter, which was installed instead
of the EMCCD. The replacement of the detector required a
significant overhaul of the instrument design. Taking advantage of
this moment, we made several key changes along the way, which made
it possible to increase the efficiency of speckle interferometric
observations. Changes in instrument design are described in
Section~\ref{sec:design}.

In Section~\ref{sec:proc}, we describe some features of the application of CMOS
in the context of speckle interferometry. We have performed a
quantitative comparison of the efficiency of EMCCD and CMOS using
our numerical measurement model (Section~\ref{sec:model}). Finally, in Section~\ref{sec:ress}
we present some results of the study of the binarity of the UX Ori type stars,
 carried out both with EMCCD and CMOS. In particular, the
binarity of BM\,And and NSV\,16694 was discovered. In this
article, we will not discuss the polarimetric mode of the instrument;
this will be done in the future paper.

\section{INSTRUMENT}
\label{sec:design}

\begin{figure}
 %%\setcaptionmargin{5mm} \onelinecaptionsfalse \captionstyle{normal}
\includegraphics[width=\linewidth] {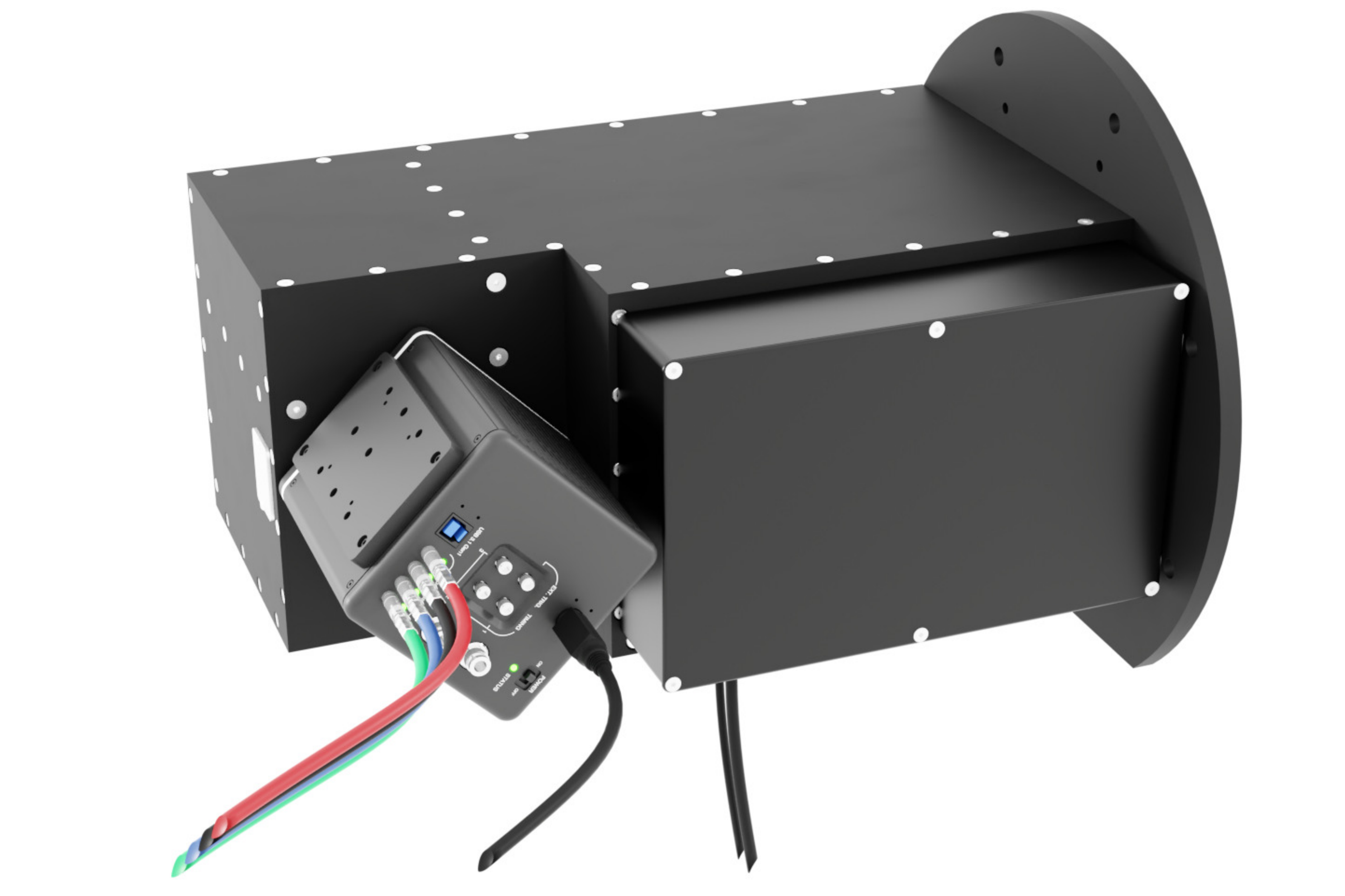}%{SpeckleCMOS_figs/scheme/SPP2_render_whole.pdf} }
%\vspace{-10pt}
\caption{General view of the instrument from the side of the main
detector (render of CAD model). Direction to the telescope---to the right, flange to
the telescope.}
\label{fig:3}
%\vspace{-20pt}
\end{figure}

The design of the speckle polarimeter was discussed in detail
earlier in the article \citep{Safonov2017}. Here we will discuss,
first of all, the changes that was applied in the context of
the installation of Hamamatsu ORCA-quest as the main detector. The
instrument consists of the following units: a prefocal unit, a
collimated beam unit, a camera unit, a control electronics unit, a
control computer (see Fig.~\ref{fig:2} and Fig.~\ref{fig:3}).

%fig3

\subsection{Prefocal unit}

\begin{figure}[t!]\vspace{3pt}
 %\setcaptionmargin{5mm} \onelinecaptionsfalse \captionstyle{normal}
\includegraphics[width=0.93\linewidth]{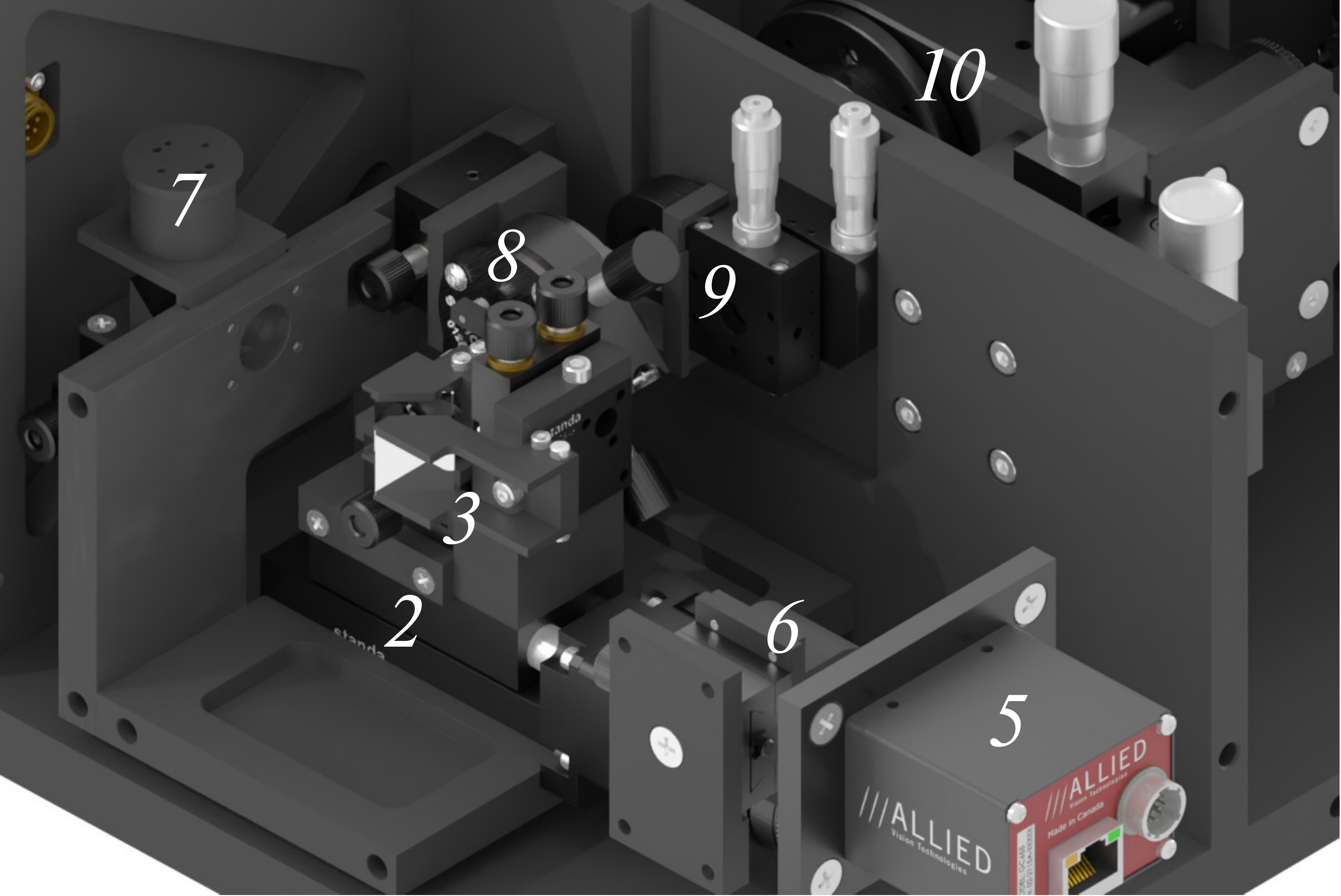} %{SpeckleCMOS_figs/scheme/SPP2_render_prefocal.eps}
%\vspace{3mm}
\caption{The prefocal unit (render of CAD model). The designations are the same as in
Fig.~\ref{fig:2}.}
\label{fig:4}
%\vspace{-20pt}
\end{figure}

The prefocal unit (Fig.~\ref{fig:4}) performs a number of auxiliary
functions. It contains an auxiliary CCD camera, onto which light
from the telescope is relayed by introducing a diagonal mirror
into the beam. The purpose of auxiliary camera is to center object when the telescope's pointing error exceeds
$15^{\prime\prime}$ and the object does not fit within the field
of view of the main detector. Since 2022, we have been using an
industrial black-and-white CCD detector Prosilica GC655 as an
auxiliary camera, which made it possible to center objects up to
$m_V=17^{\rm m}$. The camera has a field of view
$70^{\prime\prime}$.

\begin{figure}[] %\setcaptionmargin{5mm} \onelinecaptionsfalse \captionstyle{normal}
\includegraphics[width=0.87\linewidth]{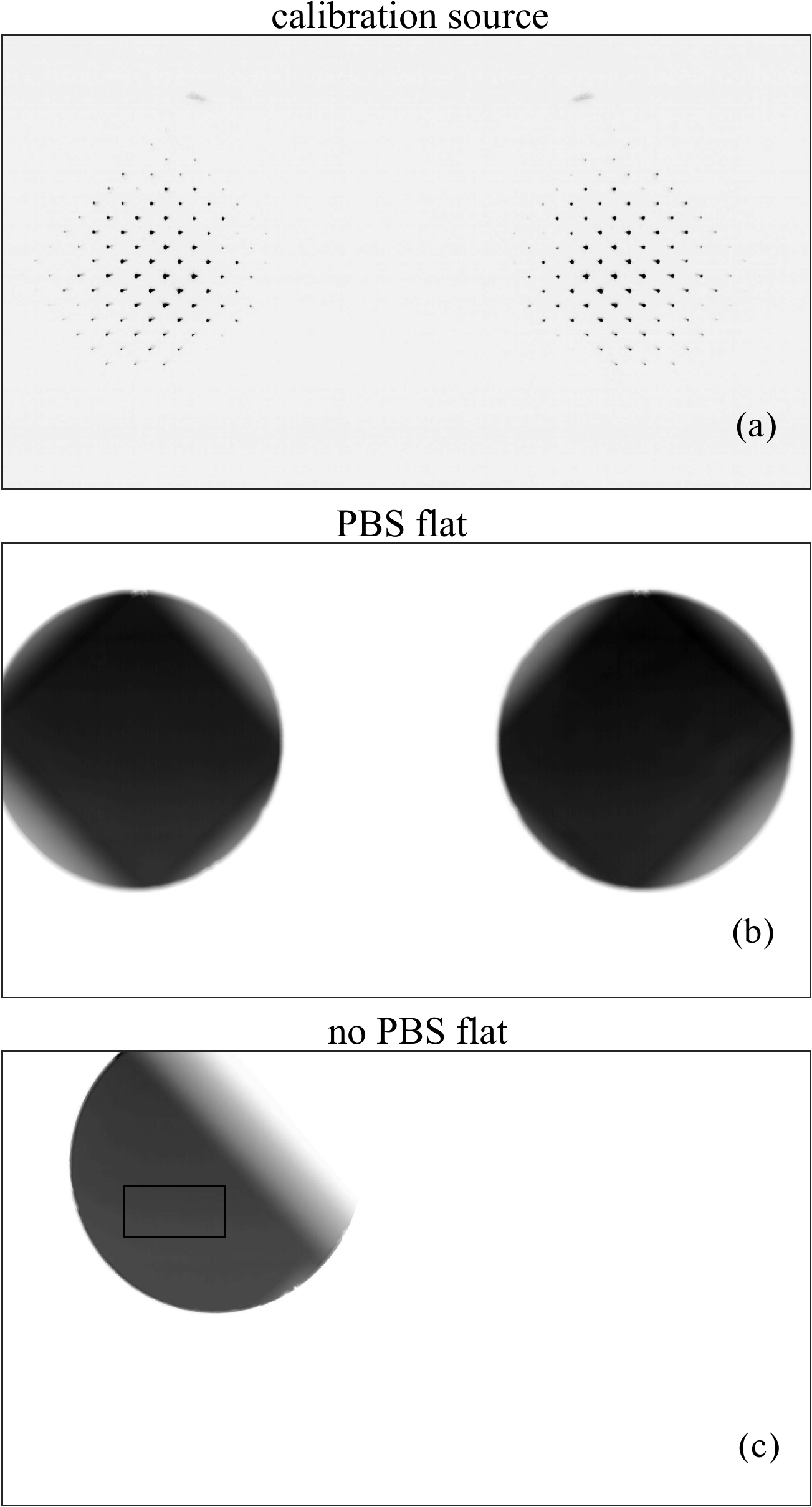}%{SpeckleCMOS_figs/flats/flats_demo.eps} }
%\vspace{-7pt}
\caption{Image examples: (a)---a calibration source; (b)---a flat
field frame when operating with the Polarizing Beamsplitter
Cube---PBS; (c)---a flat field frame when operating without the
PBS (speckle interferometry mode). The black rectangle on the
panel shows the detector area \mbox{$512\times256$}~px, which is
read in the speckle interferometry mode.} \label{fig:5}

\end{figure}

In addition, the prefocal unit includes a calibration source,
which is represented by an array of holes with a diameter of
70~$\mu$m, with a step of 800~$\mu$m, illuminated by a white superbright
LED through an additional diffuser. The image of the calibration
source is relayed into the focal plane with $3.5\times$  demagnification 
through a two-lens achromat and a moving mirror (see
%Fig.~7a for an image example).
Fig.~\ref{fig:5}a for an image example).
%fig4

%fig5

The prefocal unit also contains a linear polarizer that can be
inserted into the beam to calibrate the angle of rotation of the
half-wave plate. The relay mirrors of the auxiliary detector,
the calibration source, and the linear polarizer are located on
the Standa 8MT173-30 linear translator (stepper motor + ball
guide).

An important element of the prefocal unit is the field aperture,
which is a hole with a diameter of 3.5~mm, which corresponds to
$35^{\prime\prime}$ on the celestial sphere. The field aperture is
permanently set. Note that not the entire field of view cut out by
the field stop is non-vignetted, this issue is discussed below in
Section~\ref{sec:camera_unit}.

\subsection{Collimated beam unit}

From the main focal plane begins the main part of the optics of
the instrument, the main purpose of which is the formation of images
with specified characteristics on the detector. The key elements
in the optical design are the collimator, which forms a collimated beam,
 and the lens, which focuses this beam on the detector. The
focal lengths of the collimator and objective are 38.1 and 88.9~
mm, respectively. The scheme provides a final magnification of
$2.33\times$, which corresponds to a scale of $20.33$~mas/px for
the main detector pixel size of 4.6~$\mu$m. This magnification ensures
optimal sampling of the focal plane at wavelengths of
493~nm and more, which is a necessary condition for the
implementation of speckle interferometry. The exact value of the
scale and rotation angle of the detector is determined by
observing binary stars for which the coordinates of the components
are known from Gaia~DR3 \citep{Vallenari2022}.

Behind the collimator in the plane of the exit pupil in the Standa
8MRU-1 rotation drive, a Thorlabs SAHWP05M-700 half-wave plate is
installed, which acts as a modulator in the polarimetry mode. The
half-wave plate is driven by a stepper motor through a belt drive
(Standa 8MRU).

Behind the half-wave plate there is an atmospheric dispersion
compensator, which consists of two direct vision prisms installed
in independently rotating Standa 8MR151 drives (stepper motor +
worm gear). Each prism, in turn, consists of two prisms made of
LZOS F1 and LZOS K8 glass, with wedge angles $15\,.\!\!^\circ9$
and $18\,.\!\!^\circ65$, respectively. The aperture of the prisms
is \mbox{$15\times15$}~mm. The prisms are manufactured by
RIVoptics.

These prisms have been used by us since mid-2018. Previously, we
used prisms with much larger wedge angles, which resulted in a
significant distortion. The prisms in operation nowadays have a
distortion of 0.72\%, it is corrected during processing.

During observations, the rotation angles of the prisms are chosen
so that the total dispersion introduced by them is equal in
absolute value, and opposite in direction---to atmospheric
dispersion \citep{Safonov2017}. Atmospheric dispersion is
calculated using the formulas from \cite{Owens1967}, based on the
current parameters of the atmosphere: pressure, temperature,
humidity, which are measured by a weather station. The available
prisms in the current configuration make it possible to compensate
for atmospheric dispersion at altitudes above $28^{\circ}$ (the
most critical band is 880~nm).

A 12-position filter wheel is installed after the prisms of the atmospheric dispersion compensator. It is controlled by a Standa 8MR174 rotation drive (stepper motor + worm gear). Currently, 7 filters are used: $V$, $R_c$, $I_c$, mid-band filters
centered at wavelengths of 550, 625, 880~nm, with half-widths of
50, 50, and 80~nm, respectively. An H$\alpha$ filter with a
half-width of 8~nm is also installed. The filter bandwidths are
shown in Fig.~\ref{fig:1}.
%\endinput

\subsection{Camera unit}
\label{sec:camera_unit}

%fig6
\begin{figure}[b!] \vspace{5pt}
%\setcaptionmargin{5mm} \onelinecaptionsfalse \captionstyle{normal}
\includegraphics[width=0.96\linewidth] {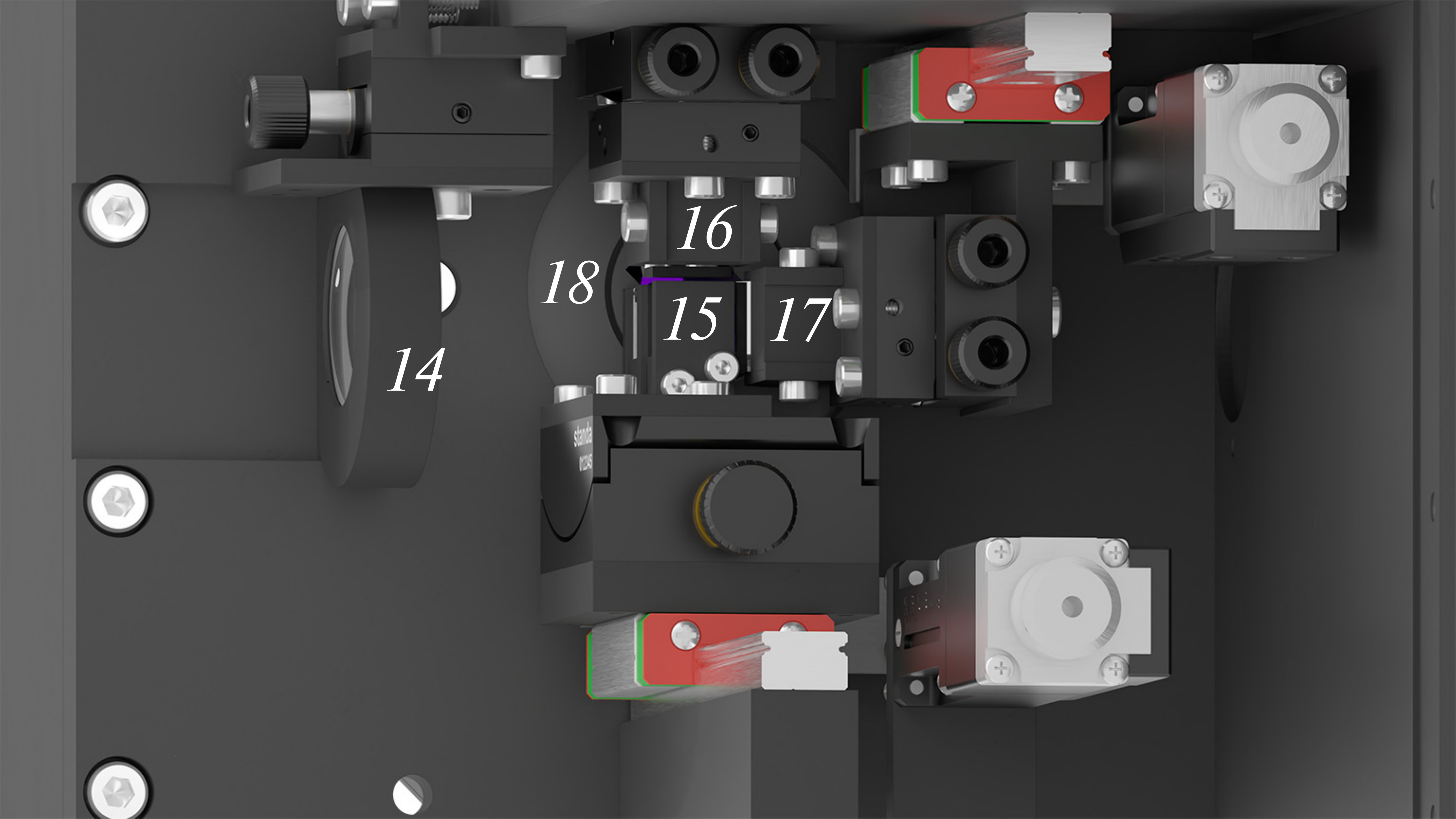}%{SpeckleCMOS_figs/scheme/SPP2_render_camera.eps}
\caption{The camera unit (render of CAD model). The designations are the same as in Fig.~\ref{fig:2}.}
\label{fig:6}

%\vspace{-20pt}
\end{figure}
%fig7
\begin{figure}[t] %\setcaptionmargin{5mm} \onelinecaptionsfalse \captionstyle{normal}
\begin{center}
\center{\includegraphics[width=0.9\linewidth] {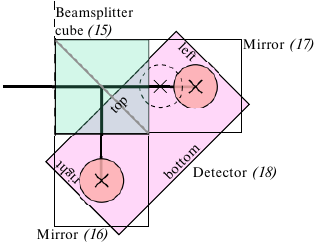}}%{SpeckleCMOS_figs/scheme/baseline_2021_v3_FOV.eps} }
%\vspace{-7pt}
\caption{Scheme of image formation on the detector (shown in pink),
view from the side of
the detector, to scale. The red circles outlined by
solid lines are the field of view in the polarimetry mode, the
circle with a dashed line is the field of view in the speckle
interferometry mode (the PBS presented by the green square is
removed from the beam). The edges
of the frame are also signed: left, right, bottom and top.}
\label{fig:7}
\end{center}
%\vspace{-20pt}
\end{figure}

Before 2022 we used a Wollaston prism
as a beam-splitting element. Although the prism has some
advantages, such as simplicity of design and polarimeter
alignment, it also has significant drawbacks. First of all,
this is dispersion, which arises due to the dependence of the
difference between the refractive indices $n_o$ and $n_e$ on the
wavelength. The dispersion smears the images along the beam
separation line, which reduces the $S/N$ ratio in the image
spectrum. We quantify this effect in Section~\ref{subs:num_compar}. The second
feature of the Wollaston prism is that it introduces differential
distortion into images. One of the images is compressed and the
other one is stretched by 1.5\% in the direction connecting the
images.

In the new design of the instrument, we tried to use a larger number
of detector pixels to realize a larger field of view. However,
this would require an increase in the beam separation angle. Since
the dispersion of the Wollaston prism is proportional to its
average separation angle, the dispersion would reach an
unacceptably high level. In this regard, we abandoned the
Wollaston prism in favor of a Polarizing Beamsplitter Cube (PBS, ({\it 15}) in Figs. \ref{fig:2} and \ref{fig:6}).
The PBS reflects light polarized in a plane perpendicular to the
plane of incidence, while light polarized along the plane of
incidence is transmitted. The PBSs lack dispersion.

The PBSs are most effectively implemented using dichroic coatings,
however, in this case, the working spectral band is not very wide. There is no commercially available PBS that would
operate in the entire wavelength range that we are interested
in. Therefore, we use
two PBSs, the first one is for 390 to 730~nm and the other---for
660 to 1160~nm. The cubes can be swapped using a motorized translator
based on the Standa 8CMA06-25/15 actuator.

The beam reflected by PBS propagates at an angle
of $90^\circ$ to the optical axis and to the beam that has passed
through the PBS. Thus, to focus both beams on the detector,
additional mirrors ({\it 16}) and ({\it 17}) are required (see
Figs~\ref{fig:2}, \ref{fig:6} and \ref{fig:7}). These mirrors are tilt-adjustable. The mirror
({\it 17}), which reflects the transmitted beam, is mounted on a
motorized translator based on a Standa 8CMA06-25/15 actuator.
Movement is performed towards to/away from the detector. This ensures
accurate relative focusing of images. The detector itself is
oriented at an angle of $45^{\circ}$ to the main optical axis
(before the PBS), which facilitates the placement of images
corresponding to horizontal and vertical polarizations on it. The
scheme for relaying images to the detector is shown in Fig.~\ref{fig:7}.

In the mode of speckle interferometry, which is the subject of
this paper, the PBS is removed from the beam, so that all light
passes to the mirror {\it (17)}. In this case, the mirror itself
shifts to the detector by 2.5~mm, eliminating the need to refocus
the telescope. Note that the center of the field of view shifts
along the detector, see Figs~\ref{fig:5} and \ref{fig:7}. A large field of view is
used to center the object, and in standard observations, an area
of $512\times256$~px is read, which corresponds to
$10\,.\!\!^{\prime\prime}4\times5\,.\!\!^{\prime\prime}2$ (see
Section~\ref{subs:detector_mode} below).

\subsection{Detector characteristics}

%fig8
\begin{figure}[t!] %\setcaptionmargin{5mm} \onelinecaptionsfalse \captionstyle{normal}
\begin{center}
\center{\includegraphics[width=\linewidth]{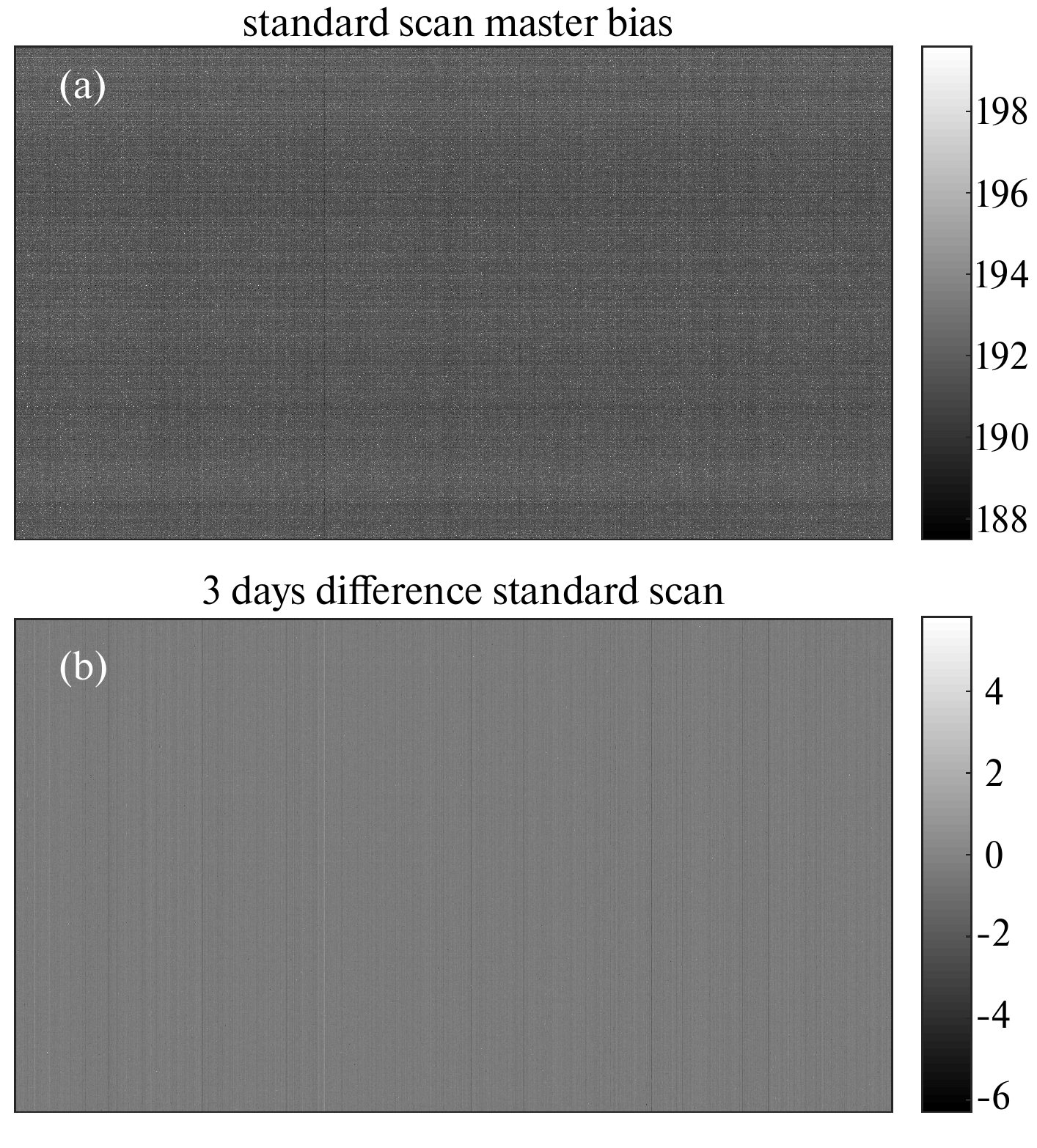}}%{SpeckleCMOS_figs/bias/master_bias_and_3daydifference2ink.eps} }
%\vspace{-7pt}
\caption{(a)---example of master bias; (b)---the difference between two master biases taken with an interval of 3 days. The width of the range of displayed values is the same. The master bias was calculated over 500 frames, while discarding 1\% of the brightest and 1\% of the dimmest values of each pixel in the series.}
\label{fig:8}
\end{center}
%\vspace{-20pt}
\end{figure}

Knowledge of the characteristics of the main detector is of great
importance for evaluating the efficiency of the instrument in solving
astrophysical problems, for determining the optimal operating mode
of the instrument, and also for developing processing methods. In this
subsection, we present the results of measuring the main
characteristics of the Hamamatsu ORCA-quest detector when
operating without illumination and with uniform illumination. The
Hamamatsu ORCA-quest detector can operate in two modes: standard
and ultra-quiet, which differ in readout speed and readout noise.
The mode parameters are shown in Table~\ref{tab:1}. The study of the
detector was carried out by us in both modes.

In CMOS detectors, as in CCDs, some constant number is added to
the signal, which is called bias. The bias value is fairly
constant, but there is a slight dependence on the position in the
frame and on time. To analyze bias stability, we obtained
several series of frames without illumination. For each series,
the per-pixel median (master bias) was calculated, an example can
be seen in Fig.~\ref{fig:8}. For series taken with an interval of 3~days,
the bias difference turned out to be less than 0.1~ADU, so the bias drift at
shorter times can be neglected.

An analysis of the dependence of the bias type on the registration
parameters showed that the following parameters should be the same
for scientific frames and for bias frames used for their
reduction: the size and position of the read area, binning,
standard/ultra-quiet detector operation mode, and exposure. The
effect of exposure dependence of bias is especially noticeable
when observing faint  sources in the ultra-quiet mode.

The bias frames show the vertical and horizontal structure (see
Fig.~\ref{fig:8}). This structure has a random component, that is, its
realization changes from frame to frame. Because of this, it
appears as an increase in the level of the power spectrum at
frequencies $|f_x|\approx0$ or $|f_y|\approx0$. The method of
 elimination of this effect in speckle interferometric
  processing is presented in Section~\ref{subs:reduction}.

\begin{figure}[t!] %\setcaptionmargin{5mm} \onelinecaptionsfalse \captionstyle{normal}
\begin{center}
\center{\includegraphics[width=1\linewidth] {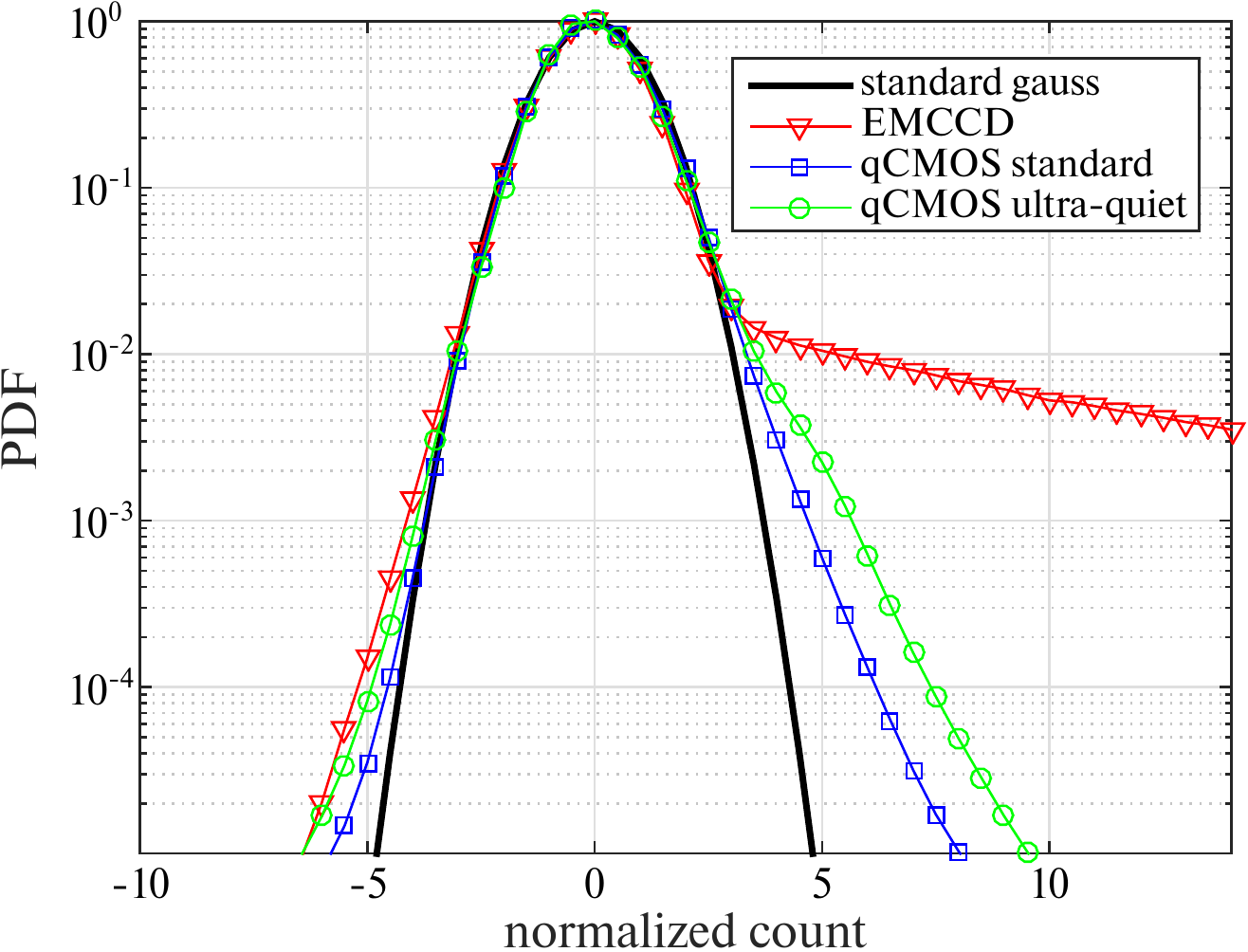}%{SpeckleCMOS_figs/bias/normalized_hist2.eps}
}
%\vspace{-7pt}
\caption{
Distribution of the deviation of bias from the mean (the deviation
is normalized to the RMSD of the readout noise). The blue line with
squares is the CMOS standard mode, the green line with circles is
the CMOS ultra-quiet mode, the red line with triangles is
 EMCCD Andor iXon 897 operating at $G_\mathrm{EM}=500$. The solid thick line is the
standard normal distribution.}
\label{fig:9}
\end{center}
%\vspace{-20pt}
\end{figure}

Based on the series of bias frames, we estimated the root mean
square of per-pixel standard deviations (not taking into account
outliers beyond the second and 98th percentiles)
$\sigma_{\mathrm{RON}}$, it turned out to be equal to 0.241 and
0.422\,$e^{-}$, for ultra-quiet and standard modes, respectively, which is
consistent with the specifications, Table~\ref{tab:1}. Figure~\ref{fig:9} shows the bias
sample distributions normalized to the RMSD of the readout noise
(the normalization was performed individually for each pixel).
Also, for comparison, the standard normal distribution is given,
as one can see, it describes well the distributions of samples up
to the values $3\sigma_{\mathrm{RON}}$. At large normalized
deviations, a significant excess of the measured histogram begins
in comparison with the standard normal distribution. Note that a
similar effect is observed in EMCCD, it is caused by the
contribution of spurious
 charge (CIC noise) \cite{Harpsoe2012}. As can be seen, the readout noise distribution of
the Hamamatsu ORCA-quest detector is much
more ``normal'' than that of the Andor iXon 897 EMCCD. In the
simulations in Section~\ref{sec:model}, we will assume the readout noise is
normal, with the RMSD corresponding to the measurements made.

%fig9

The dark current measured by us at a detector temperature of
$-20^{\circ}$C (air cooling) turned out to be 0.011 and 0.013
$e^{-}$/px/s for the ultra-quiet and standard modes, respectively.
For typical exposures that we use, the level of dark current is
negligible.

%fig10
\begin{figure}[t!] %\setcaptionmargin{5mm} \onelinecaptionsfalse \captionstyle{normal}
\begin{center}
\center{\includegraphics[width=\linewidth]{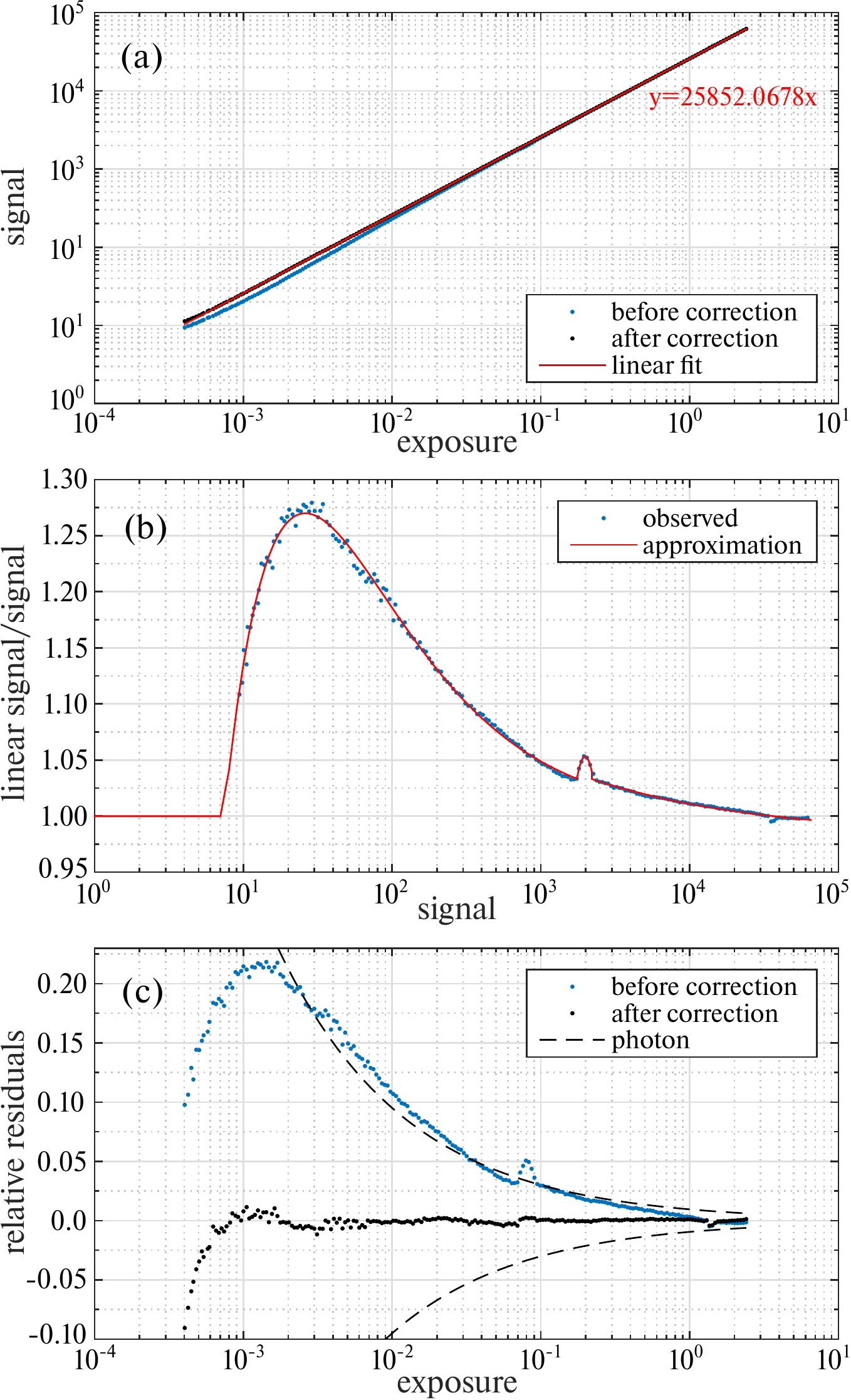}}%{SpeckleCMOS_figs/nonlin/nonlin_all.eps} }
%\vspace{-17pt}
\caption{(a)---dependence of signal on exposure before and after non-linearity correction, (b)---function approximating dependence ``ratio of linear approximation of signal to signal\,--\,signal'', (c)---comparison of relative deviations from the linear approximation of the dependence of the signal on the exposure for data with and without nonlinearity correction, the dashed line reflects the RMSD value of photon noise in a pixel.}
\label{fig:10}
\end{center}
%\vspace{-20pt}
\end{figure}

To study the nonlinearity of the detector response, we used a
stabilized light source (halogen lamp). The average signal was
25\,000~ADU/s. Frames were obtained in the standard mode with exposures $t$ from 0.4~ms to 2.4~s on a uniform logarithmic
grid, which corresponds to signal range from 10~ADU to 60\,000~ADU
(covers the entire ADC range). The constant bias structure was
eliminated by subtracting the master bias. Then we performed
photometry over an area of $128\times128$~px (relative
illumination variations over the area under consideration were
about 1\%).

Figure~\ref{fig:10}a shows the dependence of the average signal $s$ on the
exposure time. We have approximated the dependence $s(t)$ by the
$at$ function, the relative deviation \mbox{$\Delta
s=(at-s(t))/at$} is shown in the same figure below. As can be
seen, $\Delta s$ increases towards the region of weak signals and
reaches 0.2, which indicates a significant nonlinearity of the
detector. In addition, the deviation curve has a complex shape
with a jump in the 2000~ADU signal region, which is probably
caused by the junction of the ranges of the two ADCs.

We have approximated the quantity \mbox{$r(s)=at/s(t)$} by a
function of a special form; we will denote this approximation as
$r^{\prime}(s)$. The approximation is illustrated in Fig.~\ref{fig:10}b. To
correct the nonlinearity, the signal, after subtraction for bias,
was multiplied by $r^{\prime}(s)$. The accuracy of the correction
was checked over the region $512\times512$~px. It is better than
1\% for a signal greater than 100~ADU. For signals from 10 to
100~ADU, the accuracy is about 2\%. For the ultra-quiet mode, the
dependence was obtained in a similar way. When performing classic
long-exposure photometric measurements with this detector, special
attention must be paid to non-linearity correction.

\subsection{Choice of readout parameters}
\label{subs:detector_mode}

When observing, we must choose the optimal readout parameters for
the detector: its mode \linebreak (standard/ultra-quiet), field of
view, exposure time. 

First, the size of the read area must be
minimized in order to ensure fast reading, which is especially
relevant for ultra-quiet operation. However, in standard mode, fast
 reading is preferred as well as we use short exposures to prevent saturation.

It follows from previous experience with speckle interferometric
observations that a field of view of
\mbox{$5^{\prime\prime}\times5^{\prime\prime}$}, or
$256\times256$~px, is large enough. Increasing the width of
the field of view does not affect the reading speed, but allows a
more reliable estimate of the background. As a result, we chose
the rectangular field $512\times256$~px
($10^{\prime\prime}\times5^{\prime\prime}$) as the default field
of view for the speckle interferometry mode, see Fig.~\ref{fig:5}c. Such a
region can be read in 0.96~ms and 22.98~ms in standard and
ultra-quiet modes, respectively.

The Hamamatsu ORCA-quest detector has a rolling shutter mode and a
global exposure start mode, there is no global shutter mode. We
use only rolling shutter mode, so the rows are read in turn,
which ensures minimal gaps between exposures. For standard readout there is an electrical shutter mode to obtain exposures shorter than 1 frame readout time. But there is no such possibility in ultra-quiet mode. The
rolling shutter can make it difficult to achieve the isoplanatism
conditions necessary for the implementation of speckle
interferometry, we discuss this issue in Appendix~\hyperref[app:rolling]{A}.

The total vignetted field of view reaches $35^{\prime\prime}$ in diameter (see Fig.~\ref{fig:5}) and is used by us to center the object.

%fig11
\begin{figure}[b] %\setcaptionmargin{5mm} \onelinecaptionsfalse \captionstyle{normal}
\begin{center}
\center{\includegraphics[width=0.8\linewidth]{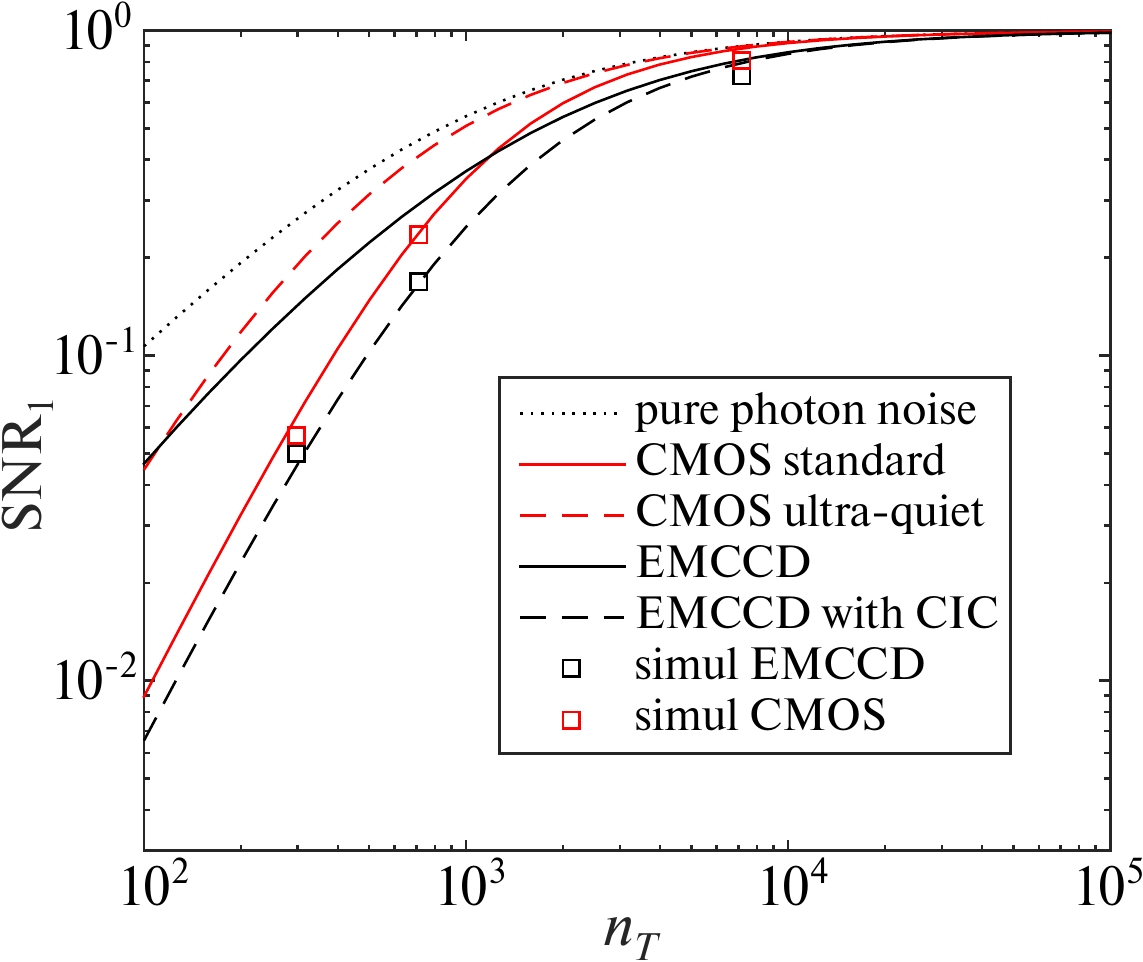}}%{SpeckleCMOS_figs/pspec/SNR_RON_ORCA.eps} }
%\vspace{-7pt}
\caption{Dependence of the $S/N$ ratio in the power spectrum per one
frame on the average number of photons per frame. The dotted
line---no read noise and no CIC noise. The red solid and dashed
lines---CMOS Hamamatsu ORCA--quest for standard and ultra-quiet
modes, respectively. The black solid and dashed lines---EMCCD
Andor iXon 897 without and  with spurious charge (CIC noise),
respectively. The black and red squares show the numerical
simulation data from Section~\ref{subs:num_compar} for EMCCD and CMOS, respectively.}
\label{fig:11}
\end{center}
%\vspace{-20pt}
\end{figure}

To choose the operating mode: standard/ultra-quiet, we consider
the contributions of photon noise and readout noise
to the S/N ratio when estimating the average power
spectrum \citep{Miller1977}:
\begin{eqnarray}
\mathrm{SNR}_1(\boldsymbol{f})\!\!\!\!&\!\!\!\! = \bigg(1 + \Big(n_T\langle|I_N(\boldsymbol{f})|^2\rangle\Big)^{-1} + \nonumber\\
                             & \dfrac{N_\mathrm{pix} p_\mathrm{CIC}}{n_T^2\langle|I_N(\boldsymbol{f})|^2\rangle} + \dfrac{N_\mathrm{pix} \sigma^2_\mathrm{RON}}{n_T^2\langle|I_N(\boldsymbol{f})|^2\rangle} \bigg)^{-1},
\label{eq:SNR}
\end{eqnarray}
where $\boldsymbol{f}$ is the spatial frequency vector, $n_T$
is the average number of photons per frame, $N_\mathrm{pix}$ is
the number of pixels used to calculate the power spectrum,
$p_\mathrm{CIC}$ is the probability of registering a CIC electron
in a pixel, $\sigma_\mathrm{RON}^2$ is the dispersion of readout
noise, $\langle|I_N(\boldsymbol{f})|^2\rangle$ is the average
signal in the absence of photon and readout noise, normalized
so that $\langle|I_N(0)|^2\rangle = 1$. Expression \eqref{eq:SNR} is
written per 1 frame. In it, the first term (equal to one) is
responsible for atmospheric noise, the second---for photon
noise, the third---for CIC noise, and the fourth---for readout
noise. In the case of CMOS, we assume that there is no CIC noise.

For the quantity $\langle|I_N(\boldsymbol{f})|^2\rangle$ at
frequencies \linebreak \mbox{$|\boldsymbol{f}|>r_0/\lambda$} the generally
accepted model is as follows:
\begin{equation}
\langle|I_N(\boldsymbol{f})|^2\rangle = 0.435 \Bigg( \dfrac{r_0}{D} \Bigg)^2 \widetilde{T_0}(\boldsymbol{f}) |\widetilde{O}(\boldsymbol{f})|^2,
\end{equation}
where $r_0/D$ is the ratio of the Fried radius to the aperture diameter, $\widetilde{T_0}(\boldsymbol{f})$ is the diffractive aperture OTF, $|\widetilde{O}|^2$ is the object visibility function module. For typical observation conditions with a 2.5-m telescope  ($r_0\!=\!0.1$\,m and a point-like object) \mbox{$\langle|I_N(f)|^2\rangle\!=\!2.7\!\times\!10^{-4}$} at $f=0.5D/\lambda$ (see Fig.~\ref{fig:20}).

%fig12
\begin{figure*}[t!] %\setcaptionmargin{5mm} \onelinecaptionsfalse \captionstyle{normal}
{\includegraphics[width=0.8\linewidth]{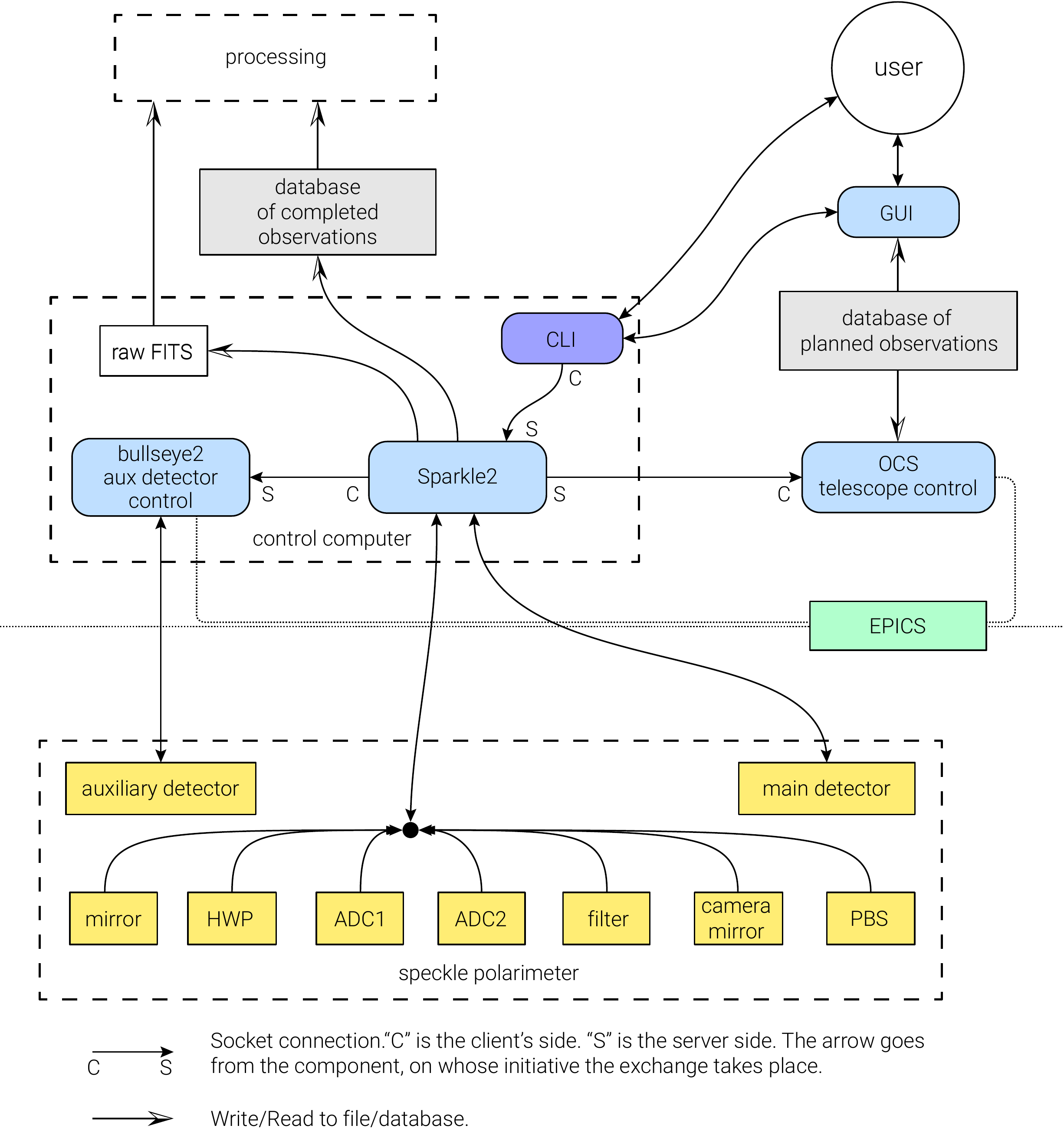}} %{SpeckleCMOS_figs/scheme/componentsDiagram_vector_colored.pdf} }
%\vspace{-7pt}
\caption{Scheme of the hardware-software complex of the instrument. }
\label{fig:12}
%\vspace{-20pt}
\end{figure*}

Figure~\ref{fig:11} shows the dependences of $\mathrm{SNR}_1$ on $n_T$ for
the size of the image spectrum calculation area $140\times
140$~px, seeing
$0\,.\!\!^{\prime\prime}73$, in the band $I_c$. The evaluation was
carried out for CMOS (standard and ultra-quiet modes), as well as
for EMCCD with and without taking into account CIC noise.

As can be seen, the EMCCD detector would have an advantage over
CMOS in the region of low fluxes ($n_T<10^3$~photons/frame) if it
were not subject to the effect of CIC noise. Accounting for
realistic CIC noise
\mbox{$p_\mathrm{CIC}=0.045$} eliminates the advantage of \mbox{EMCCD}.
Previously, \cite{Tokovinin2010} noted that CIC noise is the limiting
factor in speckle interferometric observations of faint objects with \mbox{EMCCD}.

It can also be seen from Fig.~\ref{fig:11} that it is more beneficial for
 CMOS to use the ``ultra-quiet'' readout mode if the flux is less
  than $10^4$~photons/frame. At higher fluxes, the difference
  becomes negligible and the standard mode becomes preferrable
   due to the fact that it allows faster reading. Starting from
    about $n_T=2.4\times10^5$~photons/frame, the signal in the brightest
     pixel approaches saturation. To prevent loss of information due to
      saturation, we decrease the exposure. Exposure reduction while
      maintaining 100\% time efficiency is possible up to 0.96~ms.
      Exposure reduction is also useful in suppressing the effects
       of speckle smearing due to temporal evolution and telescope mount vibration.

\subsection{Control software}

The control software consists of the following main components (diagram in Fig.~\ref{fig:12}):
 \begin{list}{}{
\setlength\leftmargin{6mm} \setlength\topsep{2mm}
\setlength\parsep{0mm} \setlength\itemsep{2mm} }
\item [1.]
The main control program {\tt Sparkle2} (implementation in C++)
interacts with the motorization drives (7 pcs.), as well as the
main detector. {\tt Sparkle2} performs recording of the raw data (see below).
\item [2.]
The control program of the auxiliary detector {\tt bullseye2} (implementation in C++). Provides object centering.
\item [3.]
The user interface {\tt Specktate} (implementation in PyQt5) is intended for convenient management of routine measurements.
\end{list}
%fig13
\begin{figure*}[t!] %\setcaptionmargin{5mm} \onelinecaptionsfalse \captionstyle{normal}
\includegraphics[width=\linewidth]{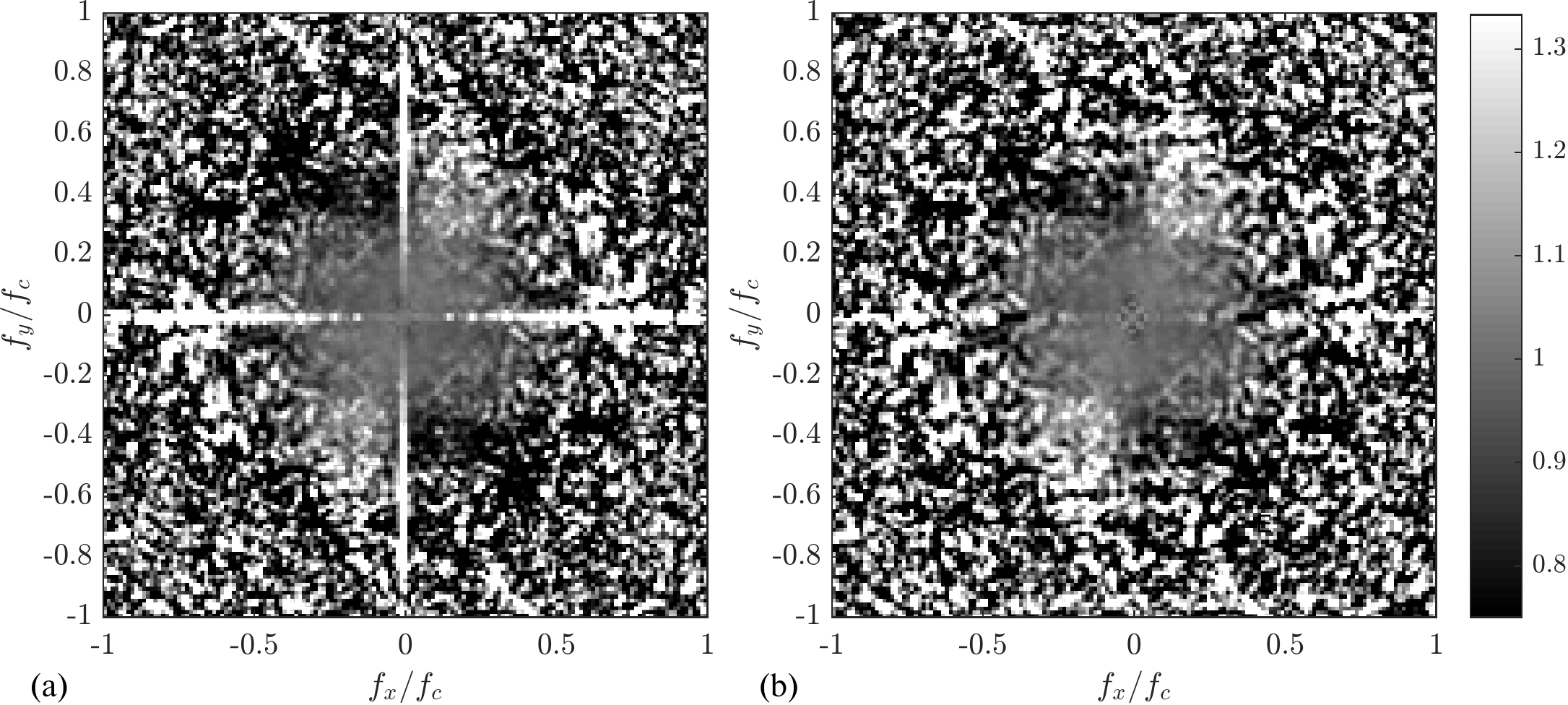}
\caption{Comparison of power spectra normalized by themselves azimuthally averaged without and with the procedure for removing the cross artifact.}
\label{fig:13}
\end{figure*}

The programs {\tt Sparkle2} and {\tt bullseye2} are constantly
running on the control computer ``in the background'', interaction
with them is carried out via a TCP/IP socket using commands that
make up a special protocol. The user can issue commands directly
over the socket connection; in this case, maximum control
flexibility is achieved (engineering mode). An alternative way is
to control via a {\tt Specktate} GUI running locally on the user's
PC. {\tt Specktate} communicates with {\tt Sparkle2} over the same
socket connection used in engineering mode. {\tt Specktate}
provides a high degree of automation for routine observations.

Interaction with the telescope control program is performed via the EPICS connection\footnote{\url{https://epics-controls.org/}}. {\tt Sparkle2} automatically requests the coordinates of the object, meteorological parameters (temperature, pressure, humidity) necessary for the correct operation of the atmospheric dispersion compensator. This information is written to the metadata of the FITS file.

{\tt Sparkle2} provides automatic object tracking (``autoguiding'') by determining the current position of the object in the frame and giving the telescope the corrections necessary to keep the object in the center of the field of view. Due to the fact that the main detector registers frames at a sufficiently high frequency (10--100 frames per second), autoguiding is performed using the image coming from it during the registration of scientific data.

The series received with the main detector are written to FITS
files. The metadata is also used to populate the observational
database {\tt sppdata} for subsequent processing. Information
about planned observations is stored in the {\tt sai2p5} database,
which has a uniform structure for all instruments of the \mbox{2.5-m}
telescope. The {\tt sai2p5} database contains information about
the coordinates of objects, time constraints on observations,
links with observations of other objects, optimal environmental
conditions (seeing, sky background, atmospheric transparency), and
also about the mode in which the instrument should operate.

\section{PROCESSING}
\label{sec:proc}

Atmospheric turbulence distorts the wavefront coming from a
distant source, resulting in blurry images. The half-width of a
long-exposure star image---seeing---is on the order of
$1^{\prime\prime}$. For binary stars with a
separation between the components that is less than the seeing,
estimating the binarity parameters: separation, position
angle, flux ratio from a long-exposure image---is difficult or
impossible at all due to the influence of photon noise and
detector noise.

However, the binarity parameters can be extracted from
short-exposure images of the star by applying the speckle
interferometry method. Classical speckle interferometry 
is based on the estimation of the average power spectrum
from a series of short-exposure images of an object and its
subsequent approximation by a model, in this case, of a binary
source. In this Section, we describe the processing of data
obtained with a speckle polarimeter using the speckle
interferometry method.

\subsection{Reduction, approximation and error estimation}
\label{subs:reduction}

The preliminary reduction of each frame consists of the following steps.
\begin{list}{}{
\setlength\leftmargin{5mm} \setlength\topsep{2mm}
\setlength\parsep{0mm} \setlength\itemsep{1mm} }
\item[1.]
First, the master bias is subtracted from the frame. It is
calculated from a series of 1000 frames obtained on the same
observational night.
\item[2.]
For areas of the frame outside the window with the object under
study, the average background value is estimated and subtracted
from the entire frame.
\item[3.]
The image of the object is centered in the window. To estimate the
required offset, a Gaussian filter is applied to the image to
reduce the effect of noise.
\item[4.]
The ORCA C15550-20UP detector exhibits an artifact in the form of
a cross in the averaged power spectrum (see Fig.~\ref{fig:13}a). We remove
this artifact in the following way. The horizontal and vertical
structure of the image is estimated from a number of rows and
columns outside the scientific data window. This structure is
subtracted from the data in the window. This operation allows one
to even the vertical and horizontal components of the cross on
the future power spectrum. After that, the image power spectrum is
calculated, the vertical and horizontal components of the cross
behind the cutoff frequency $D/\lambda$ are estimated and
subtracted from it, see the result in Fig.~\ref{fig:13}b.
\item[5.]
Next, the frame power spectrum is corrected for the distortion of
the ADC\footnote{Atmospheric Dispersion Compensator.} and the
Wollaston prism (if it is the beam-splitting element). The power
spectrum is rotated to the angle corresponding to the central
moment of the series.
\item[6.]
Further, the photon bias is subtracted
from the already averaged power spectrum. We estimate it by the
values of the power spectrum beyond the cutoff frequency.
\end{list}
After obtaining the averaged power spectrum, it is approximated by a model function.

As is known, the image $I$ is a convolution of the intensity
distribution $O$ and the point spread function $T$. For the Fourier spectrum $\widetilde{I}$
of an instantaneous image one can write:
\begin{equation}
\widetilde{I}(\boldsymbol{f})=\widetilde{O}(\boldsymbol{f})\widetilde{T}(\boldsymbol{f}),
\label{eq:imgformation}
\end{equation}
where $\boldsymbol{f}$ is the spatial frequency vector,
$\widetilde{O}$ is the object visibility, $\widetilde{T}$ is the
optical transfer function (OTF). The OTF is defined as:
%\begin{eqnarray}\label{eq:OTF}
%\lefteqn{\widetilde{T}(\boldsymbol{f})=\int_W W(\boldsymbol{x}) W^*(\boldsymbol{x}+\lambda\boldsymbol{f})\cdot}\nonumber\\
%&\cdot&\exp\{i\phi(\boldsymbol{x})-i\phi(\boldsymbol{x}+\lambda\boldsymbol{f})\} d\boldsymbol{x},
%\end{eqnarray}
\begin{equation}
    \label{eq:OTF}
\widetilde{T}(\boldsymbol{f})\!=\!\!\int_W\!\!\! W(\boldsymbol{x}) W^*(\boldsymbol{x}+\lambda\boldsymbol{f})
{e}^{(i\phi(\boldsymbol{x})-i\phi(\boldsymbol{x}+\lambda\boldsymbol{f}))} d\boldsymbol{x},
\end{equation}
where integration is performed in the pupil plane,
$\boldsymbol{x}$ is the pupil plane coordinate, $W$ is the pupil
function equal to one inside the aperture and zero outside,
$\phi$ is the instantaneous phase distribution in the pupil. As can
be seen, equation \eqref{eq:imgformation} takes into account diffraction at the edges
of the aperture, telescope aberrations, and atmospheric
distortions.

Fourier spectrum modulus is squared and averaged over the entire series of short-exposure images:
\begin{equation}
\label{eq:sq_img_form_eq}
\langle{|\widetilde{I}(\boldsymbol{f})|}^2\rangle={|\widetilde{O}(\boldsymbol{f})|}^2\langle{|\widetilde{T}(\boldsymbol{f})|}^2\rangle.\end{equation}

To estimate the square of the visibility amplitude, it is
necessary to know the value $\langle{|\widetilde{I}_{\rm
ref}|}^2\rangle$, which can be obtained from the observation of a
single star (for a point object $\widetilde{O}_{\rm ref}=1$)
carried out, if possible, either shortly before or shortly after
the observation of the scientific object:
\begin{equation}
{|\widetilde{O}|}^2\approx\dfrac{\langle{|\widetilde{I}(\boldsymbol{f})|}^2\rangle}{\langle{|\widetilde{I}_{\rm ref}(\boldsymbol{f})|}^2\rangle}.
\end{equation}

In the vast majority of cases, to obtain
$\langle{|\widetilde{I}_{\rm ref}|}^2\rangle$, instead of
observing a close reference single star, one can use an
azimuthally averaged initial power spectrum
$\langle{|\widetilde{I}|}^2\rangle$.

Let us find the square of the modulus of visibility under the
assumption that the object consists of two point--like sources. The intensity distribution
in this case is
\begin{equation}
O^0(\alpha_x,\alpha_y)=\delta(\alpha_x,\alpha_y)+\epsilon\,\delta(\alpha_x-\Delta_x,\alpha_y-\Delta_y),
\end{equation}
where $\delta$ is the delta function, $\epsilon$ is the ratio of component fluxes, $\Delta_x,\Delta_y$ is the separation, $\alpha_x,\alpha_y$ are the angular coordinates. The use of the delta function to model a star is justified because most stars have angular sizes much smaller than the diffraction limited resolution of the 2.5-m telescope. The Fourier transform of the given intensity distribution is:
%\begin{eqnarray}\label{eq:visibility}
%\begin{split}
%\lefteqn{\widetilde{O^0}\left(f_x,f_y\right)=}\nonumber\\
%&=&\sqrt{A_1}\left(1+\epsilon\exp{\left[-%i2\pi(f_x\Delta_x+f_y\Delta_y)\right]}\right),
%\end{split}
%\end{eqnarray}
\begin{equation}
    \label{eq:visibility}
%\begin{split}
\widetilde{O^0}\left(f_x,f_y\right)=
\sqrt{A_1}\left(1+\epsilon\,e^{\left(-i2\pi(f_x\Delta_x+f_y\Delta_y)\right)}\right)\!,
%\end{split}
\end{equation}
where $A_1=(1+\epsilon)^{-2}$ is the normalization factor.
Then, taking the modulus and squaring it, we have:
%\begin{eqnarray}\label{eq:squared_visibility}
%\lefteqn{{|\widetilde{O^0}\left(f_x,f_y\right)|}^2\!=}\nonumber\\
%&A_1\left(1+\epsilon^2+2\epsilon\cos{\left(2\pi(f_x\Delta_x+f_y\Delta_y)\right)}\right).
%\end{eqnarray}
\begin{equation}
   \label{eq:squared_visibility}
{|\widetilde{O^0}\!\left(f_x,\!f_y\right)\!|}^2\!\!\!=\!
A_1\!\left(1\!+\!\epsilon^2\!\!+\!2\epsilon\!\cos{\left(2\pi(f_x\Delta_x\!\!+\!\!f_y\Delta_y)\right)}\right)\!.
\end{equation}

\begin{figure}[t!] %\setcaptionmargin{5mm} \onelinecaptionsfalse \captionstyle{normal}
\begin{center}
\center{\includegraphics[width=\linewidth] {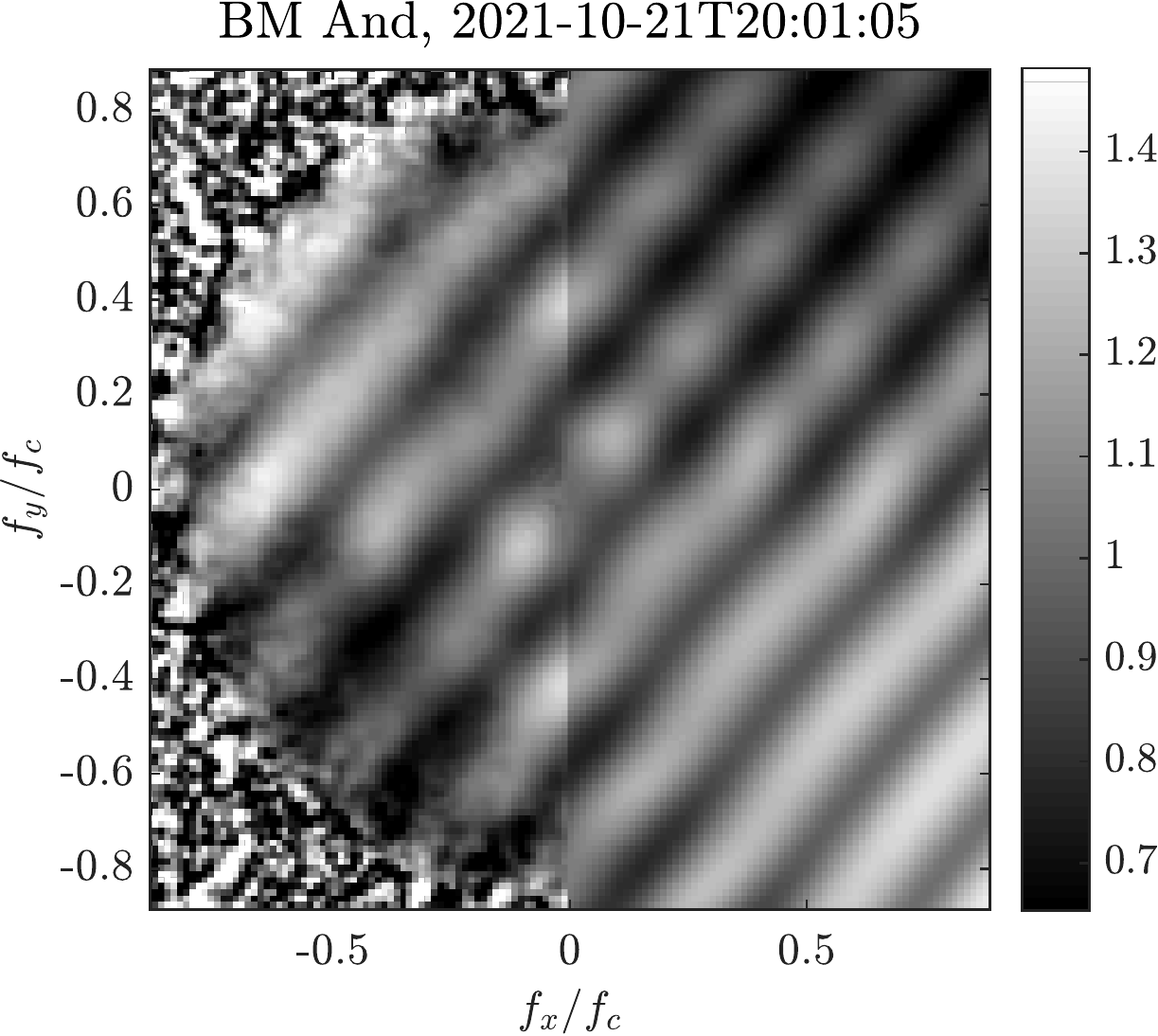}}%{SpeckleCMOS_figs/approx_example/BMAnd_2.eps} }
\vspace{-7pt}
\caption{An example of power spectrum approximation. The left half is the average power spectrum normalized by its azimuth mean, the right half is the binary source model. Note that the power spectrum has central symmetry, so the half-plane contains all the useful information.}
\label{fig:14}
\end{center}
\end{figure}

To obtain the binarity parameters, we approximate the root of the
power spectrum normalized by its azimuthal average. The
applied model for approximation is the following:
%\begin{eqnarray}
%\lefteqn{|\widetilde{O^0}\left(f_x,f_y\right)|=\Bigl|\sqrt{A_1}\bigl(1 + }\nonumber\\
 %   &&\epsilon e^{\left(-i2\pi(f_x\Delta_x+f_y\Delta_y+\phi)\right)}\bigr) + %\nonumber\\
 %   &&\left(1+\sqrt{f_x^2+f_y^2}A_a\cos{2(\theta_a-\theta_{a_{\rm ref}})}\right)\Bigr|,
%\end{eqnarray}
\begin{equation}
\begin{array}{l}
|\widetilde{O^0}\left(f_x,f_y\right)\!|\!=\!\Bigl|\sqrt{A_1}\bigl(1 + \epsilon e^{\left(-i2\pi(f_x\Delta_x+f_y\Delta_y+\phi)\right)}\bigr)  \times  \\[5pt]
        \left(1+\sqrt{f_x^2+f_y^2}A_a\cos{2(\theta_a-\theta_{a_{\rm ref}})}\right)\Bigr|, \\[-5pt]
   \end{array}\\
\label{eq:appmodel}
\end{equation}
where $A_1=(1+\epsilon)^{-2}$ is the normalization factor, $\epsilon$ is the ratio of component fluxes, $\Delta_x,\Delta_y$ is the separation. Compared to \eqref{eq:visibility}, additional parameters have been added to the expression: \mbox{$\theta_{a_{\rm ref}}={\arctan^2}(f_y,f_x)$}, $A_a$ and $\theta_a$ are the parameters characterizing the amplitude and angle of the asymmetric component, respectively, which must be taken into account due to the jitter of the telescope. The parameter $\phi$ is the shift of the power spectrum relative to the central maximum in the direction perpendicular to the fringes. It should be noted that the model function is also normalized by its azimuthal average, similarly to how it is done with the estimation of the root of the power spectrum. An example of approximation is shown in Fig.~\ref{fig:14}.
%fig14

The error of the binarity parameters is estimated by the
``bootstrap'' method \citep{Efron1993}. The sample of
indices used to average the frame power spectrum is repeatedly
randomly generated. In this case, the length of this sample is
equal to the length of the original series, and the set of frames
used is different. So, as a result of such a sample generation,
the same frame can be counted several times, while the
other---never. We generate twenty such index samples. Using these
indices, we average the power spectra. As a result, we have
20 averaged power spectra for one series. We then do exactly the
same processing for each of the 20 power spectra. And, further, we
approximate each of them by the function described above. This
results in a sample of 20 sets of binarity parameter values. For
this sample, we estimate the standard deviation of each parameter.

\subsection{Estimation of the achievable contrast curve}

In those cases where no binarity has been found in the star, the important information that we can glean from observation can be an estimate of the achievable contrast curve $\epsilon_{\rm lim}(\rho)$. The curve of achievable contrast characterizes the minimum possible registrable ratio of component fluxes for a given observation depending on the angular distance. In other words, the component (if it exists) at the angular distance $\rho$ with the flux ratio $\epsilon$ greater than $\epsilon_{\rm lim}(\rho)$ will be unambiguously registered by us.

%fig15
\begin{figure}[t!] %\setcaptionmargin{5mm} \onelinecaptionsfalse \captionstyle{normal}
\begin{center}
\vspace{4pt}
\center{\includegraphics[width=0.95\linewidth]{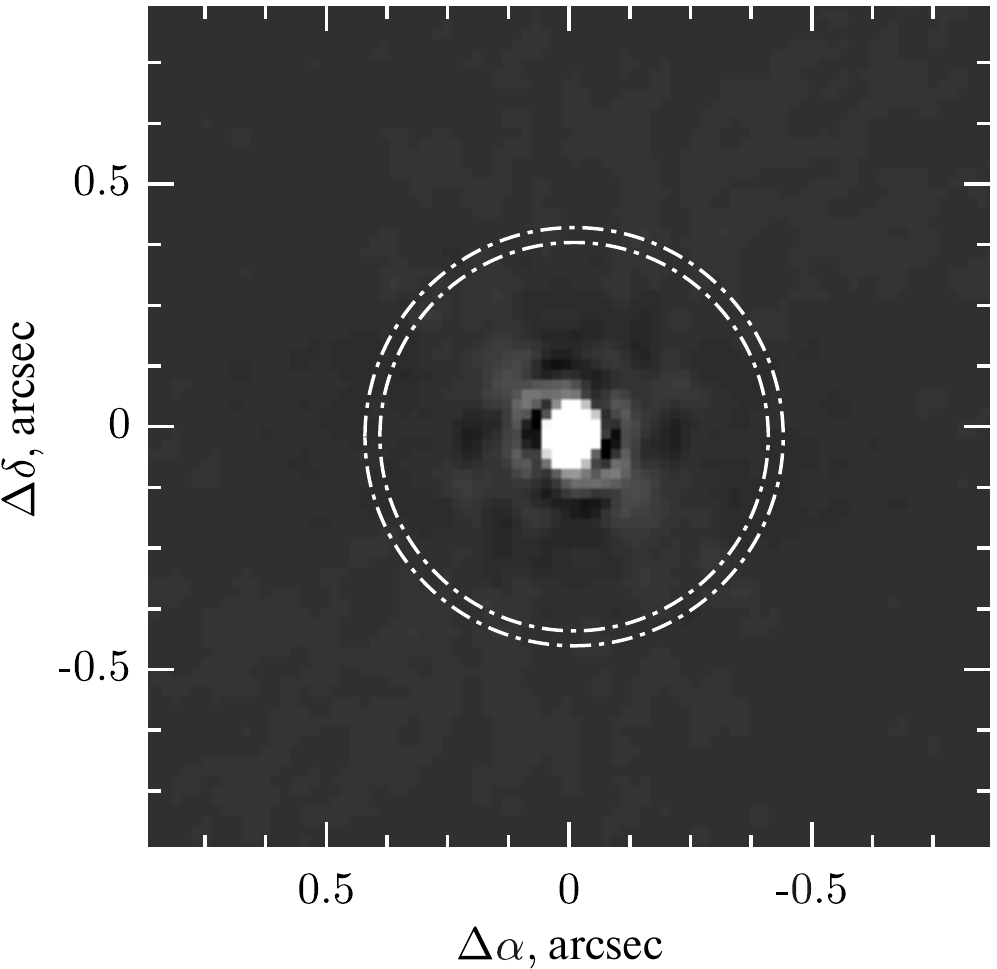}%{SpeckleCMOS_figs/acf_example/ACF_example_in_white.eps}
}
%\vspace{-15pt}
\caption{ACF example. The dashed-dotted ring shows the area in which the RMSD is calculated for the estimate $\epsilon_{\rm lim}$, in this case, $\epsilon_{\rm lim_{0.4}}$.}
\label{fig:15}
\end{center}
\end{figure}

Let us obtain the formula for determining $\epsilon$ for each point of the autocorrelation function (ACF, see Fig.~\ref{fig:15}). To do this, we take the inverse Fourier transform of the resulting function  ${|\widetilde{O^0}\left(f_x,f_y\right)|}^2$ \eqref{eq:squared_visibility}:
%\begin{eqnarray}
%\Psi\left(\alpha_x,\alpha_y\right)&\!=\!&\nonumber \\[7pt]
%=A_2A_1\big((1&\!\!+\!\!&\epsilon^2)\delta(\alpha_x,\alpha_y)+\epsilon\delta(\alpha_x\!-\!\Delta_x,\alpha_y-\Delta_y)\nonumber \\[7pt]
%&\!\!+\!&\epsilon\delta(\alpha_x\!+\!\Delta_x,\alpha_y\!+\!\Delta_y)\big).
%\end{eqnarray}
\begin{equation}
\begin{array}{l}
\Psi\left(\alpha_x,\alpha_y\right)= A_2A_1\big((1+\epsilon^2)\delta(\alpha_x,\alpha_y)\,+\\ [7pt]
 \epsilon\delta(\alpha_x\!-\!\Delta_x,\alpha_y\!-\!\Delta_y)
+\epsilon\delta(\alpha_x\!+\!\Delta_x,\alpha_y\!+\!\Delta_y)\big).
\end{array} \end{equation}

Next, we note that, by definition, $\Psi(0,0)=~1$. There\-fore, $A_2A_1(1+\epsilon^2)=1$.
That is, \linebreak $A_2=(1+\epsilon)^2/(1+\epsilon^2$).

Finally we get:
\begin{eqnarray}\label{eq:acf_for_binary}
\lefteqn{\Psi\left(\alpha_x,\alpha_y\right)=\delta(\alpha_x,\alpha_y)\,+}\nonumber\\[1pt]
&&\;\;\dfrac{\epsilon}{1+\epsilon^2}\,\delta(\alpha_x-\Delta_x,\alpha_y-\Delta_y)+\nonumber    \\[1pt]
&&\;\;\dfrac{\epsilon}{1+\epsilon^2}\,\delta(\alpha_x+\Delta_x,\alpha_y+\Delta_y).
\end{eqnarray}

This shows that the ratio of the fluxes of the binary components
can be obtained from the ratio of the intensities of the secondary
and central peaks in the ACF.

Now let us take a real ACF obtained from observations. Due to
factors introduced by the atmosphere, instrument and method, we
have some noise in the resulting ACF.

Our task is to find the dependence of the achievable contrast (flux ratio) on the distance to the central star. Let us move from the Cartesian coordinate system
$\alpha_x,\alpha_y$ to the polar one $\rho,\varphi$. Let us denote
the intensity of some ACF point as $\eta$. Moreover, as unity we
take the value of the coefficient of the Gaussian approximating
the central peak, plus the bias. Let us find for each ring
with radius $\rho$ the value of reachable $\eta_{\rm lim}$ by the
formula:
\begin{equation}\label{dispeq}
\eta_{\rm lim}(\rho)=5\sigma(\rho)+\overline{\eta(\rho,\varphi)},
\end{equation}
where
\begin{equation}
\sigma(\rho)=\sqrt{\dfrac{1}{n-1}\sum_{\varphi}\left(\eta(\rho,\varphi)-\overline{\eta(\rho,\varphi)}\right)^2}.
\end{equation}
The line $\overline{\phantom{15}}$ means averaging over the angle
in the ring, $n$ is the number of pixels in the ring. The
summation over $\varphi$ in the expression for the standard
deviation $\sigma$ means the summation over the pixels in the
ring.

Using the equation \eqref{eq:acf_for_binary}, we have:
\begin{equation}
\dfrac{\epsilon_{\rm lim}}{1+{\epsilon_{\rm lim}}^2}=\eta_{\rm lim}.
\end{equation}

%fig16
\begin{figure}[t] \vspace{1mm} %\setcaptionmargin{5mm} \onelinecaptionsfalse \captionstyle{normal}
%\vspace{10pt}
%\center{
\includegraphics[width=0.93\linewidth]{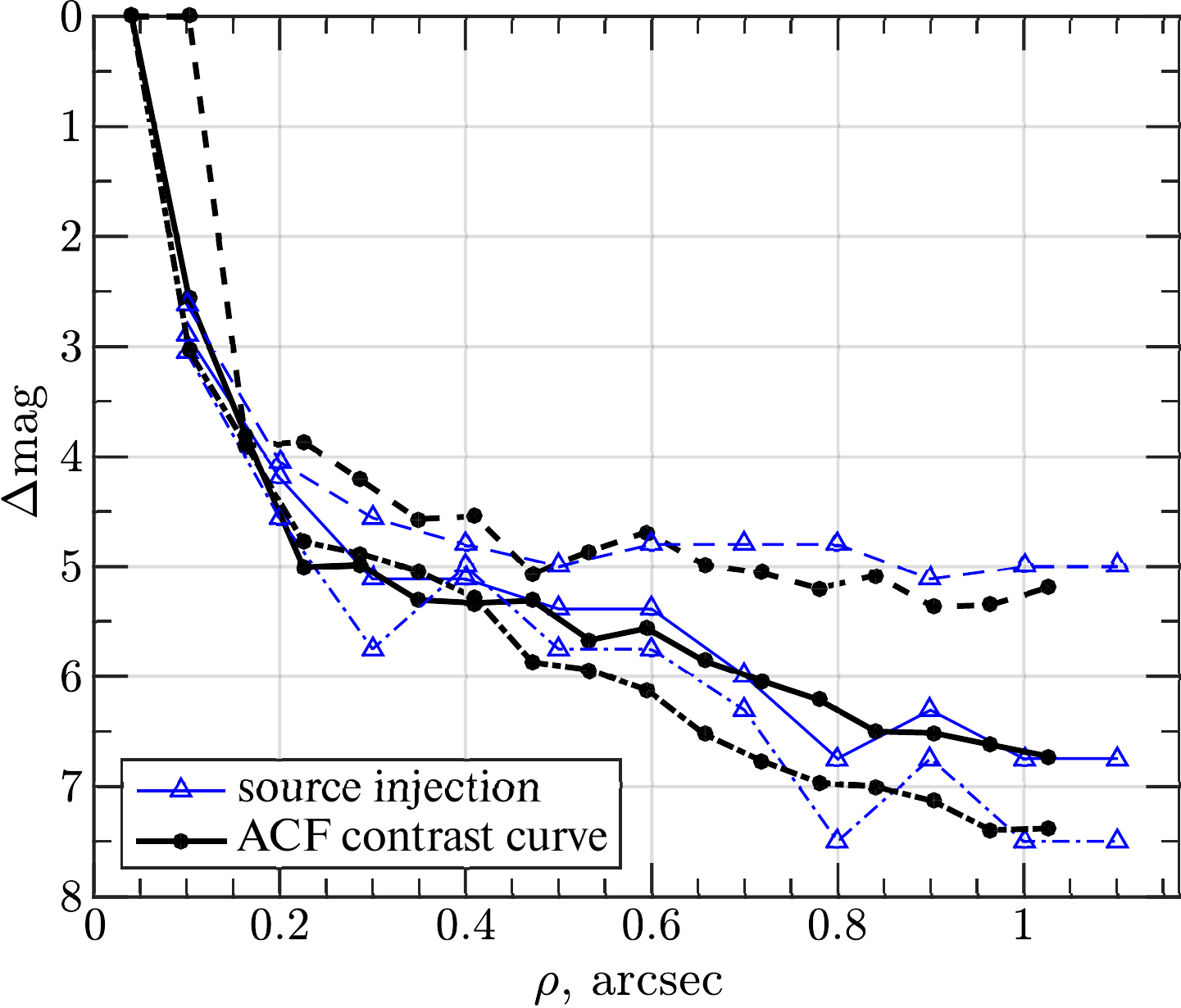}
%{SpeckleCMOS_figs/Curve_comparison/CONTR_ANALIS_ALL_NEW_ed.pdf}}
\caption{Comparison of achievable contrast curves,
obtained from ACF and by source injection. Lines of different types correspond to observations of different stars.}
\label{fig:16}

\vspace{-7pt}
\end{figure}
From here we get:
\begin{equation}\label{dispeq15}
\epsilon_{\rm lim}(\rho)=\dfrac{1-\sqrt{1-4{\eta_{\rm lim}}^2(\rho)}}{2\eta_{\rm lim}(\rho)}.
\end{equation}

Now, based on this dependence, we can argue that given star has no components
with contrast \mbox{$\epsilon>\epsilon_{\rm
lim}(\rho)$} at an angular distance $\rho$.

It is possible to verify the achievable contrast curve by adding a
secondary component with known parameters to the real data. We multiply
the power spectrum of a single star directly by expression \eqref{eq:squared_visibility} in
accordance with expression \eqref{eq:sq_img_form_eq}, where we vary the separation and
for each separation find the minimum (in terms of flux ratio)
contrast value at which the ``artificial'' injected source is still
detected at the ACF. The comparison illustrated in Fig.~\ref{fig:16} shows
that the contrast curve from the ACF method is estimated adequately.

\section{DIRECT MODEL OF SPECKLE INTERFEROMETRIC OBSERVATIONS}
\label{sec:model}

To study the capabilities of the detector and verify the
processing methods, we created a program that allows us to
generate images recorded by the detector in the focal plane of the
instrument. The model takes into account wavefront distortions in a
turbulent atmosphere, in an optical system, as well as photon
noise and detector noise.

The von K{\'a}rm{\'a}n turbulent disturbance model gives the following
expression for the phase distortion power spectrum (see \cite{Johansson1994,Xiang2014}):
\begin{equation}
\widetilde{P}(f_x,f_y)\!=\!\dfrac{4\pi^2}{G_x{G_y}}0.00058\,r_0^{-5/3}\!\left(f_x^2\!+\!f_y^2\!+\!\dfrac{1}{L_0}\right)^{\!\!-11/6}\!\!\!,\nonumber \\
\label{model_1eq}
\end{equation}
where $f_x,f_y$ are the spatial frequencies, $G_x,G_y$ are the phase screen dimensions (in meters),  $L_0$ is the outer scale of turbulence , $r_0$ is the Fried radius related to the turbulence profile $C_n^2(h)$ as follows:
\begin{equation}
r_0=\left(0.423\left(\dfrac{2\pi}{\lambda}\right)^2\mathrm{sec}\gamma\int_{0}^{H_{\rm max}}C_n^2(h)dh\right)^{-3/5},
\label{eq:r0}
\end{equation}
where $\lambda$ is the wavelength, $\gamma$ is the zenith angle, $h$ is the altitude.
The integration is carried out from the altitude of the telescope to some maximum
turbulence altitude $H_{\rm max}$.

We will model the phase screen $\Phi(x,y)$ ($x,y$ are the
coordinates in meters) as a two-dimensional array, then its
Fourier spectrum $\widetilde{\Phi}(f_x,f_y)$ is two-dimensional
array of random numbers distributed according to the normal law:
\begin{equation}
\widetilde{\Phi}(f_x,f_y)=\sqrt{\widetilde{P}(f_x,f_y)}\times({N(0,1)+iN(0,1))},
\end{equation}
where $N(\mu,\sigma^2)$ in this case means a two-dimensional array
filled with normally distributed numbers, coinciding in its
dimensions with $\widetilde{P}$.

The phase screen is determined by the inverse Fourier transform of
its spectrum  ($\mathrm{FT}^{-1}$ is the inverse Fourier
transform):
\begin{equation}
\Phi(x,y)=\text{Re}\left\{\mathrm{FT}^{-1}\left[\widetilde{\Phi}(f_x,f_y)\right]\right\},
\end{equation}
taking the real or imaginary part does not matter.

Assuming that the modulus of the complex amplitude of the incident
wave front does not depend on the coordinate, we write the complex
amplitude of the wave after passing through the turbulent layer:
\begin{equation}
A(x,y)=e^{i\Phi(x,y)}.
\end{equation}

Let us define the pupil function of the telescope:
\begin{equation}
\begin{cases}
    W(x,y)\equiv1, \;\;\; D>r(x,y)>\epsilon D; \vspace{1mm}\\
    W(x,y)\equiv0, \;\;\; r(x,y)<\epsilon D,\; r(x,y)>D;\vspace{1mm}\\ 
    r(x,y)\equiv\sqrt{x^2+y^2},\\
\end{cases}
\label{W_equation}
\end{equation}
where $D$ is the telescope aperture diameter,
$\epsilon$ is the relative central obscuration.

As is known \citep{Tokovinin1988}, the OTF
 of an optical instrument is equal to the
autocorrelation function of the distribution of the field
amplitude on the pupil, therefore, we can immediately write the
expression for the point spread function (PSF):
\begin{equation}
T(\alpha_x,\alpha_y)=\Bigl\lvert \mathrm{FT}\left[A(x,y)W(x,y)\right]\Bigr\rvert^2
\label{model_eq3}
\end{equation}

In the last equation there was an implicit transition to new
variables---($\alpha_x,\alpha_y$)---angular coordinates in the celestial sphere. The reader will find a detailed derivation in the book \cite[p.~21]{Tokovinin1988}.

We define the object function  $O(\alpha_x,\alpha_y)$ as an array, filled with zeros and  selecting single pixels from a
two-dimensional array and assigning them the desired intensity.
Further, knowing the PSF, we obtain the image $I$ of the object
distorted by the atmosphere by convolving the function of the
object with the PSF:
\begin{equation}
I(\alpha_x,\alpha_y)=O*T.
\label{model_eq4}
\end{equation}

The method of generating the phase screen described above will be
referred to as the FFT method (short for the Fast Fourier
Transform algorithm used).

\subsection{Effects taken into account}

\subsubsection{Filter bandwidth}
The influence of the finiteness of the filter bandwidth is taken
into account by adding the PSF for several wavelengths within the
filter bandwidth. The wavelength step is calculated from the
following relationship (see \cite{Tokovinin1988}):
\begin{equation}
\Delta\lambda = \dfrac{r_0}{D}\lambda.
\end{equation}
Here, the Fried radius is calculated taking into account the user-defined seeing.

\subsubsection{Finiteness of exposure and correlation of frames}
We proceed from the Taylor hypothesis of frozen turbulence, which
assumes that the turbulent layer (phase screen) is transferred  in
space without change. To simulate this process, it is enough for
us to move along the phase screen and cut out the necessary areas
from it. If two regions intersect, then there will be a
correlation of phase perturbations, which, in fact, is what we are
trying to achieve.

\begin{figure}[t!] \vspace{2mm} %\setcaptionmargin{5mm} \onelinecaptionsfalse \captionstyle{normal}
\begin{center}
\begin{minipage}[h]{\linewidth}
\includegraphics[width=0.85\linewidth]{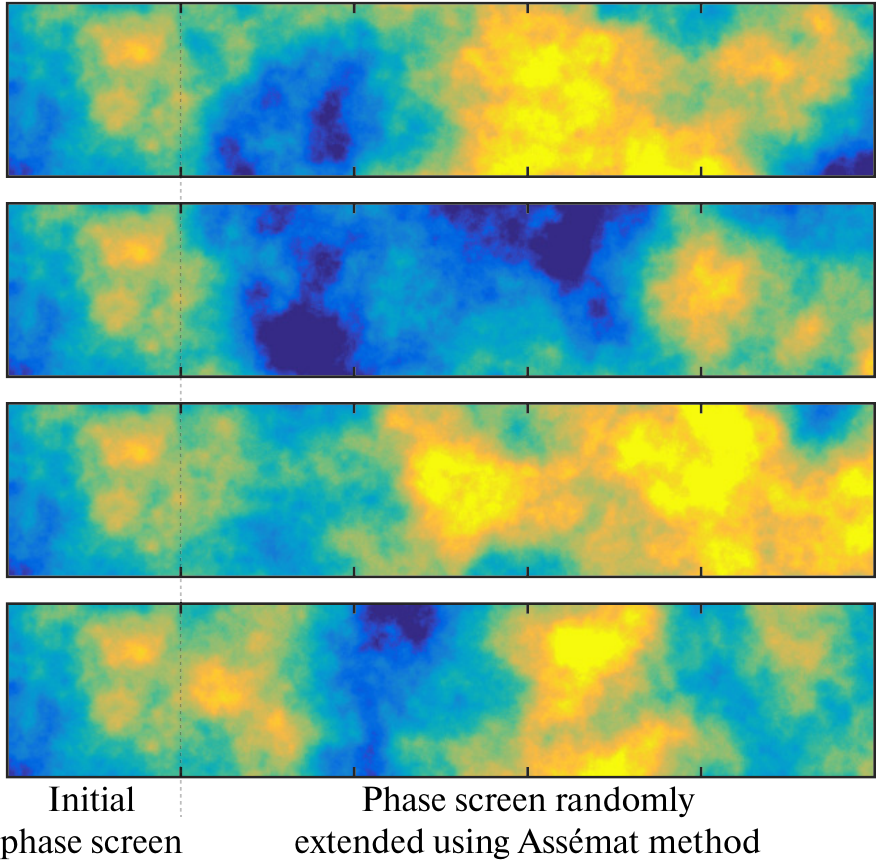}%{SpeckleCMOS_figs/Assemat/several_phase_screen_a.eps}
\caption{Demonstration of the result of the Ass{\'e}mat method
with the same initial phase screen and its various continuations
due to the randomness of the process.}
\label{fig:17}
\end{minipage}
\end{center}
\vspace{-20pt}
\end{figure}

In order to be able to cut out such areas, we need to either
create a large phase screen using the FFT, as described at the
beginning of the Section, and move along it with a cut
``aperture'', or somehow ``prolong'' the screen and cut out the
necessary areas from the extended screen. This method saves RAM.
The method of screen extension by Ass{\'e}mat \citep{Assemat2006}, which starts from an FFT generated initial screen,
has been shown as very reliable. An example of the result of the
method is shown in Fig.~\ref{fig:17}.
%fig17

%fig18
\begin{figure}[t!] %\setcaptionmargin{5mm} \onelinecaptionsfalse \captionstyle{normal}
\begin{center}
\begin{minipage}[h]{\linewidth}
\includegraphics[width=0.75\linewidth]{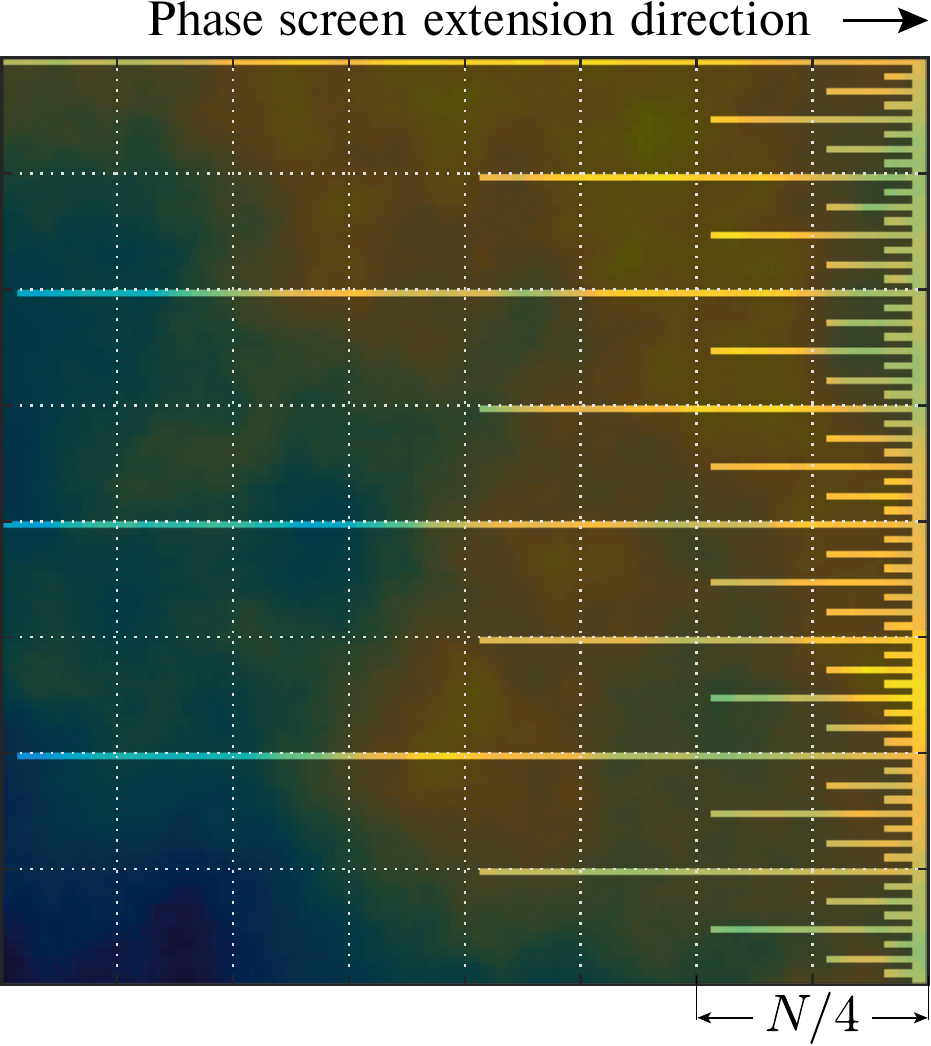}%{SpeckleCMOS_figs/Assemat/assemat_scheme_3_ill.eps}
\caption{Light areas are pixels taken from the phase screen as the input vector (i.e. the $\mathrm{Z}$ vector in \cite{Assemat2006}).}
\label{fig:18}
\vspace{-25pt}
\end{minipage}
\end{center}
\end{figure}

\begin{figure*}% \vspace{-8mm} %\setcaptionmargin{5mm} \onelinecaptionsfalse \captionstyle{normal}
\vspace{-10pt}
\begin{minipage}[h]{1\linewidth}
%\center{
\normalsize{$\ \ \ \ \ \ \ \ \ \ \:L_0=2.5$~m}\\
\vspace{-3pt}
\hspace{0.08pt}
\includegraphics[width=0.87\linewidth]{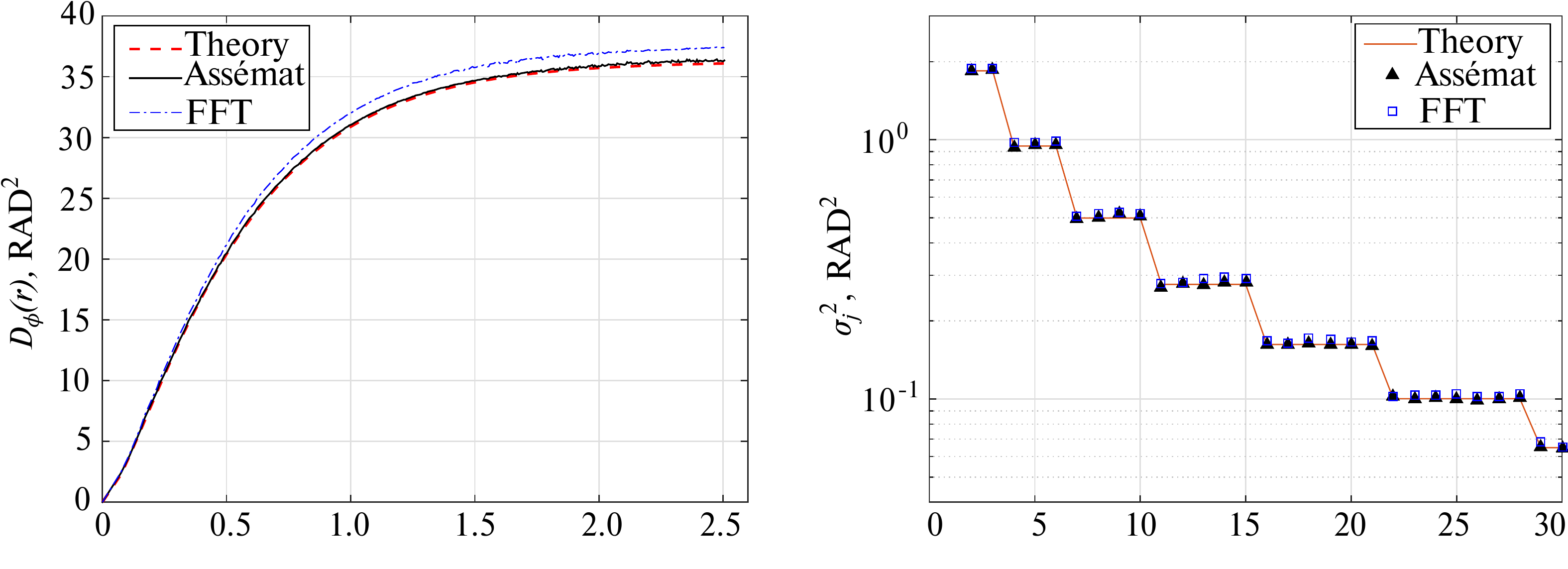}%{SpeckleCMOS_figs/Assemat/256_L0_2_5_eng.eps}
%}
\end{minipage}
\begin{minipage}[h]{1\linewidth}

\vspace{-7pt}
\normalsize{$\ \ \ \ \ \ \ \ \ \ \ L_0=10$~m}\\
\vspace{-3pt}
\hspace{0.08pt}
\includegraphics[width=0.87\linewidth]{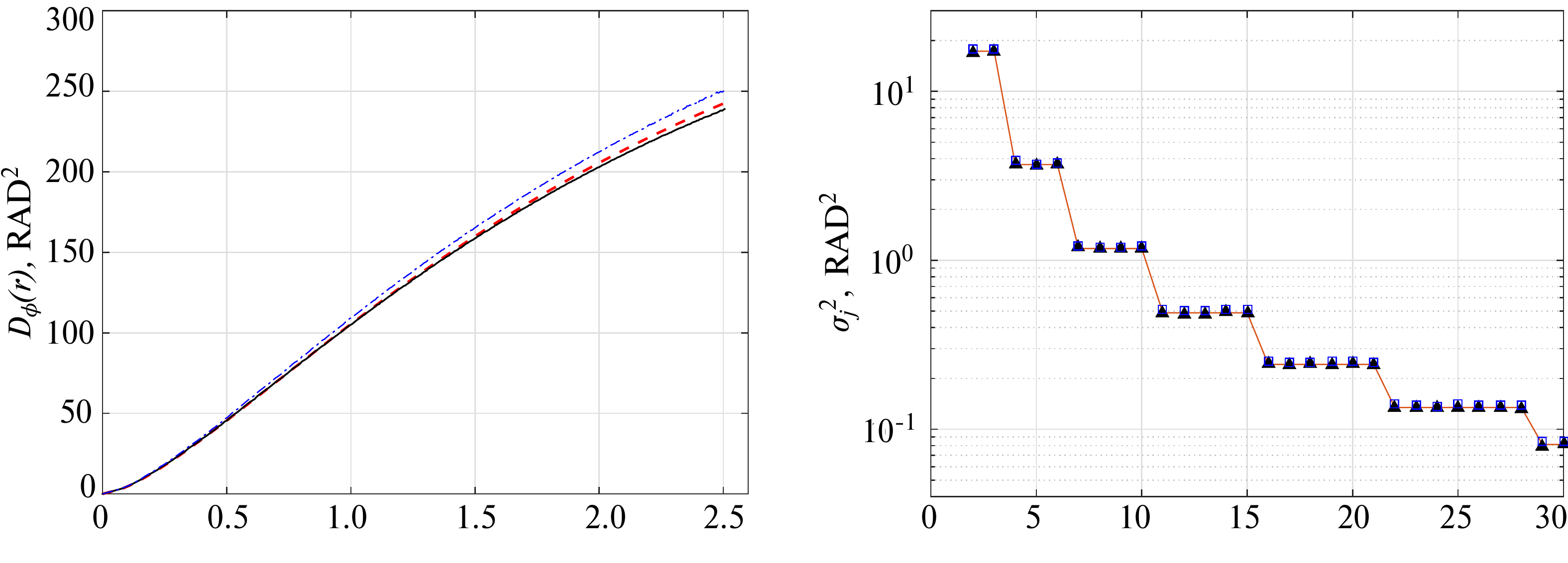}%{SpeckleCMOS_figs/Assemat/256_L0_10.eps}

\end{minipage}

\begin{minipage}[h]{1\linewidth}

\vspace{-7pt}
\normalsize{$\ \ \ \ \ \ \ \ \ \ \:L_0=25$~m}\\
\vspace{-3pt}
\hspace{0.01pt}
\includegraphics[width=0.87\linewidth]{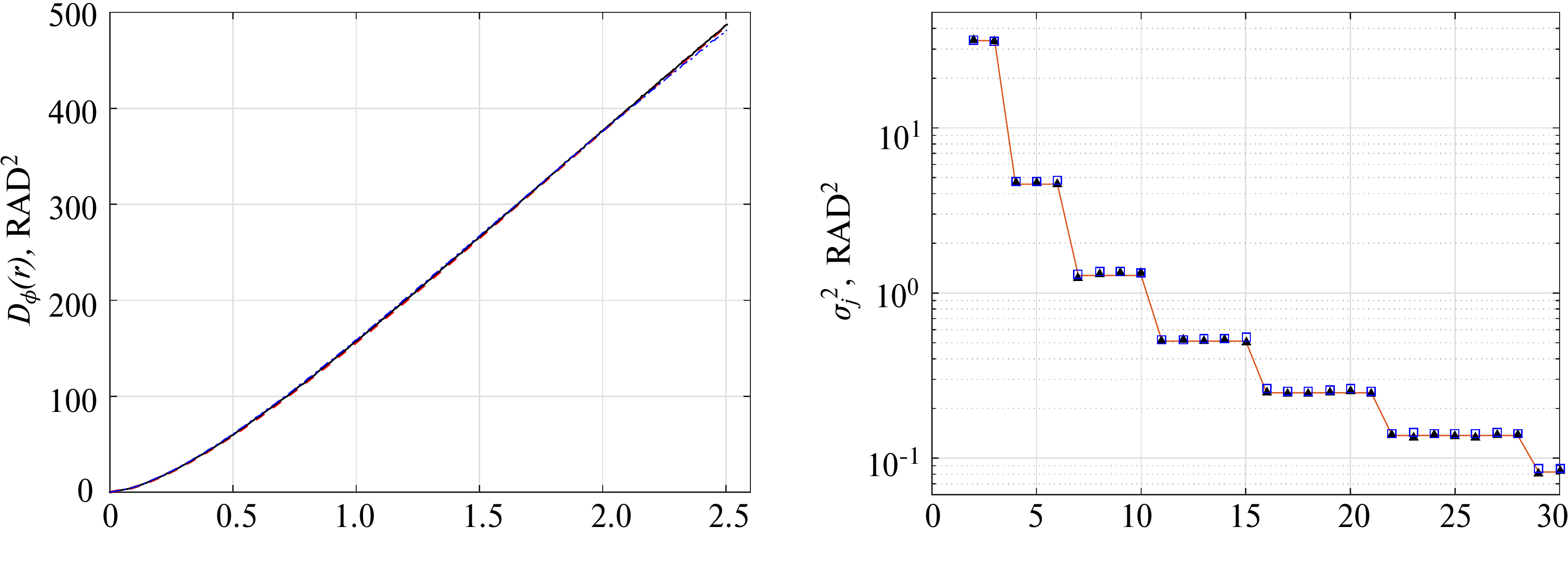}%{SpeckleCMOS_figs/Assemat/256_L0_25.eps}
\end{minipage}

\begin{minipage}[h]{1\linewidth}

\vspace{-7pt}
\normalsize{$\ \ \ \ \ \ \ \ \ \ \ L_0=100$~m}\\
\vspace{-3pt}
\hspace{3.5pt}
\includegraphics[width=0.87\linewidth]{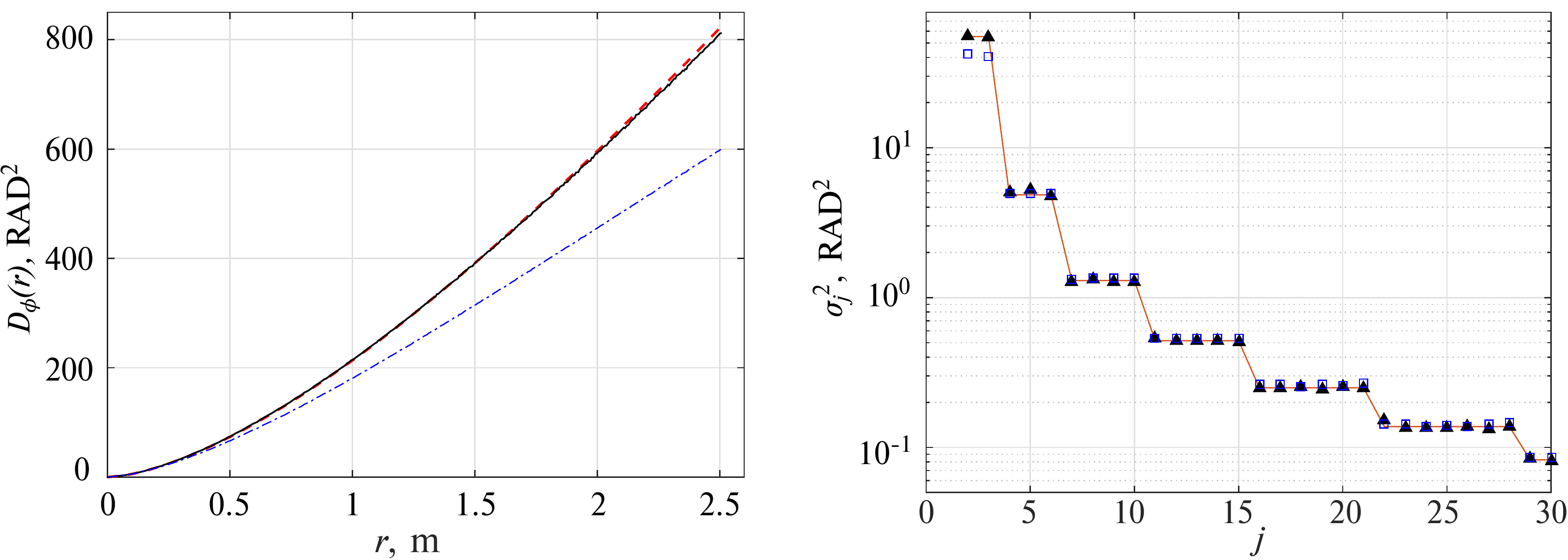}%{SpeckleCMOS_figs/Assemat/256_L0_100.eps}
\end{minipage}
\vspace{-4pt}
\caption{
%Comparison with those expected from the theory (see \cite{Winker1991,Takato1995,Assemat2006})
% of structure functions and mode variances of Zernike
% polynomials calculated from a sample of 10\, 000 phase screens
% generated by two methods for cases with $L_0$ equal to 2.5, 10,
%  25 and 100~m with a screen size of 2.5~m (256$\times$256 points).
  Comparison of structure functions and mode variances of Zernike
 polynomials with the expected theoretical values (see \cite{Winker1991,Takato1995,Assemat2006}). Calculated from a sample of 10\, 000 phase screens
 generated with Ass{\'e}mat and FFT methods. Cases for $L_0$ equal to 2.5, 10,
  25 and 100~m are considered. Screen size is 2.5~m (256$\times$256 points).
  }
\label{fig:19}
\end{figure*}

We slightly modified the method by changing the selection of
pixels of the initial phase screen to cover large scales, while
not greatly increasing the amount of used RAM. The following
selection algorithm was chosen: the two columns of the phase
screen closest to the new column being added are taken unchanged,
in each of the next four columns only half of the pixels (every
second) are selected, in the next eight columns every fourth is
selected, and so on. An illustration of this algorithm is shown in
Fig.~\ref{fig:18}. With this selection method, it turned out to be
sufficient to take into account only the first $N/4$ columns. Even
with this number, the structure function and the variances of the
Zernike polynomials of the resulting phase screens agree quite
well with the theory.
Moreover, the screens generated by this method showed better
theory fit than the FFT method (see Fig.~\ref{fig:19}).

%fig19

In this model, in order to approximate the reality better,
we generate two phase screens and move the cut sections along them
(or extend them) in perpendicular directions. The phase for these
two cutouts is then summed. This is done to obtain a realistic
characteristic time of phase variations: $r_0/V$, where $V$ is the
wind speed. Otherwise, when only one phase screen is used, the
characteristic time of variations will be $D/V$, where $D$ is the
diameter of the telescope aperture.

The speed of movement of the cut out area does not always
correspond to the whole number of pixels of the phase screen. In
this case, there are two options: one can linearly interpolate the
image and shift along it by the required amount, or one can use
the Fourier transform, which allows one to shift the image by a
fractional number of pixels, since it is known that the Fourier
images of the original and shifted images are interconnected by a
linear multiplier in frequency space.

\subsubsection{Anisoplanatism}

Light coming from different points in the sky will pass through
different parts of the phase screen, thus resulting in 
different distortion of the image, which is called anisoplanatism.

%fig20
\begin{figure*}[t!] %\setcaptionmargin{5mm} \onelinecaptionsfalse \captionstyle{normal}
%\begin{center}
\begin{minipage}[h]{0.342\linewidth}
\center{\includegraphics[width=\linewidth] {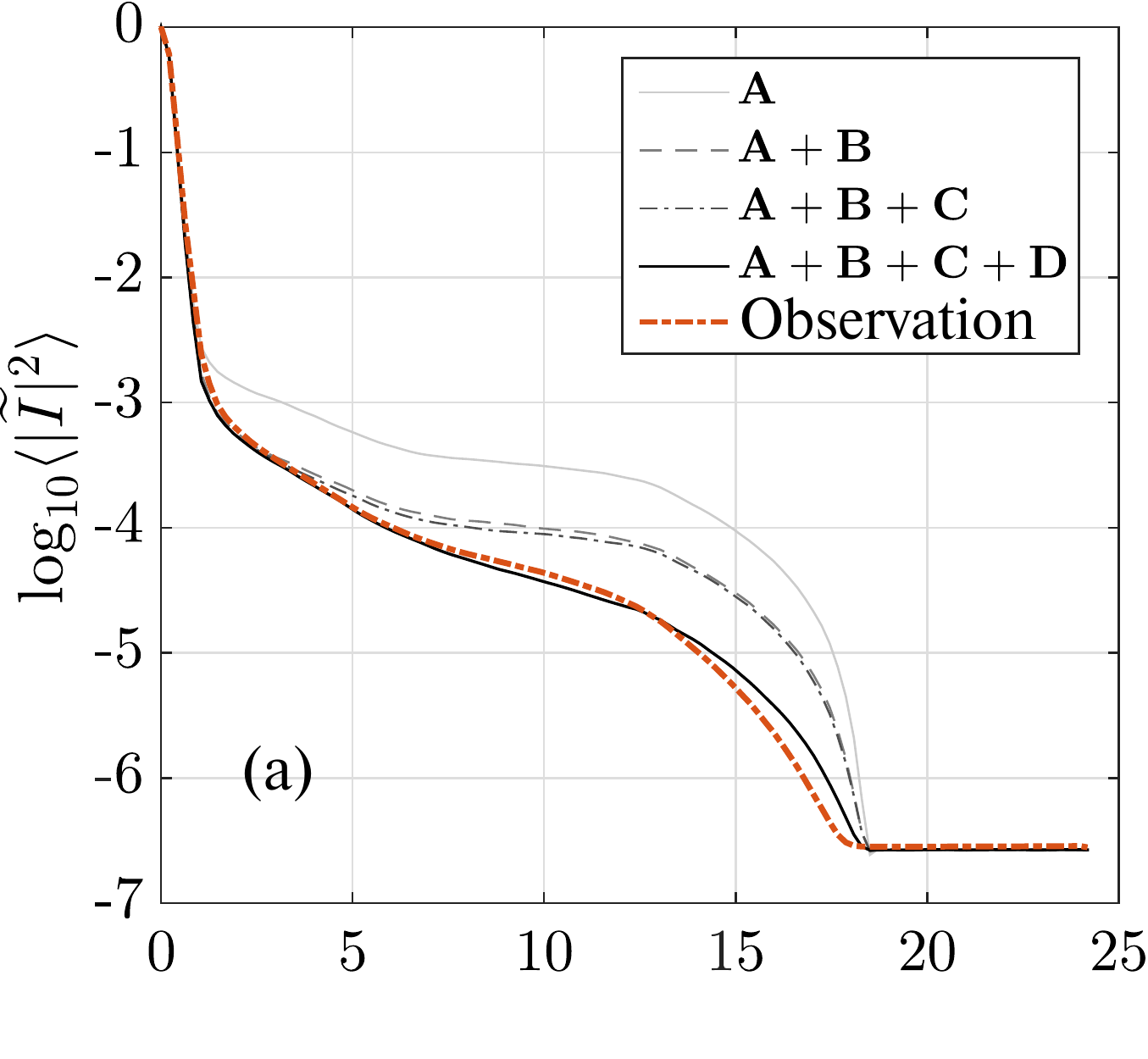}} \\
\end{minipage}\hspace{2.5pt}
\begin{minipage}[h]{0.316\linewidth}
\center{\includegraphics[width=\linewidth] {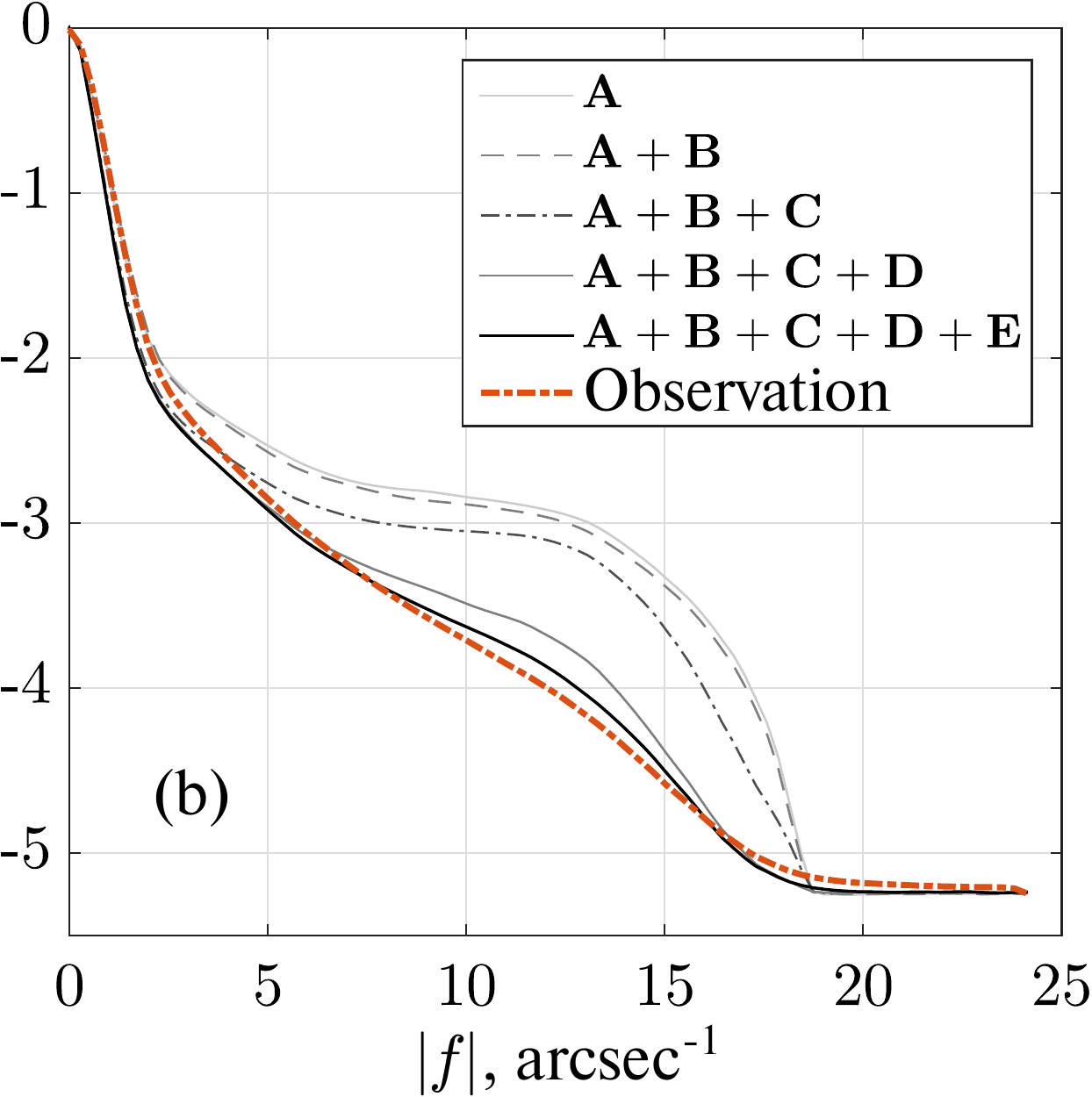}} \\
\end{minipage}\hspace{2.5pt}
\begin{minipage}[h]{0.31\linewidth}
\center{\includegraphics[width=\linewidth] {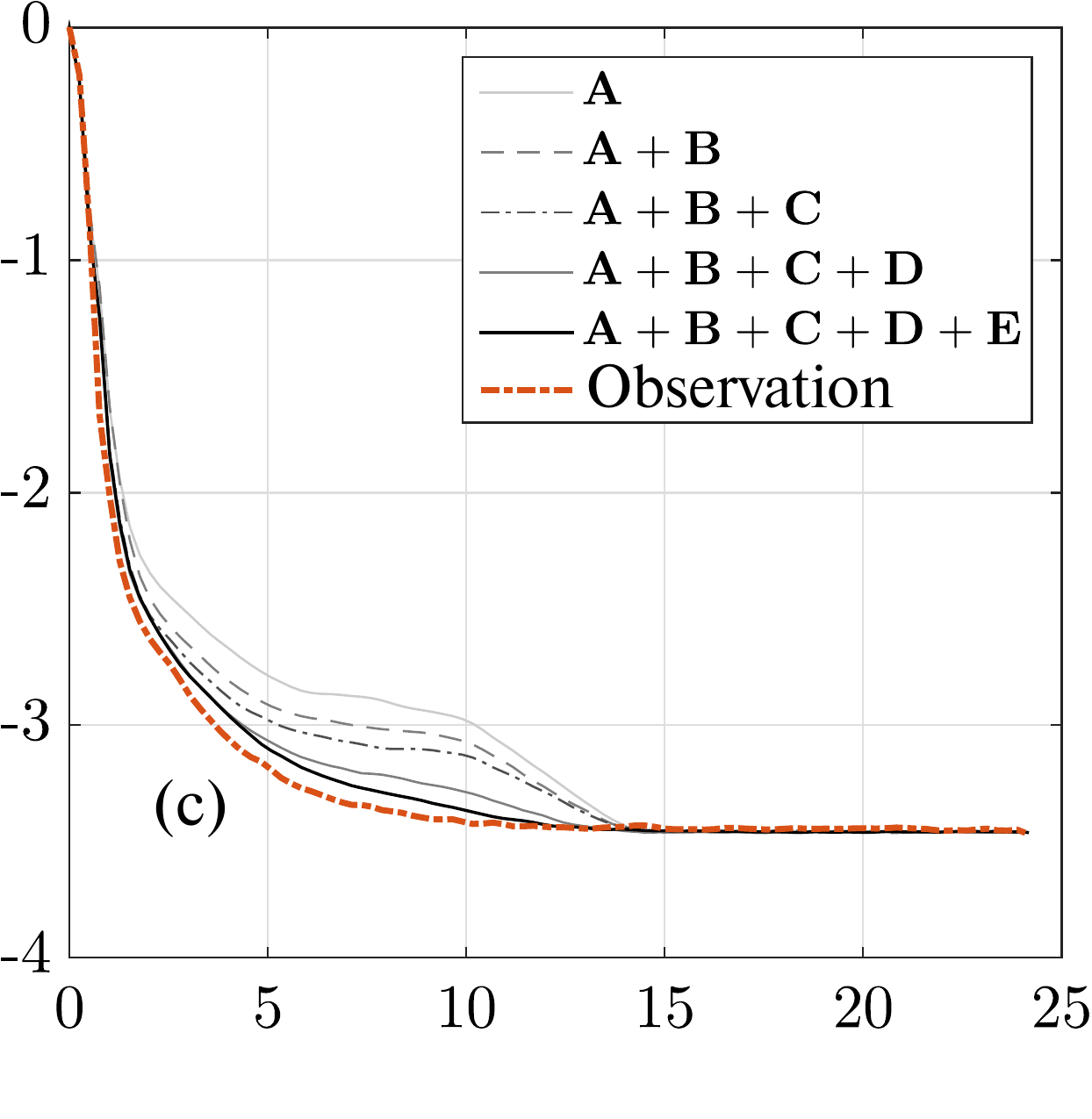}}
\end{minipage}
\caption{{The logarithms of slices of azimuth-averaged normalized model power spectra with successive inclusion of various factors taken into account are compared with real ones (from observations). The horizontal axis shows the modulus of the spatial frequency. Considered factors in the model: {A}---the CIC noise, photon noise, readout noise, defocus, atmospheric seeing $\beta$, motion of turbulent layers, filter wavelength, {B}---the finiteness of the exposure value,
{C}--- the telescope aberrations, {D}---the telescope jitter, {E}---the finiteness of the filter bandwidth. Information about series and simulation parameters:  {(a)}---the H${\alpha}$ filter, stellar magnitude $m_{\rm H\alpha}=-1\,.\!\!^{\rm m}1$, %$mag=-1\,.\!\!^{\rm m}1$,
seeing $\beta=1\,.\!\!^{\prime\prime}5$, wind speed $v_{\rm wind}=10$~m\,s$^{-1}$, the value of the telescope jitter $j_{\rm tel}=0\,.\!\!^{\prime\prime}08$, the value of the defocus in the displacements of the secondary mirror $\varepsilon_{M2}=11$~microns, {(b)}---the $R$ filter, $m_R=6\,.\!\!^{\rm m}5$, %$mag=6\,.\!\!^{\rm m}5$,
$\beta=0\,.\!\!^{\prime\prime}62$, $v_{\rm wind}=6$~m\,s$^{-1}$, $j_{\rm tel}=0\,.\!\!^{\prime\prime}085$, $\varepsilon_{M2}=-10$~microns, {(c)}---the $I$ filter, $m_I=10\,.\!\!^{\rm m}2$, %$mag=10\,.\!\!^{\rm m}2$,
$\beta=1^{\prime\prime}$, $v_{\rm wind}=8$~m\,s$^{-1}$, $j_{\rm tel}=0\,.\!\!^{\prime\prime}08$, $\varepsilon_{M2}=11$~microns.
}}
\label{fig:20}
%\end{center}
%\vspace{-20pt}
\end{figure*}

To take into account anisoplanatism, we assume that one of the
phase screens is located at an altitude of 10 km, and the second---at
zero altitude. Then, for example, if two stars are at an angular
distance of $5^{\prime\prime}$ from each other, then at an altitude
of 10~km, apertures for each star will be cut out from the first
phase screen at a distance approximately equal to 24~cm from each
other . On the other hand, the same aperture for both stars will be cut
from the second screen.

Based on the object function $O(\alpha_x,\alpha_y)$ specified as an array, we calculate the relative position of
the centers of the apertures cut out from the phase screen. Then
we get the PSF for each star. Every star in object function is then isolated (so that the other sources in object function except the chosen one are ``turned off'') and convolved with its PSF to get a set of images for each star in the field. Each of these images is an
atmospherically distorted image of a star if there were no other
stars. Further, we add up all images from this set
and obtain final image, taking into account anisoplanatism.

\subsubsection{The telescope jitter}

With the help of special experiments, described in Appendix~\hyperref[app:jitter]{B}, it
was found that the image of a star is ``blurred'' by about
\mbox{$j_{\rm tel}\approx0\,.\!\!^{\prime\prime}08$} due to the
jitter of the telescope mount. This affects the power spectra in
the form of a decrease in the high-frequency region (see Fig.~\ref{fig:20}).
Analysis shows that the telescope oscillations are dominated by
harmonics at frequencies of 20, 40, 60~Hz. Naturally, harmonics
above 30~Hz will significantly suppress the power spectrum at high
spatial frequencies, since typical exposures are 23~ms.

The telescope jitter is taken into account by convolving the image
with a segment whose brightness distribution is determined by the
ratio of the exposure length and typical amplitude of the jitter.

\subsubsection{Other factors}

The model takes into account the aberrations of the telescope
measured by the Shack--Hartmann sensor \citep{Potanin2017}, and
it is also possible to set the defocus value. In addition to atmospheric and optical distortions, we also
take into account photon noise, readout noise, and CIC noise
(Clock-induced charge, exponentially distributed). We assume that
the photon noise follows a Poisson distribution, and the readout
noise follows a Gaussian distribution. All the necessary
information for their determination (readout noise variance,
electron multiplication gain, exposure, detector sensitivity) is taken from
real data cubes (FITS observation files).

\subsection{Comparison with real data}

To validate the model, we compare the power spectra obtained in real observations
with the power spectra generated in the model. We sequentially include into consideration various ``refinements'' of the model, such as the finite filter width, the finite exposure value, and so on. Three observations were selected for comparison in different filters and with different magnitudes of objects.

It can be seen from the graphs (see Fig.~\ref{fig:20}) that with all the
factors taken into account, our model predicts the shape of the
power spectra quite accurately. Therefore, we can proceed directly
to the study and theoretical comparison of the speckle
interferometric capabilities of detectors.

\subsection{Comparison of EMCCD and qCMOS}
\label{subs:num_compar}

In this Section, we use the model just described to estimate the expected performance of the upgraded instrument compared to the previous version. As a metric, we will consider the ratio $S/N$ in the power spectrum.
%fig21
\begin{figure}[t!] \vspace{2mm}%\setcaptionmargin{5mm} \onelinecaptionsfalse \captionstyle{normal}
\begin{center}
\center{\includegraphics[width=0.92\linewidth] {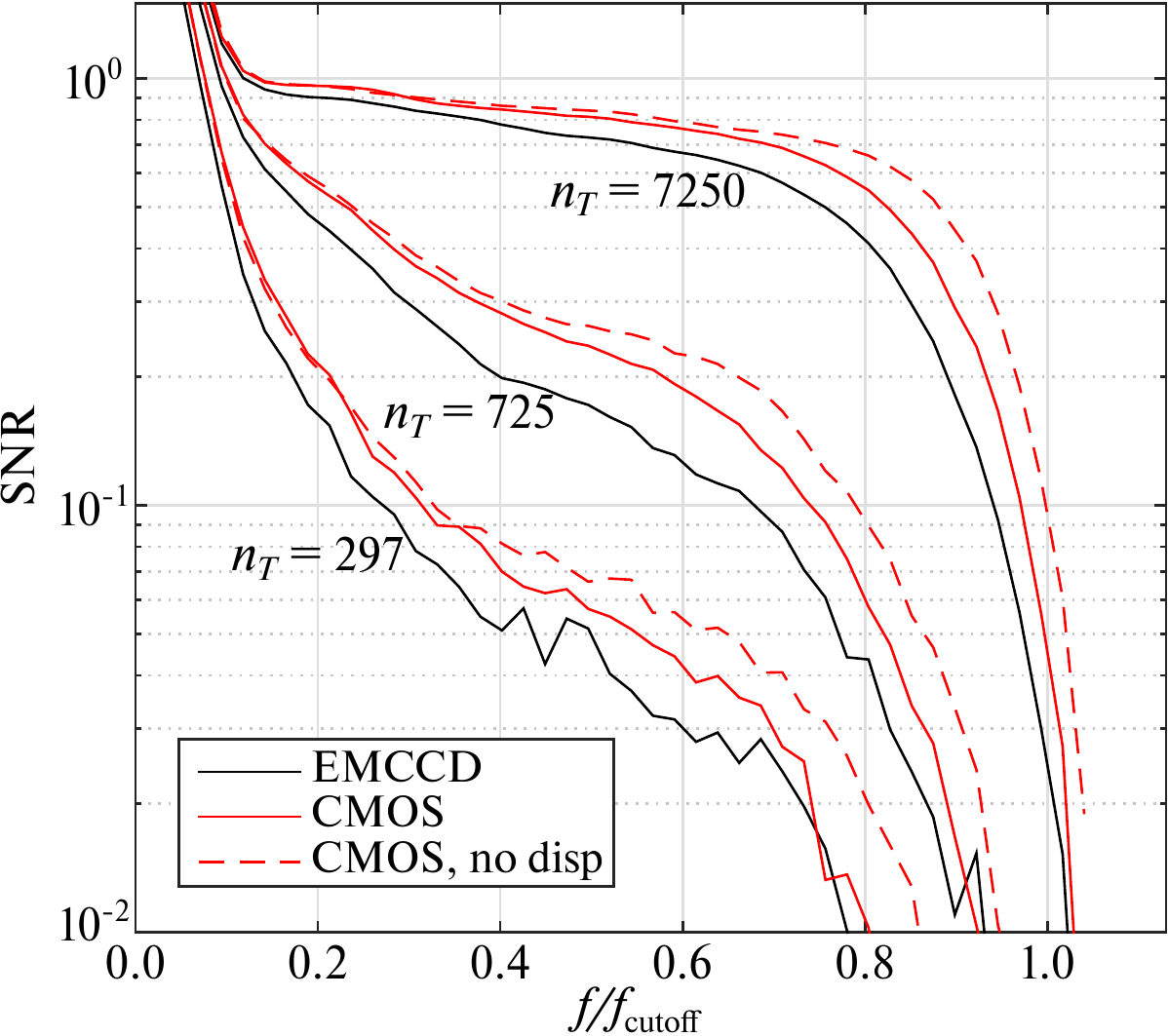}}%{SpeckleCMOS_figs/pspec/pspec_EMCCD_CMOS.eps}}
\vspace{0pt}
\caption{
Comparison of the $S/N$ ratio in power spectra per frame for EMCCD, CMOS and CMOS without Wollaston prism dispersion. The horizontal axis shows the modulus of the spatial frequency, normalized to the cutoff frequency $D/\lambda$. The groups of curves correspond to the different average number of photons in the frame $n_T$ (indicated in the figure).
\label{fig:21}}
\end{center}
\vspace{-10pt}
\end{figure}

The simulation was performed for EMCCD and CMOS, the parameters of
which are shown in Table~\ref{tab:1}. The spectra were calculated over 2000
frames in the region of $140\times140$~px, which corresponds to
$2\,.\!\!^{\prime\prime}8\times2\,.\!\!^{\prime\prime}8$. The
angular scale for ease of comparison was chosen to be the same (as
can be seen from Table~\ref{tab:1}, it differs only slightly). Again, for
ease of comparison, we assumed the same number of registered
photons per frame, and in both cases the dispersion of the
Wollaston prism was taken into account.

The seeing was chosen to be $0\,.\!\!^{\prime\prime}73$,
atmospheric model: two layers of equal intensity moving in
perpendicular directions at a speed of 10~m\,s$^{-1}$.
Exposure---30~ms (the temporal evolution of the image was taken
into account), the band $I_c$ (the dependence of the image on the
wavelength was taken into account).

Figure~\ref{fig:21} shows the $S/N$ ratio (SNR) in the power spectrum for
objects of three different brightnesses. CMOS provides SNR better
than EMCCD by about 1.2--1.6 times. This effect is especially
noticeable in the case of faint objects, since the EMCCD efficiency
for them is limited by the CIC noise. Figures~\ref{fig:11} and \ref{fig:21} confirm
the results of the theoretical evaluations carried out in
Section~\ref{subs:detector_mode}. For bright objects, the difference between CMOS and
EMCCD is less pronounced, since atmospheric effect dominates the noise.

\section{RESULTS}
\label{sec:ress}
The classic problem solved by speckle interferometry is to test
stars for binarity and, if the latter is detected, to determine
the parameters of binarity: separation, position angle, and
contrast. The study of binarity is also relevant for UX\,Ori type variables. The UX\,Ori type stars are a
subclass of young variable stars with sharp chaotic brightness
variations, the amplitude of which reaches $3^{\rm m}$. Flux variations
in stars of this type are due to the variable absorption of direct
stellar radiation in the circumstellar envelope \citep{Grinin1995}. Absorption can occur either in the protoplanetary disk or in
dust clouds above the disk---in the dusty disk wind or in the
remnants of the protostellar cloud falling onto the disk.

The presence of a stellar component has a significant effect
on the protoplanetary disk around a young star. The outer parts of
the disk are swept out of the system, the lifetime of the disk is
reduced \citep{Kraus2012,Zagaria2022}. The statistics
of exoplanets in binary star systems indirectly indicate the
disk-component interaction. Thus, it was found that in stars with
a hot Jupiter, the probability of detecting a stellar component with
 a separation from 50 to 2000~AU is significantly
larger than for field stars \citep{Ngo2016}.

Thus, information on binarity in young stars is of great value
both for statistical studies and for the astrophysical
interpretation of individual objects. From an observational point
of view, information about the stellar  component must be
taken into account when interpreting observations of the UX\,Ori
stars, since the component can distort the flux and polarization
measurements. For example, in the very deep eclipses of 2014-2015,
the secondary component dominated in the total flux of RW\,Aur \citep{Dodin2019}.

The Gaia DR3 survey \citep{Vallenari2022} has more than 50\%
completeness only for systems with a separation greater than
$0\,.\!\!^{\prime\prime}54$ \citep{Fabricius2021}. At the same
time, the diffraction limited resolution of ground-based telescopes in the optical range, which is achievable, in
particular, using the speckle interferometry method, is much
better. In case of 2.5-m telescope, the diffraction limited resolution is
\mbox{$0\,.\!\!^{\prime\prime}05$--$0\,.\!\!^{\prime\prime}08$}.
Thus speckle interferometric data are a valuable addition to the
Gaia survey.

In the period from 2018 to 2022, speckle interferometric
observations of 25 UX\,Ori stars from the list in \cite{Poxon2015} were carried out with the speckle polarimeter at the 2.5-m
telescope. \cite{Poxon2015}, in turn, adopted this
list from the Internet resource of the American Association of
Variable Star Observers AAVSO \citep{Watson2014}. Among the
selected stars, only one is present in the Gaia DR3 Part 3
Non-single stars survey \citep{Vallenari2022}---the BO Cep star
assigned to the Single Lined Spectroscopic binary model with
a period of 10.7 days, which gives a separation that lies deep
below the diffraction limited resolution of the telescope and undetectable
by speckle interferometry.

\subsection{Single UX\,Ori stars---achievable contrast}

As a result of observations, binarity was not found in 23 stars
from the list. Table~\ref{tab:2} shows the results of processing these
observations---achievable contrast at a distance of
$0\,.\!\!^{\prime\prime}2$ and $1^{\prime\prime}$.

\renewcommand{\baselinestretch}{1.0}
\begin{table*}
%\setcaptionmargin{0mm}\onelinecaptionstrue \captionstyle{normal}
\caption{UX\,Ori variables for which binarity has not been found.  $\beta$---seeing, $n_p$---the number of registered photoelectrons per second; $\epsilon_{\rm lim_{0.2}}$ and $\epsilon_{\rm lim_{1.0}}$---estimates of extreme contrast at a distance of $0\,.\!\!^{\prime\prime}2$ and $1^{\prime\prime}$ respectively.} \label{tab:2}
\medskip
%\footnotesize
\setlength\extrarowheight{-0pt}
%\medskip
%\setlength\extrarowheight{-0pt}
\renewcommand{\arraystretch}{1.65} % Default value: 1
\begin{tabular}{l|c|c|c|c|c|c}
\hline
\multicolumn{1}{c|}{Object}& Time, UT & Filter & $\beta$, arcsec & $n_p$ & $\epsilon_{\rm lim_{0.2}}$ & $\epsilon_{\rm lim_{1.0}}$\\
\hline
ASAS\,J055007+0305.6 & 2018/12/02 22:23:37 & $I_c$ & 1.24 & 8.69e+04 & 3.63 & 4.79\\
& 2018/12/07 00:14:03 & $I_c$ & 1.56 & 8.21e+04 & 3.18 & 5.07\\
& 2019/01/25 21:02:47 & $V$ & 1.50 & 6.54e+04 & 2.16 & 3.86\\

%\hline
BF\,Ori & 2018/12/02 23:00:30 & $I_c$ & 1.51 & 2.43e+05 & 3.82 & 5.32\\
& 2018/12/02 23:09:27 & $I_c$ & 1.38 & 1.55e+05 & 3.55 & 5.50\\
& 2018/12/07 00:23:15 & $I_c$ & 1.89 & 1.18e+05 & 3.15 & 4.40\\
%\hline
BH\,Cep & 2018/12/02 18:50:59 & $I_c$ & 1.12 & 1.30e+05 & 2.96 & 5.07\\
%\hline
BO\,Cep & 2018/12/02 18:56:27 & $I_c$ & 1.13 & 5.81e+06 & 4.61 & 7.45\\
%\hline
CQ\,Tau & 2019/10/27 22:13:18 & $I_c$ & 0.83 & 9.66e+05 & 4.73 & 7.46\\
%\hline
GM\,Cep & 2018/12/02 19:09:51 & $I_c$ & 1.04 & 3.48e+04 & 2.44 & 4.39\\
& 2019/01/20 16:51:46 & $I_c$ & 1.27 & 3.84e+04 & 3.38 & 4.49\\

%\hline
GSC\,05107-00266 & 2018/07/31 20:39:51 & $I_c$ & 1.56 & 1.36e+05 & 3.58 & 5.23\\
%\hline
GT\,Ori & 2018/12/02 23:27:43 & $I_c$ & 1.24 & 8.27e+04 & 2.86 & 4.75\\
%\hline
HQ\,Tau & 2018/12/02 20:24:51 & $I_c$ & 1.13 & 1.76e+05 & 3.13 & 5.13\\
%\hline
IL\,Cep & 2018/12/02 19:26:34 & $I_c$ & 1.10 & 9.23e+05 & 4.22 & 7.20\\
%\hline
LO\,Cep & 2018/12/02 19:38:49 & $I_c$ & 1.15 & 2.79e+04 & 0.34 & 3.44\\

%\hline
PX\,Vul & 2017/03/09 01:49:04 & $I_c$ & 0.89 & 2.69e+05 & 3.11 & 6.00\\
& 2018/08/29 20:39:41 & $I_c$ & 1.23 & 1.25e+05 & 2.36 & 4.76\\
& 2018/08/29 20:42:55 & $V$ & 1.28 & 5.66e+04 & 2.02 & 3.64\\
& 2020/04/26 01:42:22 & $I_c$ & 1.06 & 1.94e+05 & 3.31 & 4.89\\
& 2020/06/10 00:15:59 & $I_c$ & 0.71 & 1.86e+05 & 3.17 & 7.03\\

%\hline
RR\,Tau & 2018/04/01 17:53:05 & 880 & 1.13 & 1.73e+04 & 1.95 & 3.62\\
& 2018/04/01 17:46:49 & $I_c$ & 1.04 & 7.63e+04 & 3.63 & 5.38\\

%\hline
RZ\,Psc & 2018/12/02 19:56:27 & $I_c$ & 1.05 & 1.03e+05 & 2.88 & 5.37\\
& 2019/01/20 17:00:29 & 880 & 1.18 & 1.83e+04 & 2.19 & 4.11\\

%\hline
SV\,Cep & 2018/12/02 18:40:33 & $I_c$ & 1.04 & 1.48e+05 & 3.85 & 6.10\\
%\hline
T\,Ori & 2018/12/02 22:53:45 & $I_c$ & 1.34 & 2.22e+05 & 3.26 & 4.99\\
%\hline
V1012\,Ori & 2018/12/02 21:51:04 & $I_c$ & 1.37 & 4.23e+04 & 2.65 & 4.60\\
%\hline
V1977\,Cyg & 2018/12/02 18:05:48 & $I_c$ & 1.24 & 2.14e+05 & 3.04 & 5.48\\
%\hline

V517\,Cyg & 2018/12/02 18:26:04 & $V$ & 1.22 & 8.28e+03 & 0.19 & 2.83\\

%\hline

VV\,Ser & 2018/07/31 20:31:22 & $I_c$ & 1.41 & 1.12e+05 & 3.63 & 4.71\\

%\hline
VX\,Cas & 2018/12/02 20:02:56 & $I_c$ & 1.05 & 8.03e+04 & 2.94 & 5.11\\
%\hline
WW\,Vul & 2018/08/29 20:31:37 & $I_c$ & 1.17 & 1.64e+05 & 3.21 & 5.33\\
& 2018/08/29 20:34:52 & $V$ & 1.47 & 1.29e+05 & 2.31 & 4.04\\

%\hline

XX\,Sct & 2018/07/31 20:22:21 & $I_c$ & 1.46 & 5.11e+04 & 1.47 & 3.87\\
\hline
\end{tabular}
\renewcommand{\arraystretch}{1.0} % Default value: 1
\end{table*}
\renewcommand{\baselinestretch}{1.0}

\subsection{BM\,And}

BM\,And is a young T~Tau-type star and a UX~Ori-type variable
whose magnitude varies from \mbox{$m_V=11.5$} to $m_V=14.0$
\citep{Grinin1995}.

\begin{figure}[]\vspace{2mm} %\setcaptionmargin{5mm} \onelinecaptionsfalse \captionstyle{normal}
\includegraphics[width=0.88\linewidth] {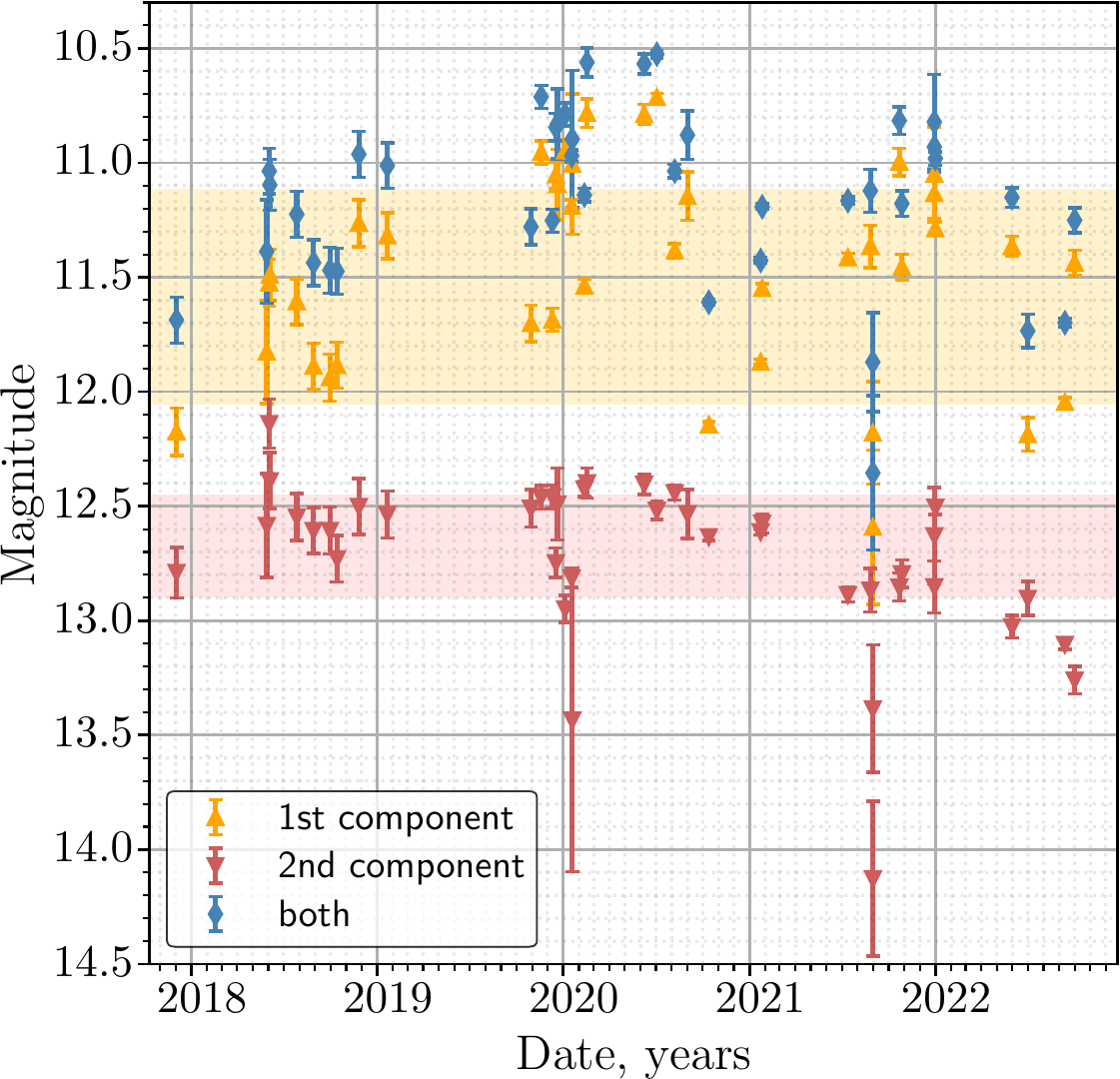}%
\caption{Light curve of BM\,And and its components in the $I$ filter. The center of the translucent band corresponds to the weighted average of the magnitude of the component in this filter. The half-width of the band is equal to the weighted standard deviation.
\label{fig:22}}
%\vspace{-7pt}
\end{figure}

\begin{figure}[] \vspace{2mm} %\setcaptionmargin{5mm} \onelinecaptionsfalse \captionstyle{normal}
\begin{center}
%\center{\includegraphics[width=\linewidth]{SpeckleCMOS_figs/BMAnd/bmand_orb.pdf}}
%\vspace{-20pt}
\begin{minipage}[h]{\linewidth}
%\hspace{0.7cm}
\center{\includegraphics[width=\linewidth]{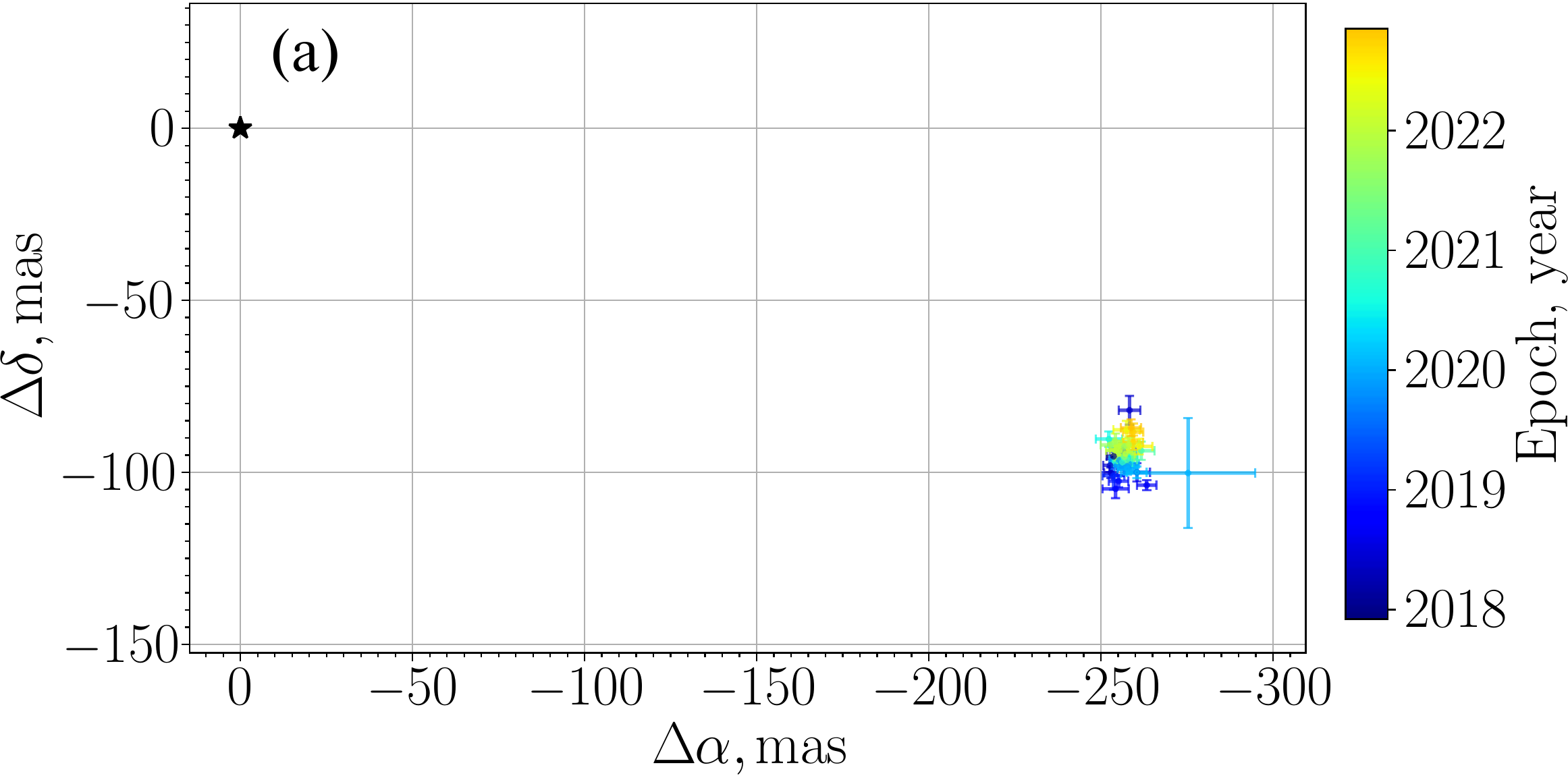}}%
%\center{\includegraphics[width=\linewidth]{SpeckleCMOS_figs/BMAnd/bmand_orb.pdf} \\ }
\vspace{0.1cm}
\end{minipage}
\begin{minipage}[h]{\linewidth}
%\hspace{0.7cm}
\center{\includegraphics[width=\linewidth] {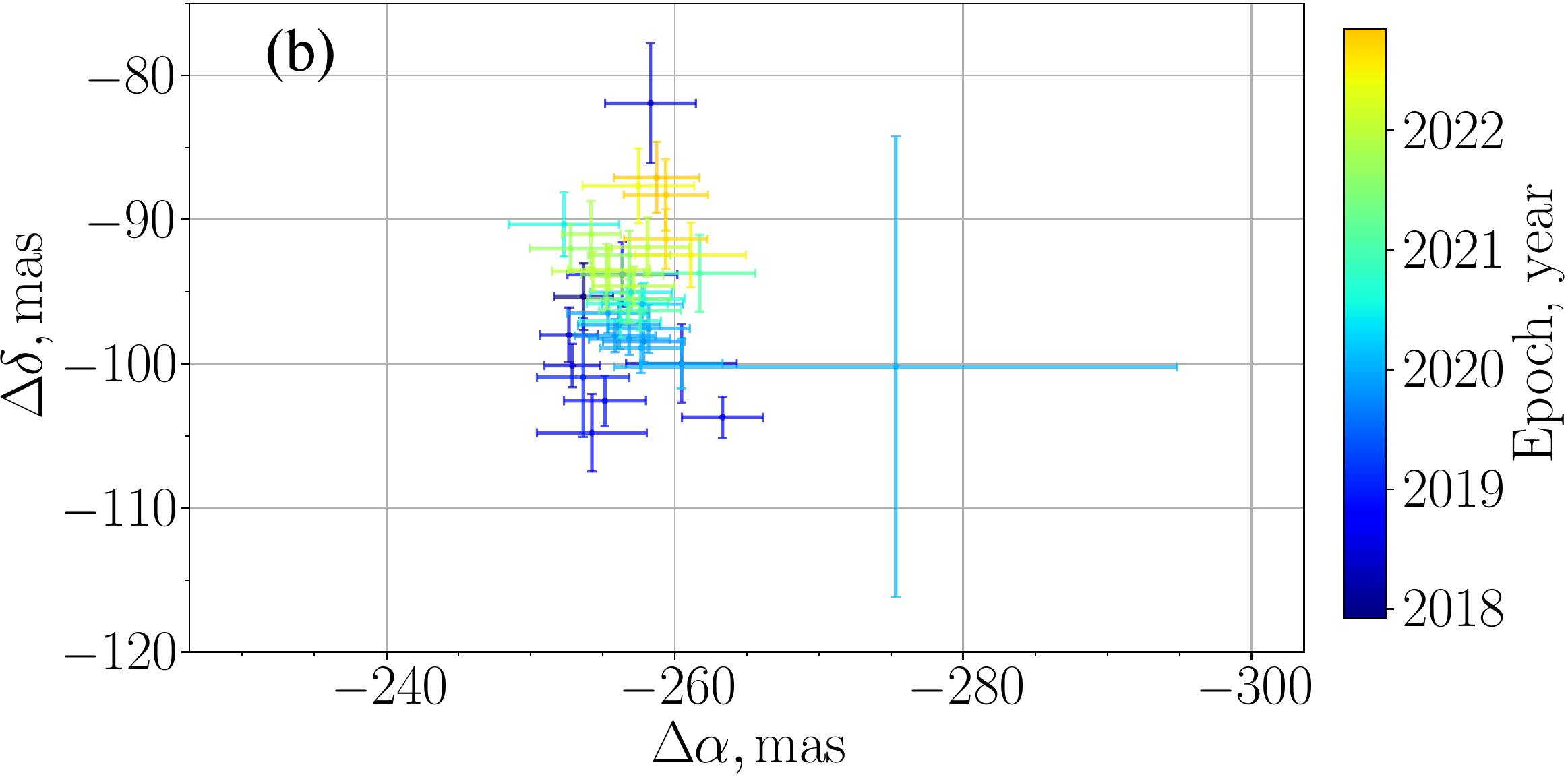}}%{SpeckleCMOS_figs/BMAnd/bmand_orb_zoom.pdf} \\ }
\vspace{0.1cm}
\end{minipage}
\begin{minipage}[h]{\linewidth}
%\hspace{-0.7cm}
\includegraphics[width=\linewidth]{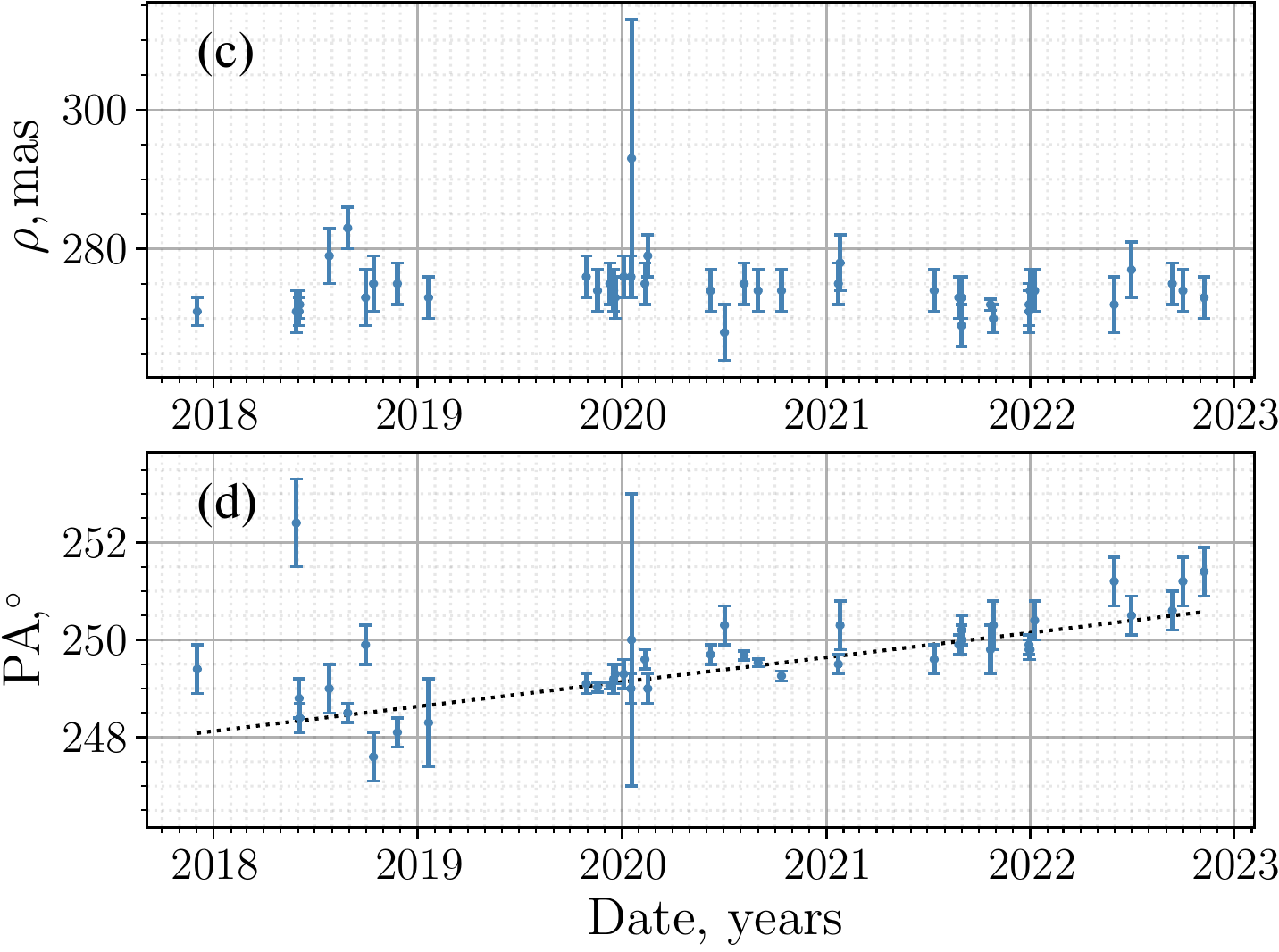}%{SpeckleCMOS_figs/BMAnd/BMAndseppa.pdf} \\
\end{minipage}
\vspace{-10pt}
\caption{Relative positions of the BM\,And components over all observations in the $I$ filter. A trend is observed in the position angle, which may be due to the orbital motion of the component. The linear weighted least squares method (LSM) gives the position angle change rate $0\,.\!\!^\circ50\pm{0\,.\!\!^\circ05}$~per year.
\label{fig:23}}
\end{center}

\end{figure}

The star was found to have binarity with parameters $\rho \approx
273$~mas, $\mathrm{PA} \approx 249^{\circ}$ (see Figs~\ref{fig:22} and \ref{fig:23}).
For BM\,And, a total of about 100 observations were made in the
$I_c, R_c, V$ filters in the period from 2017 to 2022. In the
$I_c$ filter: 40 magnitude measurements, 42 measurements of binary
parameters, $R_c$: 38 magnitude measurements, 39 measurements of
binary parameters, $V$: 18 magnitude measurements, 23 measurements
of binary parameters. The position angle of the star shows a
systematic change at a rate of
$0\,.\!\!^\circ50\pm{0\,.\!\!^\circ05}$ per year, no trend was
detected in the separation (see Fig.~ \ref{fig:23}).
The restoration of the orbit according to our observations is not possible due to the short interval on which they were performed.

According to Gaia\,DR3 catalogue \citep{Vallenari2022} excess astrometry
error---5.8~mas---is greater than the parallax value
$1.3\pm0.6$~mas. This is probably due to the simultaneous
influence of the binarity and variability of the object. In
addition, the scattering envelope, the presence of which is indicated
by the variable polarization of the object \citep{Grinin1995},
can also cause a shift in the object's photocenter \citep{Dodin2019}. Thus, estimating the distance and hence the magnitude of
the separation projection onto the celestial sphere in units of
length is difficult.

In addition to measuring the contrast between the components, the
total magnitude of the system was also estimated by photometry
using the standards observed before and after each observation of
the object. This made it possible to estimate the magnitudes of
both stars in the system (see Fig.~\ref{fig:22}).

The amplitude of the change in the brightness of the second
component obtained from observations turned out to be much smaller
than that for the main star of the system. So, in the $I_c$
filter, the ratio of the weighted standard deviations of the
stellar magnitudes of the components turned out to be 2.1, in the
$R_c$ filter---1.43, in the $V$ filter---1.4.

The following mean magnitudes of the second component in the $I,R,V$ filters are obtained:  \begin{list}{}{
\setlength\leftmargin{7mm} \setlength\topsep{0mm}
\setlength\parsep{0mm} \setlength\itemsep{0mm} }\vspace{0.2mm}
\item [] $m_2^I=12.67\pm0.23$, \mbox{$m_2^R = 13.90\pm0.44$},\vspace{0.4mm}\item [] $m_2^V = 14.96\pm0.38$. \end{list} %$\mathrm{I,R,V}$: $m_2^I~=~12.66~\pm~0.23$, $m_2^R = 13.90\pm0.44$, $m_2^V = 14.97\pm0.38$.

\subsection{NSV\,16694}
\label{sec:results}
%fig22
%fig23
%fig24
\begin{figure}[t!] %\setcaptionmargin{5mm} \onelinecaptionsfalse \captionstyle{normal}
%\begin{center}
\center{\includegraphics[width=0.92\linewidth]{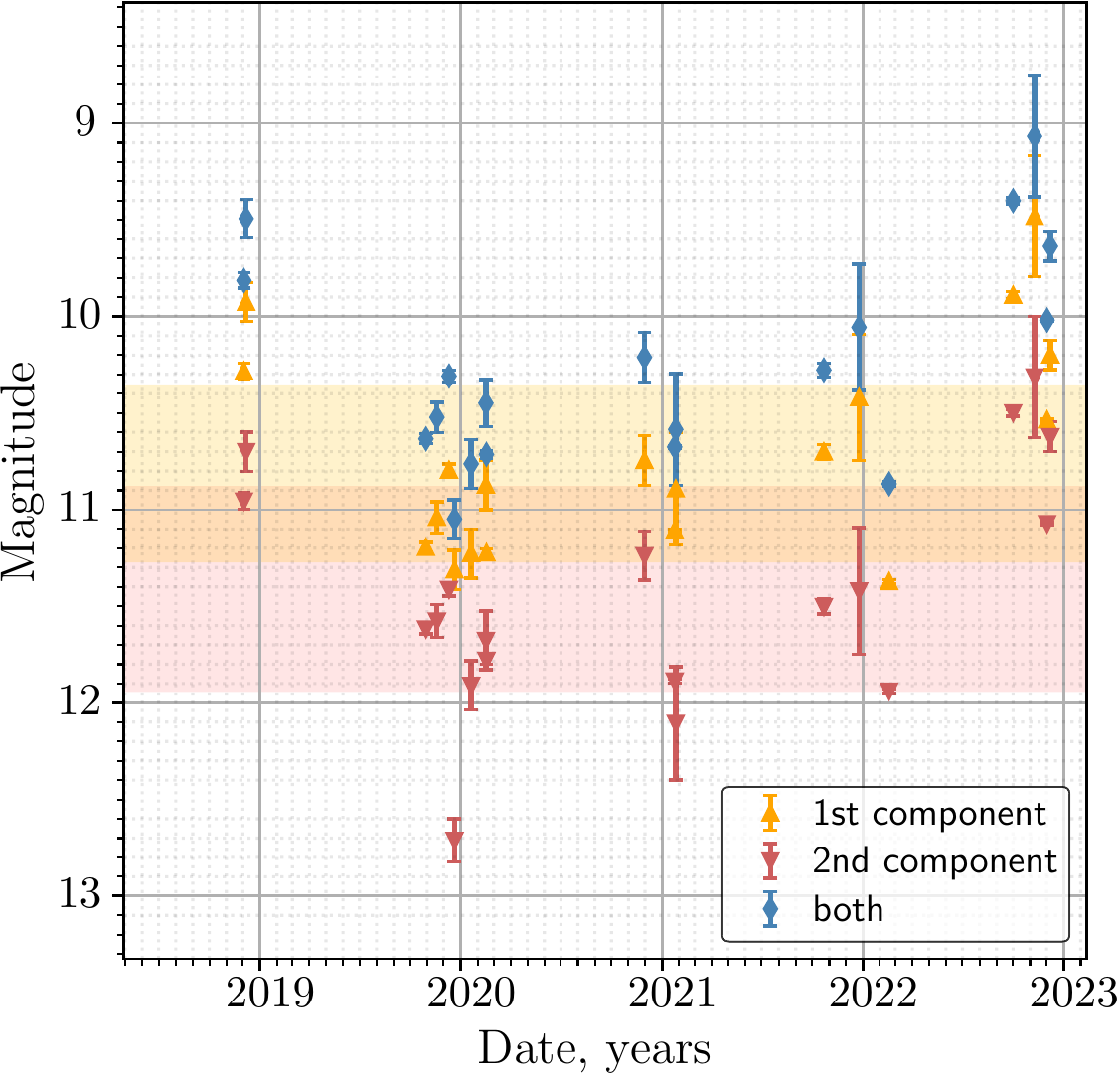}}%{SpeckleCMOS_figs/NSV16694/nsv16694.pdf}}
%\vspace{-15pt}
\caption{Light curve of NSV\,16694 and its components in the $I$ filter. The center of the translucent band corresponds to the weighted average of the magnitude of the component in this filter. The half-width of the band is equal to the weighted standard deviation.
\label{fig:24}}
%\end{center}
%\vspace{-7pt}
\end{figure}

%fig25
\begin{figure}[t!] %\setcaptionmargin{5mm} \onelinecaptionsfalse \captionstyle{normal}
\begin{center}
%\center{\includegraphics[width=\linewidth]{SpeckleCMOS_figs/BMAnd/bmand_orb.pdf}}
%\vspace{-20pt}
\begin{minipage}[h]{\linewidth}
%\hspace{0.7cm}
\center{\includegraphics[width=\linewidth]{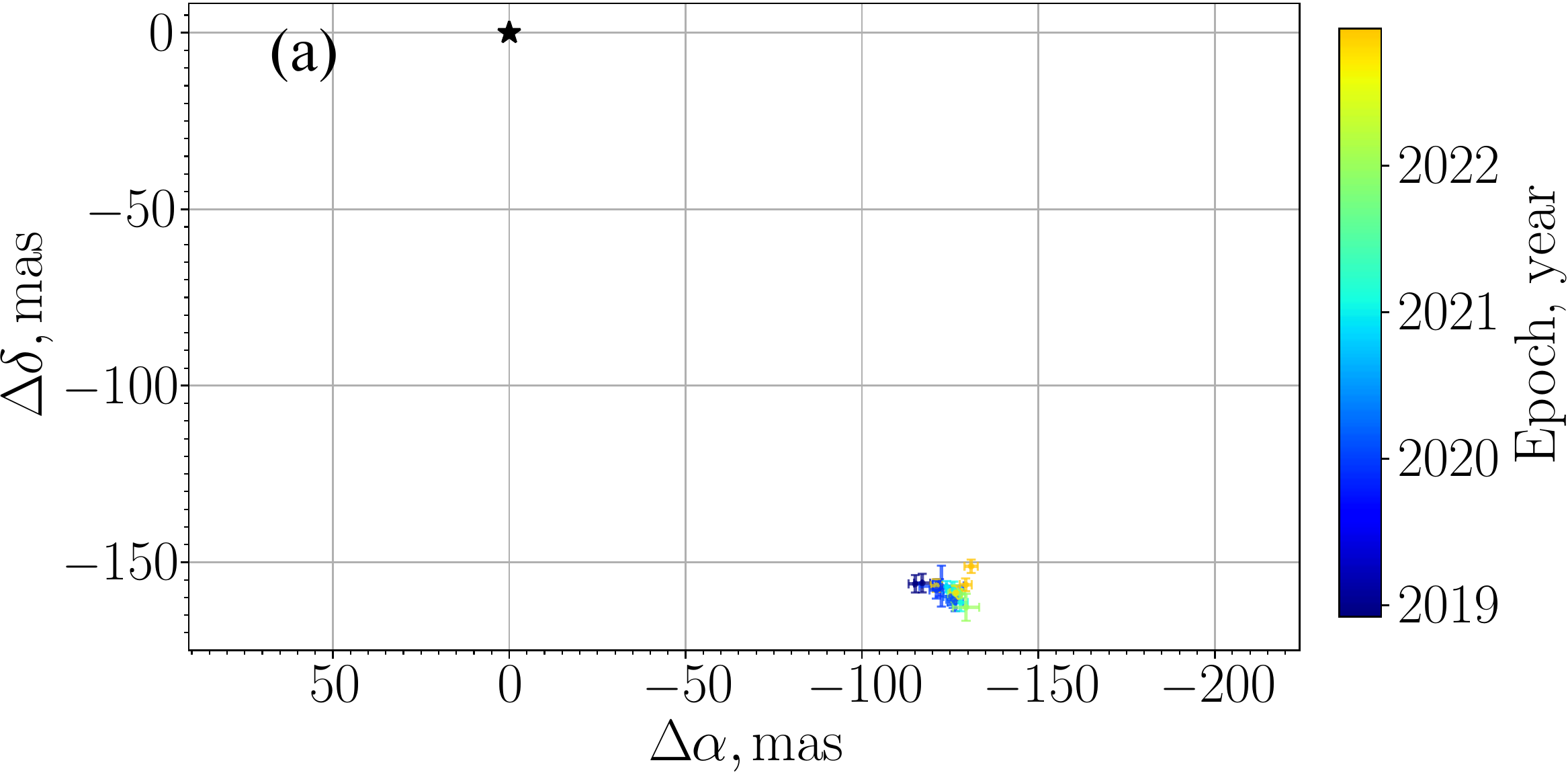}%{SpeckleCMOS_figs/NSV16694/nsv16694_orb.pdf}
\\ }
\vspace{0.1cm}
\end{minipage}
\begin{minipage}[h]{\linewidth}
%\hspace{0.7cm}
\center{\includegraphics[width=\linewidth]{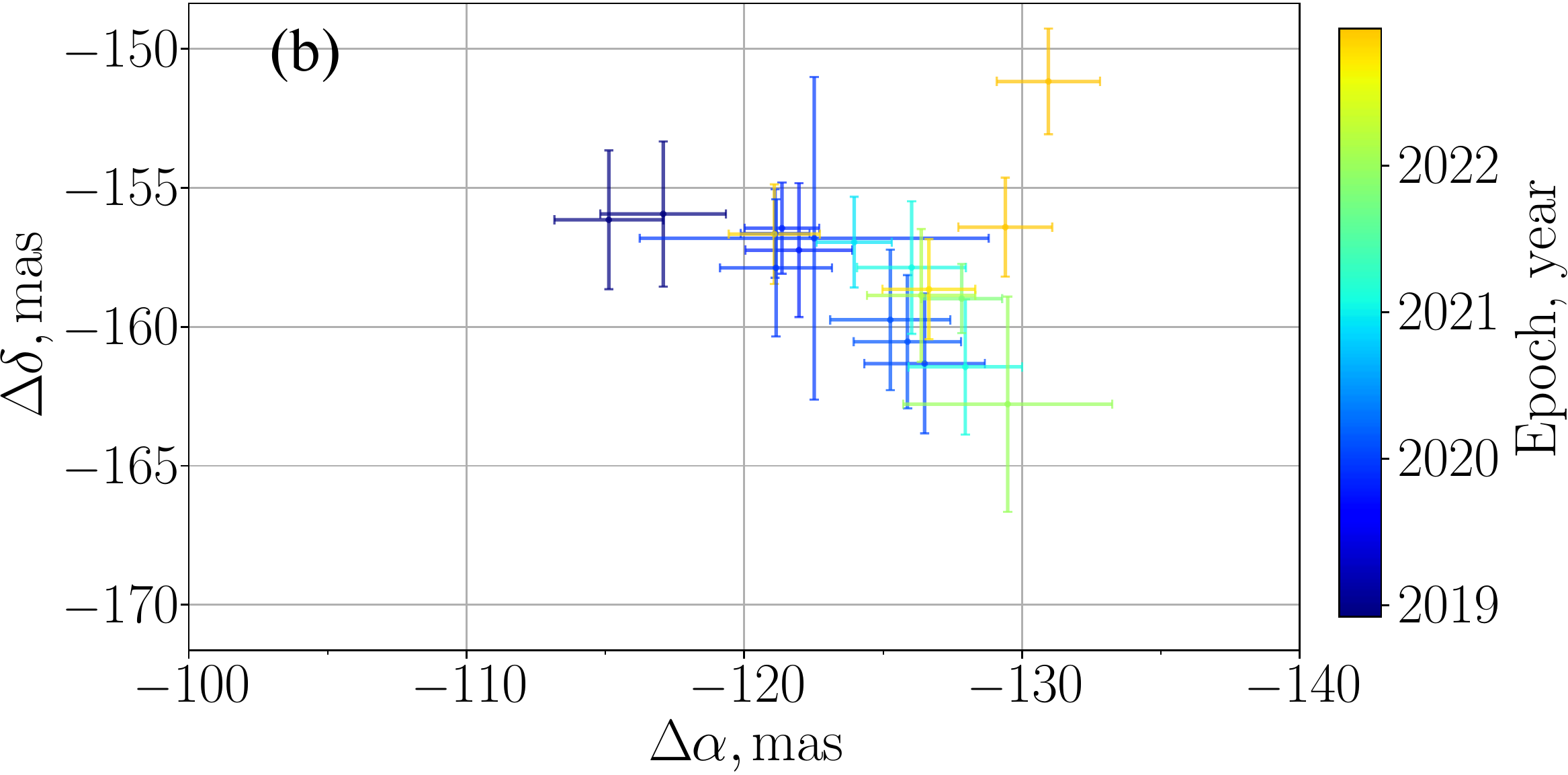}%{SpeckleCMOS_figs/NSV16694/nsv16694_orb_zoom.pdf}
\\ }
\vspace{0.1cm}
\end{minipage}
\begin{minipage}[h]{\linewidth}
%\hspace{-0.7cm}
\includegraphics[width=0.95\linewidth]{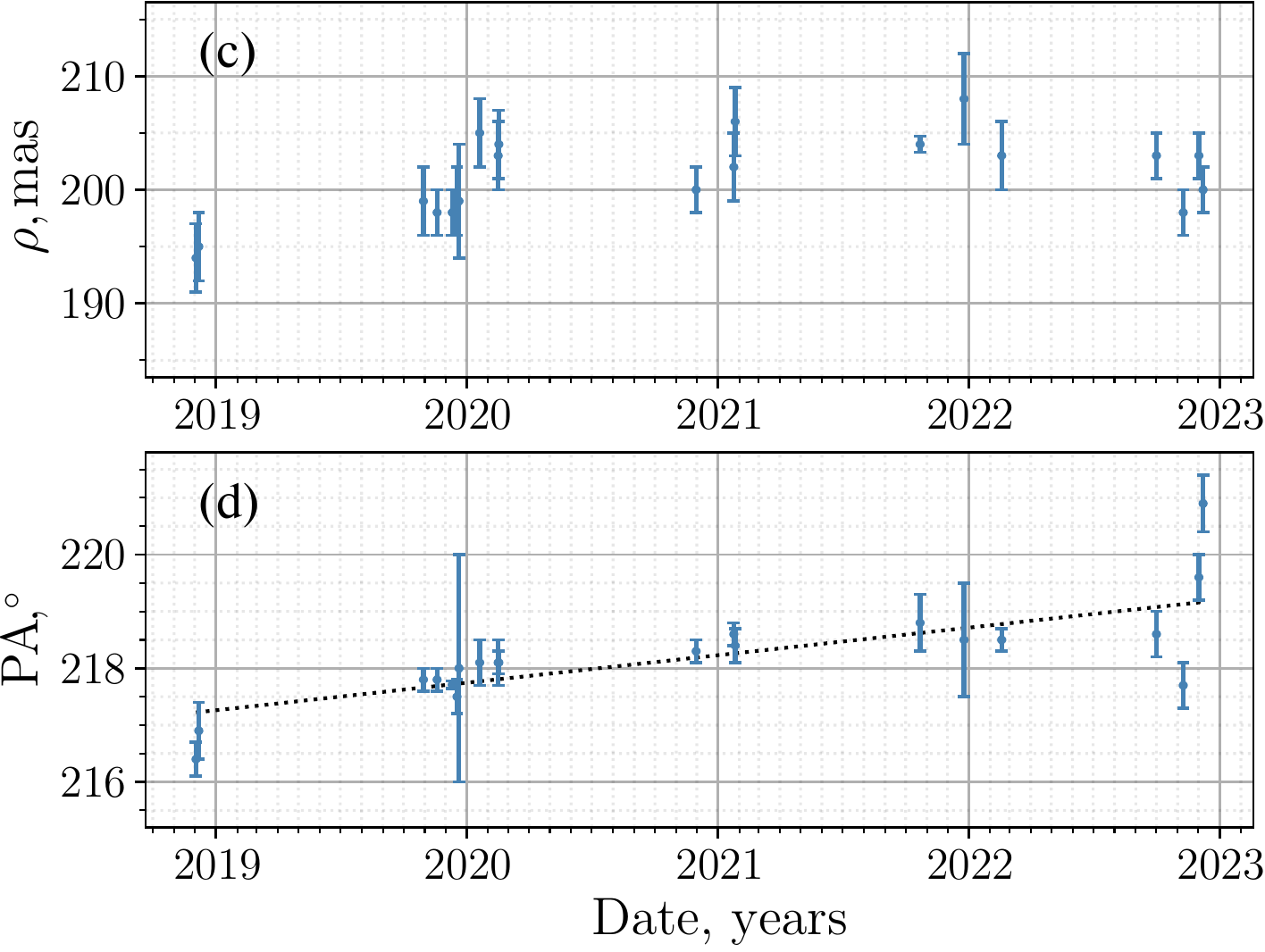} %{SpeckleCMOS_figs/NSV16694/nsv16694seppa.pdf} \\
\end{minipage}
\vspace{-10pt}
\caption{Relative positions of the NSV\,16694 components over all observations in the $I$ filter. There is a trend in position angle. A weighted linear least squares method gives the rate of change of the position angle $0\,.\!\!^\circ48\pm{0\,.\!\!^\circ07}$ per year.
\label{fig:25}}
\end{center}
\end{figure}

NSV\,16694 (TYC\,120-876-1, IRAS\,05482+0306) is a young stellar
% object with coordinates $\alpha=05^{\rm h}50^{\rm m}53\,.\!\!^{\rm s}714$,
 object with coordinates (J2000)\linebreak  ${\rm RA}=05^{\rm h}50^{\rm m}53\,.\!\!^{\rm s}714$,
% \mbox{$\delta=+03^{\circ}07^{\prime}29\,.\!\!^{\prime\prime}346$}
 \mbox{${\rm Dec}=+03^{\circ}07^{\prime}29\,.\!\!^{\prime\prime}346$}
 \citep{Vallenari2022} located in the direction of the molecular
 cloud complex in Orion. In 30--60$^{\prime\prime}$ to the north of the object lies a nebula, probably associated with the object.
  In the optical range, the object shows irregular variability \citep{Pojmanski2002}. The Gaia\,DR3 catalogue \citep{Vallenari2022} for
  NSV\,16694 lacks parallax and proper motion estimates, and the
  excess astrometry noise is 51~mas.

For NSV\,16694, we found binarity with parameters $\rho
\approx 202$~mas, ${PA} \approx 218\,.\!\!^\circ2$ (see Figs.~\ref{fig:24}
and \ref{fig:25}). From 2019 to 2022, observations were made in the $I$
filter: 19 magnitude measurements, 20 measurements of the binarity
parameters, in the $R$ filter: 11 magnitude measurements, 12
measurements of the binarity parameters, in the $V$ filter: 7
magnitude measurements, 8 measurements of the binarity parameters.

In the period from 2019 to 2022, the position angle changed
systematically at a rate of
$0\,.\!\!^\circ48\pm{0\,.\!\!^\circ07}$ per year. The magnitudes
of the components were estimated independently, similarly to how
it was done for BM\,And. It is noteworthy that in 2019 and the end
of 2022, both components of the system were significantly brighter
than in 2021 and most of 2022. Coincidence cannot be ruled out to
explain such synchronization, especially given the small time frame
on which our observations were made. However, a more natural
explanation is the hypothesis that either one or both of the
components are compact reflection nebulae, the brightness of which
is proportional to the brightness of some variable source screened
by a dust cloud. This hypothesis is supported by the fact that
\cite{Magnier1999} concluded that NSV\,16694 is a group of
young stellar objects embedded in a protostellar cloud with
complex morphology.

\section{CONCLUSIONS}

In this paper, we investigate the applicability of the Hamamatsu
ORCA-quest low-noise CMOS detector in speckle interferometry. The
work was carried out in the context of the upgrade of the
``speckle polarimeter'' instrument mounted on the 2.5-m telescope.
We present a detailed description of the instrument, as well as
the speckle interferometry technique we use to characterize binary
sources. We have carried out a study of the main characteristics
of the detector---readout noise, dark current, registration rate.
These characteristics have been shown to be within the limits
declared by the manufacturer. The readout noise distribution is
normal up to three standard deviations.

Previously, the Andor iXon 897 EMCCD detector was used in the
speckle polarimeter, so the CMOS Hamamatsu ORCA-quest was
considered in comparison with this detector. In speckle
interferometry, the main observed quantity is the image power
spectrum averaged over a certain series of frames. The ratio $S/N$
in the average power spectrum was chosen as the main metric for
comparing detectors. A simple analytical model shows that for
bright objects there is practically no difference between CMOS and
EMCCD, since in this case the dominant contribution to the noise
comes from atmospheric noise, also called speckle noise.
Atmospheric noise does not depend on the properties of the
detector. For faint
 objects, CMOS provides a 1.5--2 times better
$S/N$ ratio than EMCCD due to the absence of CIC noise (spurious
 charge) in CMOS.

We also carried out a detailed quantitative analysis of the gain
from the use of a CMOS detector using numerical simulation of data
acquisition by the Monte Carlo method. The numerical model takes
into account all the main effects that affect the operation in the
speckle interferometry mode: light propagation through turbulent
atmosphere and telescope, readout noise, CIC noise,
photon noise, amplification noise. The model was verified by us using the data
obtained with a speckle polarimeter. In the process of
verification, telescope vibrations were detected, leading to image
jitter with an amplitude of up to $0\,.\!\!^{\prime\prime}08$ and a
characteristic frequency of 40--60~Hz. Vibrations significantly
reduce the $S/N$ ratio in the power spectrum in speckle
interferometric observations.

Many CMOS detectors, in particular, the detector considered in
this study, operate in a rolling shutter mode, that is, rows are
read in turn. Using numerical simulations, we show that the use of
a rolling shutter does not significantly affect speckle
interferometry measurements.

Some features of the Hamamatsu ORCA--quest that we have found
deserve special mention. Thus, the detector shows a significant
non-linearity, reaching 15--20\% in the region of low fluxes. The
correction procedure makes it possible to reduce the non-linearity
to 1--2\% at signal levels greater than $1~e^{-}$. The readout
noise in different pixels is not statistically independent, which
causes an artifact in the average power spectrum that needs to be
corrected.

As an example of an astrophysical problem, we present a survey of
25 young variable stars with the purpose of characterizing their
binarity. Of these, 23 objects turned out to be single; the limits
of detection of secondary components are given. Binarity was found
in BM\,And and NSV\,16694. BM\,And has a separation of 273~mas, a
position angle of $249^{\circ}$, and the binary is probably a
gravitationally bound pair. The variability of BM\,And is due to
variations in the brightness of the main component. For
NSV\,16694, the separation turned out to be 202~mas, the position
angle $218\,.\!\!^\circ2$. The components of NSV\,16694 show
synchronous brightness fluctuations, suggesting that at least one
of them is a compact reflection nebula.

Among other problems solved using speckle interferometry at the
instrument is the refinement of the orbit of the binary asteroid
Kalliope-Linus \citep{Emelyanov2019}, refinement of
the orbit of the young binary ZZ\,Tau \citep{Belinski2022},
search for stellar components of exoplanets host stars
 identified by TESS spacecraft \citep{Cabot2021,Knudstrup2022}.

%\onecolumngrid
\twocolumngrid
\appendix
\section*{APPENDIX A}
\section*{THE INFLUENCE OF THE ROLLING SHUTTER EFFECT ON THE CONTRAST ESTIMATION}
\label{app:rolling}

The rolling shutter effect is a consequence of the
non-instantaneous sequential reading of detector rows. Even and odd
rows are read at the same time, but each pair of such rows takes
some time to read, and each next pair of rows is read with a
slight delay compared to the previous pair. In the standard and
ultra-quiet reading modes, the time spent on a pair of rows is
7.2~microseconds and 172.8~microseconds, respectively.

For example, taking the angular scale equal to
$0\,.\!\!^{\prime\prime}02/$px and the separation between two
stars equal to $1^{\prime\prime}$, we get that the image of one
component will lag behind from the image of the second one by
4.3~ms in the ultra-quiet reading mode if the stars are oriented
parallel to the reading direction. This value has the same order
of magnitude as the atmospheric coherence time. Therefore, it is
necessary to simulate the influence of this effect on the
efficiency of speckle interferometric contrast estimation.

To do this, we generated 20 series for three different separations
between the components by varying the reading time of a row pair.
For each series, standard speckle interferometric processing was
applied, from which contrast estimates were obtained. The
parameters of the model used to generate the series are as
follows:

 \begin{list}{}{
\setlength\leftmargin{7mm} \setlength\topsep{1mm}
\setlength\parsep{0mm} \setlength\itemsep{1mm} }
\item[$\bullet$] Separations between components are $0\,.\!\!^{\prime\prime}35$, $0\,.\!\!^{\prime\prime}7$, $1\,.\!\!^{\prime\prime}05$, the component flux ratio is 0.05. The stars are oriented in the direction of reading.
\item[$\bullet$] Two phase screens at distances of 0~m and 10\,000~m, moving in perpendicular directions.
\item[$\bullet$] Wind speed is 10~m\,s$^{-1}$.
\item[$\bullet$] Seeing is $1^{\prime\prime}$.
\item[$\bullet$] No telescope jitter.
\item[$\bullet$] Zero defocus.
\item[$\bullet$] Filter with an infinitely narrow bandwidth.
\item[$\bullet$] The considered wavelength is 822~nm.
\item[$\bullet$] Exposure time is 22~ms.
\item[$\bullet$] The interval between frames increases with the increase in the reading time.
\item[$\bullet$] Telescope aberrations are taken into account.
\item[$\bullet$] Photon noise and readout noise are taken into account. RMSD of the readout noise is 0.27$~e^{-}$. Conversion factor is $0.11~e^{-}/\mathrm{ADU}$.
\item[$\bullet$] Frame size is $256\times256$. Angular scale is   \linebreak $0\,.\!\!^{\prime\prime}0205$/px.
\item[$\bullet$] Sampling in time is equal to the time of reading a pair of lines.
\item[$\bullet$] 1000 frames were generated for each series.
\end{list}

%fig27
\renewcommand{\thefigure}{A.1}
\begin{figure}[]\vspace{0.3mm} %\setcaptionmargin{5mm} \onelinecaptionsfalse \captionstyle{normal} \renewcommand{\thefigure}{A.1}

\includegraphics[width=1.00\linewidth]{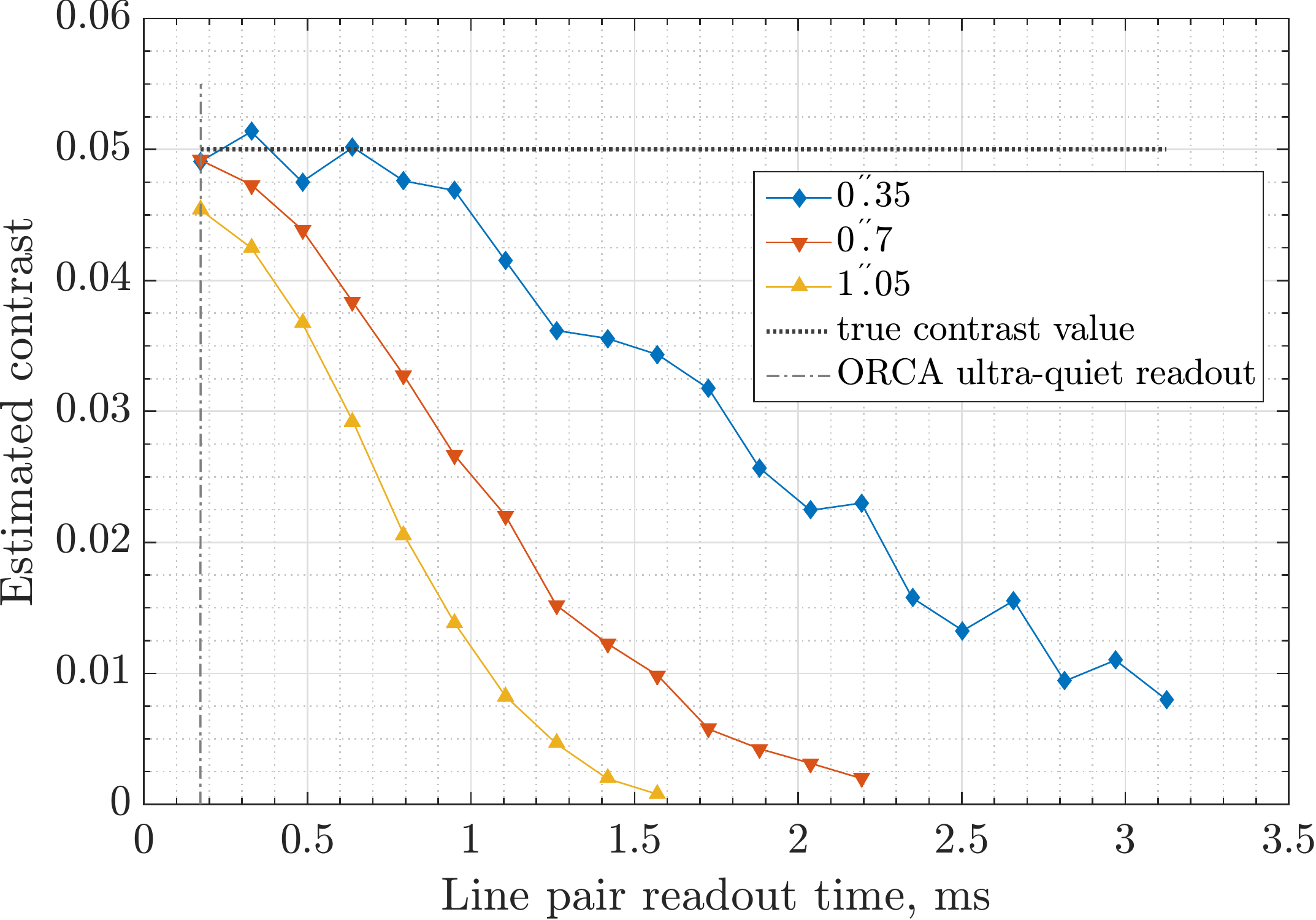}%{SpeckleCMOS_figs/rolling_shutter/rolling_rotime_seps_ed.eps}
\caption{
Estimation of contrast from processing of model series at
different reading times of a pair of lines and different
separations between components. The vertical dash-dotted line
corresponds to the detector's ultra-quiet reading mode.
}
\label{fig:26}%\vspace{-5mm}
\end{figure}
\renewcommand{\thefigure}{\arabic{figure}}

It can be seen from Fig.~\ref{fig:26} that the rolling shutter effect
clearly influences the speckle contrast estimation. The longer the
readout time and the greater the distance between the components,
the more the contrast is underestimated (secondary component appears fainter than it is). However, even if we assume a scenario where the separation between the components is
$1^{\prime\prime}$ and the ultra-quiet readout mode is used, then
 the underestimation of the contrast will be only
10\%. For larger separations, the contrast can be obtained
without using speckle interferometry by approximating the 
the averaged image.

\section*{APPENDIX B}
\section*{TELESCOPE JITTER}
\label{app:jitter}
In the course of comparison (see Fig.~\ref{fig:20}) of real and model power
spectra, it turned out that the high-frequency region of the model
spectra, other things being equal, is higher than in real
observations. We assumed that this excess is due to the fact that
the model does not take into account the vibrations of the
telescope. We tested this hypothesis in the following way.
In good seeing conditions ($\beta<1^{\prime\prime}$) on different
 dates at different altitudes and azimuths,
bright stars were observed with a low exposure and a high frame rate.
 The value of the electron multiplication was chosen
depending on the brightness of the object. Thus, according to our
estimates, the frame rate and exposure required to test the
hypothesis should be 500~Hz and 0.002~s, respectively.
%fig26

\renewcommand{\thefigure}{B.1}
\begin{figure}[t!] \vspace{1mm} %\setcaptionmargin{5mm} \onelinecaptionsfalse \captionstyle{normal}
\begin{minipage}[h]{\linewidth}
\includegraphics[width=1\linewidth]{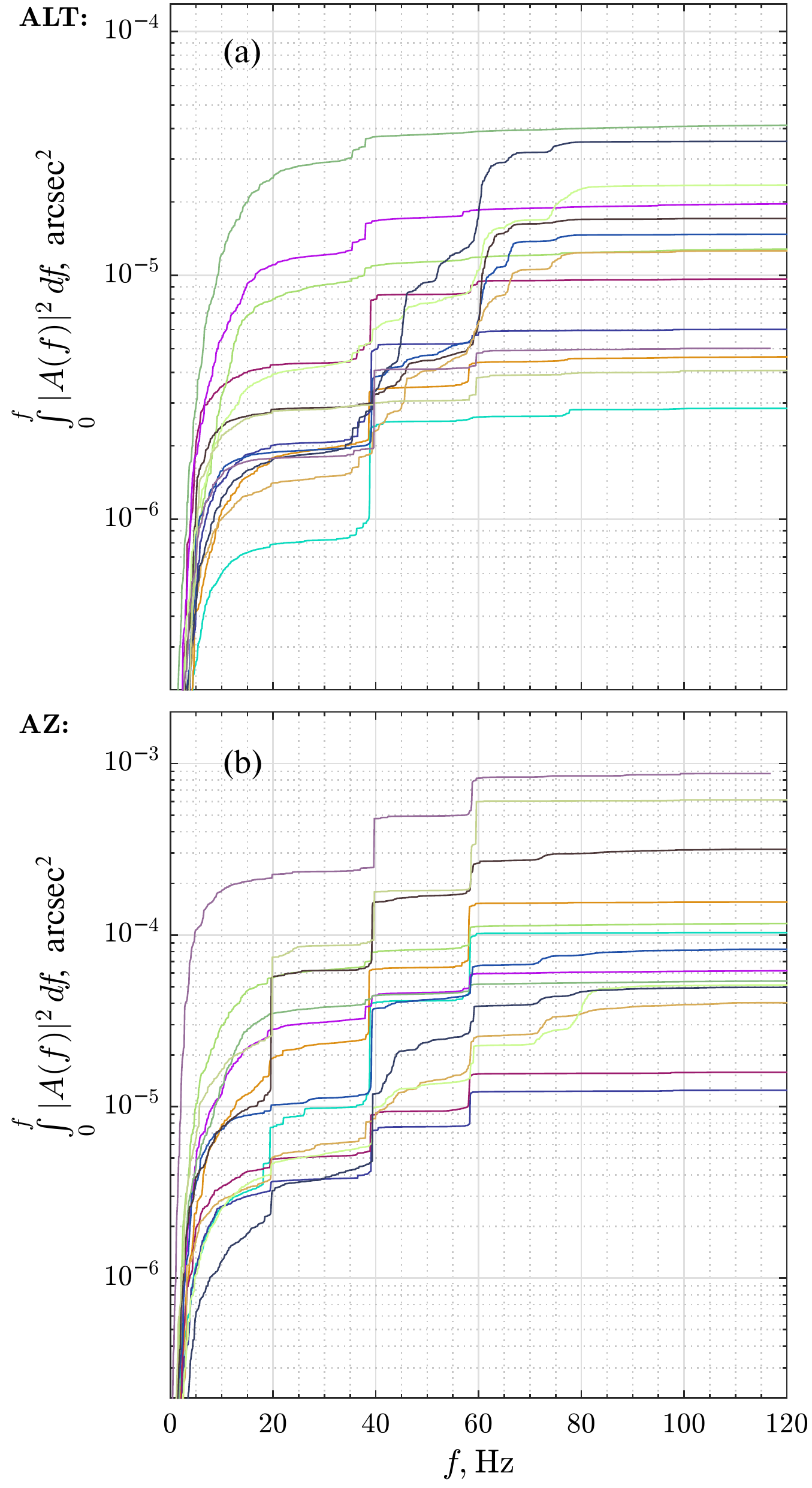}%
\end{minipage}
\vspace{-7pt}
\caption{
Cumulative power spectra (that is, the integral of the power
spectrum $|A(f)|^2$ over frequency) from zero to the frequency of
the telescope jitter in altitude and azimuth (in the plane of the
sky), units are the arc seconds squared. 14 graphs are
superimposed on each other for different series with different
conditions and exposures (from 1~ms to 4~ms).
}
\label{fig:27}
\end{figure}
\renewcommand{\thefigure}{\arabic{figure}}
The following processing was applied to the obtained series.
Cross-correlation of each ($i\!+\!1$)-th frame with $i$-th was
performed. From here we got the offset of each next frame relative
to the previous one, that is, the dependence of the offset on
time. Next, filtering was performed by moving average with a
window size of 0.1~s, that is, by 50~measurements. Then, the
filtered ones were subtracted from the original ones in order to
eliminate the offsets associated with atmospheric effects.

Spectral analysis of the data obtained (see \linebreak Fig.~\ref{fig:27}) shows that the vibration amplitude in azimuth is indeed greater than in altitude, and reaches $j_{\rm tel}\approx0\,.\!\!^{\prime\prime}08$. The graph is dominated by harmonics at frequencies of 20, 40, 60~Hz. No significant correlation of these amplitudes with wind speed was found. Moreover, the observations were made on four different days with different weather conditions.

\twocolumngrid

\begin{acknowledgments}

Employees of the Caucasian Mountain Observatory of the SAI MSU
facilitated the assembly, adjustment and use of the speckle
polarimeter on the \mbox{2.5-m} telescope. The authors are grateful to
Anastasia Fedotova and Anastasia Baluta for their help with
alignment of the speckle polarimeter. The
reviewer's comments improved the presentation of the results.

\end{acknowledgments}

\section*{FUNDING}

This work was supported by the RSF grant No. \mbox{20-72-10011} and the development program of \linebreak  Moscow State University.

\section*{CONFLICT OF INTEREST}
The authors declare no conflict of interest.

\bibliographystyle{aa}
\bibliography{Strakhov_nv_en}

\onecolumngrid
\begin{flushright}
{\it Translated by T.~Sokolova}
\end{flushright}
%\endinput
\end{document}